\documentstyle[prl,multicol,epsf,epsfig,aps]{revtex}
\newcommand{\BEQ}{\begin{equation}}
\newcommand{\EEQ}{\end{equation}}
\newcommand{\BEA}{\begin{eqnarray}}
\newcommand{\EEA}{\end{eqnarray}}
\newcommand{\nn}{\nonumber \\}
\newcommand{\xinfl}{{x_{\rm if}}}
\newcommand{\g}{\gamma\,}
\newcommand{\am}{{\alpha_{\rm m}}}
\newcommand{\amsq}{{\alpha_{\rm m}^2}}
\newcommand{\hc}{{k}}
\newcommand{\Del}{}
\renewcommand{\d}{{\rm d}}
\newcommand{\p}{\partial}
\newcommand{\eps}{\varepsilon}

\renewcommand{\O }{{\cal O}  }

\renewcommand{\H}{{\cal H}}
\newcommand{\K}{{\cal K}}
\newcommand{\N}{{\cal N}}
\newcommand{\W}{{\cal W}}
\newcommand{\Q}{{\cal Q}}

\newcommand{\minfty}{{-\infty}}

\newcommand{\V}{{\cal V}}
\newcommand{\taum}{{\sigma_{\rm min}}}
\newcommand{\wb}{\bar w}

\newcommand{\half}{\frac{1}{2}}
\renewcommand{\thesection}{\arabic{section}}

\renewcommand{\theequation}{\thesection\arabic{equation}}

\def\dbarrm {{\mathchar'26\mkern-11mu{\rm d}}}                       %
\begin{document} 
\draft
\title{Statistical thermodynamics of quantum Brownian motion:
\\ Birth of perpetuum mobile of the second kind}
\author{Th.M. Nieuwenhuizen$^{1)}$ and A.E. Allahverdyan$^{2,1,3)}$}
\address{$^{1)}$ Institute for Theoretical Physics,
University of Amsterdam,\\
Valckenierstraat 65, 1018 XE Amsterdam, The Netherlands; \\ 
$^{2)}$ S.Ph.T., CEA Saclay, 91191 Gif-sur-Yvette cedex, France;\\
$^{3)}$Yerevan Physics Institute,
Alikhanian Brothers St. 2, Yerevan 375036, Armenia. }
\date{\today}
\maketitle

\begin{abstract}

The Brownian motion of a quantum particle in a harmonic confining
potential and coupled to a harmonic quantum thermal bath is exactly solvable. 
Though this system presents at large temperature a pedagogic example to
explain the laws of thermodynamics, it is shown that at low enough 
temperatures the stationary state is non-Gibbsian due to an entanglement 
with the bath. In physical terms, this happens when the cloud of bath modes
around the particle starts to play a non-trivial role, namely when the
bath temperature $T$ is smaller than the coupling energy. 
Indeed, equilibrium thermodynamics of the total system,
particle plus bath, does not imply standard equilibrium 
thermodynamics for the particle itself at low $T$.

Various formulations of the second law are found to be invalid at low $T$. 
First, the Clausius inequality can be violated, because 
heat can be extracted from the zero point energy of the cloud of bath modes.
Second, when the width of the confining potential is suddenly changed, 
there occurs a relaxation to equilibrium during which
the entropy production is partly negative. In this process the
energy put on the particle does not relax monotonously,
but oscillates between particle and bath, 
even in the limit of strong damping. 
Third, for non-adiabatic changes of system parameters the rate of 
energy dissipation can be negative, and, out of equilibrium, cyclic
processes are possible which extract work from the bath.
Conditions are put forward under which perpetuum mobile of the 
second kind, having one or several work extraction cycles, 
enter the realm of condensed matter physics.
Fourth, it follows that the equivalence between different
formulations of the second law (e.g. those by Clausius and Thomson) 
can be violated at low temperatures.

These effects are the consequence of quantum entanglement in the presence 
of the slightly off-equilibrium nature of the thermal bath, and
become important when the characteristic quantum time scale $\hbar/k_BT$ 
is larger than or comparable to other timescale of the system.
They show that there is no general consensus between standard
thermodynamics and quantum mechanics. 
The known agreements occur only due to the weak coupling limit, 
which does not pertain to low temperatures.

Experimental setups for testing the effects are discussed.

\end{abstract}


\renewcommand{\thesection}{\arabic{section}}
\section{ Introduction}
\setcounter{equation}{0}\setcounter{figure}{0} 
\renewcommand{\thesection}{\arabic{section}.}

The faith in the laws of thermodynamics has been
strengthened time and again because numerous counterarguments 
and perpetuum mobile setups failed. It was summarized in the
classical statement of Arthur Eddington in 1948~\cite{Eddington}:
{\it ``The law that entropy always increases - 
the second law of thermodynamics - holds, I think, 
the supreme position among the laws of Nature. 
If someone points out to you that your pet theory of the
universe is in disagreement with Maxwell's equations - then so much
the worse for Maxwell's equations. If it is found to be contradicted
by observation, well, these experimentalists do bungle things
sometimes. But if your theory is found to be against the second law of
thermodynamics I can give you no hope; there is nothing for it but
collaps in deepest humiliation. }
Nevertheless, what we intend to do in this paper is to show 
that {\it several formulations of the second 
law need not apply to systems coupled to a bath in the quantum regime}. 
This paves the way for a new, modest definition of the most despised 
objects of modern physics, {\it perpetuum mobile of the second kind}.
We shall propose realizations wherein they can make a few or even many
cycles, though not infinitely many. A short version of the material 
appeared already~\cite{ANQBMprl}, which was discussed in the 
scientific press~\cite{AIP}~\cite{ScienceNews}.

The laws of equilibrium thermodynamics apply both to (quasi) 
closed quantum and classical systems, and to open classical subsystems 
\cite{landau}. This can all be traced back to the general character 
of the Gibbs distribution that describes the equilibrium state.
The same laws are {\it believed} to apply as well to 
open quantum subsystems. Our aim will be to show that,
though this belief is proper for weak coupling, it is not justified
for non-weak coupling between system and bath. 
Non-weak coupling means physically that a cloud of bath modes has been formed
around the particle, which we shall still consider as part of the bath.
Although one could defend the opinion that | as an example | 
for Kondo-type problems there is 
no need or no sense to describe the impurity spin (``the subsystem'') 
separately from the Kondo-cloud of conduction electrons (``the bath''), such
a viewpoint is not obvious for an array of mesoscopic Josephson junctions,
where the ``bath'' is the electro-magnetic field, to which they may be 
strongly coupled. Since that regime shows up in many systems at low enough 
temperature even for a small but fixed coupling, there is an important 
case to make.

When deriving the Gibbs distribution 
for a (quasi)closed system \cite{landau,klim}, the underlying 
statistical assumptions do not depend much on the quantum or 
classical nature. An open subsystem is usually considered as 
being in contact with an initially equilibrium bath. Under some
general statistical conditions
concerning the bath \cite{klim,ny,ford,weiss,gardiner}, 
which are again the same for the quantum and classical cases, 
one derives a Langevin equation. The general behavior of the 
classical Langevin equation
is well known \cite{klim}: The stationary distribution is 
Gibbsian, and, at least for the white noise case, the equivalent
Fokker-Planck equation is the main tool to describe the 
non-equilibrium statistical theory \cite{risken}.

Much less is known about the quantum Langevin equation
\cite{klim,ny,ullersma,leggett,weiss,gardiner,ka}, 
first proposed by Senitzky \cite{sen}
in the weakly-damped (weakly coupled) case, and in a more general form
by Ford, Kac and Mazur \cite{ford}. Its stationary distribution
has been obtained for the harmonic potential, in which case 
it depends explicitly on the coupling constant, becoming Gibbsian
only in the limit of weak coupling \cite{ka,klim-rev,gardiner}.

Let us recall that the situation with
a particle interacting with an equilibrium bath is known as 
Brownian motion, and the particle as well is referred to as a
Brownian particle.
As one of the paradigms of non-equilibrium statistical
mechanics, the theory of Brownian motion has numerous applications
in condensed matter physics
\cite{klim,weiss,risken,van,likho,leggett}, atomic physics
\cite{klim,atom}, quantum optics and chemistry \cite{gardiner,3}. 
It is believed that some of those practical realizations 
can be understood as being weakly-coupled with their thermal 
baths, and then standard methods of statistical physics can be applied
\cite{blum,gardiner}. However, there are nowadays well-known experimental 
situations, which are essentially far from the weak-coupling regime.
The main example is the case of weak links between superconductive
regions,  so-called Josephson junctions, in their overdamped regime 
\cite{likho,van}, where the relevant ranges of parameters 
were achieved already twenty years ago. Even in quantum optics, which has
often been satisfactorily described by weak-coupling  theories 
\cite{gardiner}, there are recent experiments showing the 
necessity of moderate and strong coupling approaches 
(see, e.g., \cite{oe}). In NMR experiments, on the other hand, very weak
coupling occurs, but it leads to a $T_2$ relaxation time of transverse
(non-classical) correlations, that can reach up to minutes. 
During that time related non-thermodynamic effects can occur.

The cause of the crucial differences between quantum and classical 
Brownian motion lies in quantum entanglement: A complete description 
in terms of a wave function is possible only for a closed system;
subsystems are necessarily in a mixed state. Since the quantum Gibbs 
distribution tends to the pure vacuum state for $T\to 0$, it is in 
that limit not an adequate candidate for the description of the
quantum subsystem non-weakly interacting with its thermal bath. 
Where typically researchers have just guessed that thermodynamics 
would apply anyhow, we shall provide the opposite answer 
by analyzing exactly solvable models.

Another important property is embodied in the structure
of the quantum Langevin equation \cite{klim-rev}: 
As predicted by the quantum fluctuation-dissipation theorem, but
in contrast to the classical case, the time-scales of
fluctuations and dissipation are different, and even in the limit of
instantaneous friction, the noise does not become white, but has 
the characteristic quantum timescale $\hbar /k_BT$.
During this period the noise has a memory and thus has the
possibility to cause non-statistical but ``mechanical'' or ``quenched'', 
in any case non-thermodynamic, behavior
of the system on which it acts~\cite{landau,ny}. 
The physical meaning of the classical 
fluctuation-dissipation theorem thus is stronger, since
it insists on equal timescales of friction and noise, which induces
the standard thermodynamic behavior.

The new properties become non-effective if the interaction with the thermal
bath is weak or if temperature is high, in which case the subsystem 
relaxes to its own quantum or classical Gibbs distribution. 
Both these cases will be referred to as the Gibbsian limit.
We recall that its existence is typically not connected with very low 
temperatures, since even for a small but generic coupling between
the system and the thermal bath, one always goes out of the weak
coupling limit by taking the temperature low enough. 

It is a crucial fact that a non-Gibbsian stationary state 
implies the inadequacy of equilibrium thermodynamics. In the present paper
we  propose a generalized thermodynamical description of a quantum
Brownian particle in a harmonic confining potential.
This description is self-consistent, and does not use {\it a priori} 
the concepts of the equilibrium (Gibbsian) statistical 
thermodynamics. Instead we employ the methods of nonequilibrium 
statistical thermodynamics developed recently for glasses
and applied to black holes, where a separation of timescales allows 
for a two-temperature approach~\cite{1,Nhammer,4,Nbh}.

The universal character of equilibrium thermodynamics led to the 
general expectation that in one way or another, thermodynamics
 will be applicable to the full quantum domain. 
A somewhat stronger point was expressed by Landau and 
Lifshitz\cite{landau}, namely that the proper formulation of 
equilibrium thermodynamics must be based on quantum mechanics. 
For the strongly coupling quantum situation one might, however, 
not be convinced. Let us give three principle arguments that
question standard wisdom:
\\ 1) The bath modes are correlated during the quantum 
timescale $\hbar/T$, even when damping is instantaneous. 
When this timescale is larger than the (largest) relaxation time of 
the system, the bath acts more like a quenched disorder than as 
a white noise.
Thus the standard condition for going from a Langevin equation to a 
Gibbs distribution is not fulfilled and new behavior should be expected. 
\\ 2) Assume that the overall system (the Brownian particle plus the bath)
is in equilibrium at a low temperature. One of the many formulations
of the second law states that no heat can be extracted from the bath.
This just follows from the Clausius inequality: 
$\dbarrm {\cal Q}\le T\d S$ (here $\dbarrm{\cal Q}$ is the heat flowing
from the bath to the subsystem, the Brownian particle, while $S$ is 
the entropy of the subsystem) for $T\to 0$. A naive argument in
support of this statement will be that the bath is close to its ground 
state, and therefore its energy cannot be lowered. However, this 
argument is invalidated by quantum entanglement: Since the bath is not an
isolated system itself, it cannot be in a definite energy eigenstate.
In particular, it cannot be in the ground state, and its energy fluctuates 
even for $T\to 0$. 
\\ 3) If a closed quantum system is its ground state, the only change
can be to do work on it. Now suppose that this system consists of 
a subsystem coupled to a bath, and that the external coupling 
connects to the subsystem, and not to the bath. Then typically the
action of doing work will reshuffle also the separate energies of 
the subsystem and the bath. As the direction of the exchange depends on
the question whether externally work was added or subtracted, in one of the 
cases the subsystem receives energy from the bath. 
Since this comes from the unobserved 
bath modes, it cannot be identified as work, and must be identified 
as heat extracted from the bath, even at $T=0$.

Because these arguments question standard wisdom, 
the only way to investigate the practical situation
is to start from first principles, namely from standard quantum mechanics. 
This is the general strategy of the present paper.

Statements on violations of certain formulations of the second law 
in the quantum micro world already appeared in literature. Capek and his
coworkers \cite{capek} noticed such effects in certain biophysically 
inspired models, and Nikulov \cite{Nikulov} reported on violations of
the second law in mesoscopic superconductivity. 
The latter author bases his view on results for permanent
currents in inhomogeneous superconducting rings
\cite{DuboKuznNiku}.

Since the subject of violating the second law and introducing
perpetuum mobile has  such a notorious history, 
new works in this field should be as convincing as possible.
Therefore we present now a somewhat extensive, but self-contained
exact analysis that leads to our conclusions, partly already
presented in~\cite{ANQBMprl}. 
Our methods are exact since the case of a quantum particle in a
harmonic confining potential and coupled to a bath of harmonic oscillators
with bilinear couplings is exactly solvable. Notice that
in a previous work ~\cite{ANQBMprl}  we also entered the discussion of 
the approximate solution for a weakly anharmonic force; this will
not be touched here.

Our paper is organized as follows. In section 2 we recall
the derivation of the quantum Langevin equation. 
In section 3 we solve the statics of the total system (subsystem plus bath)
by elementary techniques. In section 4 we show that the thermodynamics
of adiabatic changes can be described through two effective temperatures, 
and that analogies with the usual two temperature
thermodynamics can be stated much further:
The generalized relations will have the effective form of the 
first and second law. 
Next we first present details on the violation of the Clausius inequality. 
In section 5 we consider the dynamics of the system for the case where
the initial state is Gibbsian, and for which the spring constant 
is slightly modified at time zero. In section 6 we use those results
to derive the energy relaxation and the entropy production.
In section 7 we consider work done on the system for that situation of 
an instantaneously changed spring constant, and we also consider work 
for smooth, cyclic  changes. In section 8 we mention a number of experiments
where our results have applications. In section 9 we view those results
from the foundations of thermodynamics and the definition of perpetuum mobile.
Finally in section 10 we close with a discussion.

\renewcommand{\thesection}{\arabic{section}}
\section{The Quantum Langevin equation}
\setcounter{equation}{0}\setcounter{figure}{0} 
\renewcommand{\thesection}{\arabic{section}.}\label{langevin}

It is standard wisdom that the analysis of Brownian motion of
non-interacting particles may be restricted to a single 
Brownian particle. This insight goes back to Szilard in his somewhat
related work on Maxwell's demon~\cite{Szilard}, for a translation see
~\cite{rex}. In our analysis we shall also
make this simplification, but insist that the bath has many degrees of
freedom. Therefore it is equally equipped to couple to a gas of 
$N$ non-interacting Brownian particles, and our results for energy,
work, heat, entropy, etc. must just be multiplied by $N$. 
Because of this, our results yield, without say, the intrinsic 
variables of a large Brownian system in its thermodynamic limit. 

\subsection{The Hamiltonian}
\label{secthamiltonian}
The quantum Langevin equation
is derived from the exact Hamiltonian description of a subsystem
(Brownian particle) and a thermal bath, by tracing out the degrees 
of freedom of the bath. The influence of the particle on the bath is 
assumed to be sufficiently small. Thus, only the linear modes of the 
bath are excited, and the interaction of the particle
with the bath is assumed to be linear. To be as pedagogic as possible,
we first take a definite model for the bath, namely a collection of 
harmonic oscillators; later we will relax this assumption. 
For the total Hamiltonian we thus assume~\cite{weiss}
\BEA
\label{hamiltonian}
\H_{\rm tot}&=&\H+\H_B+\H_I\nn
\H&=&\K(p)+\V(x),\qquad \K(p)=\frac{p^2}{2m},\qquad \V(x)=\half a x^2\\
\H_B&=&\sum_{i}\left [
\frac{p_i^2}{2m_i}+\frac{m_i\omega_i^2}{2}x_i^2\right]\nn
\H_I&=&\sum_i\left[-c_ix_ix+\frac{c_i^2}{2m_i\omega_i^2}\,x^2\right ],
\nonumber
\EEA
where $\H$ is the Hamiltonian of the particle, consisting of 
the kinetic and potential energies $\K$ and $\V$, 
$\H_B$ is the Hamiltonian of the bath, 
and $\H_I$ is the interaction Hamiltonian.
$p$, $p_i$, $x$, $x_i$ are the momenta and coordinate operators 
of the particle and the linear modes of the bath. 
$\V(x)$ is the confining potential of the particle, 
and $m$ and $m_i$ are the corresponding masses. 

Notice that our total system is closed and its energy is conserved, 
except for the periods where work is done on it by externally changing a 
system parameter such as $m$ or $a$.
When we later on take as initial density matrix the Gibbs distribution
$\exp(-\beta \H_{\rm tot})/Z$, this still refers to our closed system;
in particular it is not part of a larger thermodynamic 
system, with which heat exchange would be possible~\cite{superbath}.

The $x_i$-terms of $\H_B+\H_I$ form a complete square,
since $\H_I$ includes a self-interaction term proportional to $x^2$.
This guarantees that 
the total Hamiltonian $\H_{\rm tot}$ will be positive definite.
In certain physical situations (e.g. electromagnetic interaction), 
such a term is generated automatically by the coupling\cite{leggett}.
Indeed, under a canonical transformation:
\BEA
\label{kush}
x_i\to \frac{p_i}{m_i\omega _i}, \qquad
p_i\to - x_im_i\omega _i
\EEA
$\H_{\rm tot}$ becomes 
\begin{equation}
\label{kush2}
\H_{\rm tot}'=\frac{p^2}{2m}+\half a x^2+
\sum_{i}\left [
\frac{1}{2m_i}(p_i - \frac{c_i}{\omega _i}x)^2
+\frac{m_i\omega_i^2}{2}x_i^2
\right],
\end{equation}
which corresponds to the minimal coupling (or subtraction) scheme. 

In other situations (such as certain cases in nuclear and atomic 
physics, see \cite{leggett} for more details) the self-interaction
term is absent, and one has ~\cite{ullersma}
\BEA
\label{hamiltonian'}
\tilde \H=\frac{p^2}{2m}+\tilde \V(x),\qquad 
\tilde \H_I=-x\sum_i c_ix_i
\EEA
while $\H_B$ is unchanged. For a harmonic system one will have 
$\tilde \V(x)=\half bx^2$. In general, the potential energy 
will have a minimum only if $b\equiv \tilde\V''(0)$ is large enough. 
This happens when the combination 
\BEQ a=b- \sum_i \frac{c_i^2}{m_i\omega _i^2}\EEQ
is positive.
In the case we shall consider below, with the $c_i$ given by
(\ref{omi=iom}), this sum can be evaluated,
leading to  $a=b-\gamma\Gamma$, where $\gamma$ is the coupling
strength (damping constant) and $\Gamma$ is a large cut-off frequency. 
So this system can be mapped on previous one provided we define
$\tilde \V(x)=\V(x)+\half\gamma\Gamma x^2$.
In doing so we identify with Hamiltonian of the subsystem
the expression $\H=\K+\V$ of (\ref{hamiltonian}), rather than 
$\K+\tilde \V$, and with $\H_I$ the combination 
$\tilde\H_I+\half\gamma\Gamma x^2$.
To give some motivation for this, let us notice that
we shall take  $\Gamma$ large and $\gamma$ finite. In that case
the expectation value $\langle \tilde \V\rangle$ is large, 
proportional  to $\gamma\Gamma$, but this is almost fully compensated by 
an opposite term arising from $\langle\tilde \H_I\rangle$. 
These cancellations have been accounted for by the mapping to $\H$
and $\H_I$, leaving at most a $\ln \Gamma$ divergence 
for large $\Gamma$, which actually arises at small temperatures 
from  $\langle \tilde K\rangle$. We shall come back to pro's 
and contra's of this identification in section ~\ref{procontra},
where we notice that it is already needed to obtain the standard
thermodynamics at very large $T$.

Some word of nomenclature is called for. The case of a harmonic
potential ${\cal V}(x)=\half a x^2$ is often called ``the linear case'' in
literature, of course referring to its linear force. 
The expressions ``linear potential'' and 
``non-linear potential'', that are sometimes found in literature, are
misnomina, and will be avoided by us.

\subsection{Derivation of general quantum Langevin equation}

The operator equations of motion for the bath modes read
\begin{eqnarray}
\label{vavilon1}
&& \dot{x}_i= \frac{1}{m_i}p_i,
\\
&& \dot{p}_i= - x_im_i\omega _i^2 +c_i x
\label{vavilon2}
\end{eqnarray}
After introducing the creation and annihilation operators by
\begin{eqnarray}
\label{vavilon3}
x_i= \sqrt{\frac{\hbar }{2m_i\omega _i}} (a_i+a^{\dagger}_i),
&& \qquad p_i= i\sqrt{\frac{\hbar m_i\omega _i }{2}} (a^{\dagger}_i-a_i)
\label{vavilon4}
\end{eqnarray}
we can write Eqs.~(\ref{vavilon1}, \ref{vavilon2}) in the form
\BEA
\dot{a}_i(t) = -i\omega _i a_i(t)
+i\sqrt{\frac{c_i^2}{2\hbar m_i\omega _i }} x(t).
\EEA
This equation is solved readily:
\BEA
a_i(t) = e^{-i\omega _it} a_i(0)
+i\sqrt{\frac{c_i^2}{2\hbar m_i\omega _i }} \int _{0}^{t}\d s 
e^{-i\omega _i(t-s)}x(s),
\label{chairman-mao}
\EEA
yielding
\BEA \label{xit=}
x_i(t)&=&x_{i}(0)\cos\omega_i t+\frac{p_{i}(0)}{m_i\omega_i}\sin\omega_it +
\frac{c_i}{m_i\omega_i}\int_0^t\d s\,\sin\omega_i(t-s)x(s)\\
\label{pit=}
p_i(t)&=&-m_i\omega_i x_{i}(0)\sin\omega_i t+
p_{i}(0)\cos \omega_it +
{c_i}\int_0^t\d s\,\cos\omega_i(t-s)x(s)\EEA

The Heisenberg equations of motion for the Brownian particle read
\begin{eqnarray}
\label{vavilon6}
&& \dot{x}= \frac{1}{m}p,
\\
&& \dot{p}= - ax +\sum _ic_i x_i - x\sum_i\frac{c^2_i}{m_i\omega ^2_i}
\label{vavilon7}
\end{eqnarray}
Combined with Eq.~(\ref{xit=}) the last equation becomes
\BEQ\label{xddm} 
m\ddot x(t)=-ax(t)+\eta(t)-\int_0^t\d t'\dot
\gamma(t-t')x(t')-\gamma\Gamma x(t)\EEQ
where
\BEQ \label{etat=}\eta(t)=\sum_i
c_i[x_{i}(0)\cos\omega_it+\frac{p_{i}(0)}{m_i\omega_i}
\sin\omega_i t]=
\sum_i\sqrt{\frac{\hbar c^2_i}{2 m_i\omega _i}}[a^{\dagger}_i(0)e^{i\omega _it}
+a_i(0)e^{-i\omega _it}],
\EEQ
\BEQ \label{hh}\gamma (t)=\sum_i\frac{c^2_i}{m_i\omega ^2_i}
\cos (\omega_i t),
\EEQ
are the noise related to the {\it unperturbed} bath,
and the friction kernel, respectively. 
A partial integration brings
\BEA\label{pdot=}
m\ddot x(t)= - ax(t) - x(0)\gamma (t)-
\int _{0}^{t}\d s \gamma (t-s)\dot{x}(s)+\eta (t),
\EEA
Notice that in this derivation
the back-reaction of the bath on the particle has been taken into account
in an exact manner. 
It is described by the integrals in eqs. (\ref{xit=}), (\ref{pit=}), 
and brings the damping terms $x(t)\gamma(0)-x(0)\gamma (t)-
\int _{0}^{t}\d s \gamma (t-s)\dot{x}(s)$ in going from
eq. (\ref{vavilon7}) to (\ref{pdot=}).

\subsection{Drude-Ullersma spectrum}

For some, but not all, of our applications it is benefitable to
consider a fully explicit case for the bath.
The bath is assumed to have uniformly spaced modes 
\BEQ \label{omi=iom}
\omega_i=i\Delta\qquad i=1,2,3,\cdots \EEQ 
and for the couplings we choose the Drude-Ullersma 
spectrum \cite{weiss,ullersma}
\BEQ \label{ci=}
c_i=\sqrt{\frac{2\gamma m_i\omega_i^2\Delta}{\pi}\,
\frac{\Gamma^2}{\omega _i^2+\Gamma^2}}
\EEQ 
Here $\Gamma$ is the characteristic Debye cutoff frequency of the bath,
and $\gamma$ stands for the coupling constant; it has dimension 
$kg/ s$. Our parameter $\gamma$,
related to another one ($\gamma'=\gamma/m$) sometimes employed, 
see, e.g., \cite{weiss}, allows to consider changes in the effective 
mass $m$ at fixed coupling to the bath.

The thermodynamic limit for the bath is taken  by sending 
$\Delta\to 0$, which induces relaxational behavior.
As usual, the ``Heisenberg'' timescale $1/\Delta$ will be extremely large, 
implying that in the remaining approach the limit of ``large times''
always means the quasi-stationary
non-equilibrium state where time is still much less than $1/\Delta$. 
In the limit $\Delta\to 0$ each coupling $c_i\sim\sqrt{\Delta}$
is very weak. The fact that the bath has many modes nevertheless
induces its non-trivial influence. 
At finite but small $\Delta$ the system would have an initial relaxational 
behavior, which at times of order $1/\Delta$ is changed in a 
recurrent behavior.

It is customary to define the spectral density
\BEQ J(\omega)=\frac{\pi}{2}\sum_i \frac{c_i^2}{m_i\omega_i}
\delta(\omega-\omega_i)=
\frac{\gamma \omega\Gamma^2}{\omega^2+\Gamma^2} \EEQ
It has the Ohmic behavior $ J\approx\gamma\omega$ for
$\omega\ll\Gamma$,
and $\gamma$ is called the interaction strength or damping constant. 
As $J(\omega)$ is cut off at the ``Debye'' frequency $\Gamma$, 
it is called a quasi-Ohmic spectrum. 

For many applications only the spectral density needs to be defined.
However, for our further calculations
it is advantageous to stick to the fully specified linear bath,
with its frequencies (\ref{omi=iom}) and couplings (\ref{ci=}).
It can then be shown that the friction kernel (\ref{hh}) becomes 
\BEQ \label{cisum=}
\gamma (t)=\frac{2\gamma }{\pi}\,
\int_0^\infty \d\omega\,
\frac{\Gamma^2}{\omega^2+\Gamma^2}\,\cos\omega t=
\gamma\Gamma\,e^{-\Gamma \,|\,t|}
\EEQ
It is non-local in time, but on timescales much larger than $1/\Gamma$
it may be replaced by $\gamma\delta_+(t)$.

 Finally we wish to mention that there are alternative ways
to derive the quantum Langevin equation \cite{weiss}, since
many of its properties are rigidly determined by general statements
like the quantum fluctuation-dissipation theorem \cite{landau}.
Nevertheless, we choose to focus on concrete models, 
because they show in detail how the quantum Langevin 
equation arises from first principles, and thus are better 
suited for pedagogical purposes.

\renewcommand{\thesection}{\arabic{section}}
\section{Gibbsian state for a harmonic confining potential}
\renewcommand{\thesection}{\arabic{section}.}
\setcounter{equation}{0}

The case of an oscillator subject to a parabolic confining potential 
is a celebrated exactly solvable problem in quantum mechanics,
\BEQ \label
{Hpq=}
\H(p,x)={\cal K}(p)+{\cal V}(x)=\frac{p^2}{2m}+\half ax^2 \EEQ 
The eigenfrequency is already know from the classical treatment,
\BEQ \omega_0=\sqrt{\frac{a}{m}} 
\label{omega0=} \EEQ
When this oscillator is coupled to an oscillator bath with bilinear
coupling, as was done in Eq. (\ref{hamiltonian}),
the problem remains exactly solvable. It is in the true sense 
``the harmonic oscillator model'' for quantum Brownian motion.

It is well known that, besides its direct physical meaning, the harmonic 
oscillator can be interpreted as an LC circuit \cite{klim-rev}.
Then $x$ may correspond to the charge $Q$ on a capacitor, $1/a$ to its
capacitance $C$, $m$ to an inductance $L$, $p$ to a flux $L\dot{Q}$,
$\gamma $ to a resistance $R$, and $\eta (t)$ to a random 
electro-motoric force.
Although we will not use this language explicitly, it is
useful to keep it in mind, especially when considering 
variations of parameters. Indeed, in this setup there should be
nothing very difficult in varying $L$ or $C$, in our 
notation $m$ and $a$.

The theory of the dissipative harmonic oscillator
is considered in many works (see \cite{weiss} and Refs. there, as
well as a recent work for the driven case \cite{zerbe}). We will now be
primarily interested in thermodynamical aspects of this problem.

\subsection{Shift of the bath frequencies due to coupling with 
the central particle}

In Fourier space the equation of motion of the particle may 
be written as
\begin{equation} (a -m\omega^2+\sum_i\frac{c_i^2}{m_i\omega_i^2})x
=\sum_i c_i x_i,
\end{equation}
and for the bath
\begin{equation} (-m_i\omega^2+m_i\omega_i^2)x_i=c_ix
\end{equation}

From these relations one derives a condition for 
the eigenfrequencies $\nu_k$
\BEA
\label{jsum} 
\frac{a}{m}-\nu^2&=&{\nu^2}
\sum_{i\ge 1}\frac{c_i^2}{m\,m_i\omega_i^2(\omega_i^2-\nu^2)}
=\frac{2\g \Gamma^2\nu^2}{\pi }\sum_{i\ge 1}\frac{\Delta}
{(\omega_i^2+\Gamma^2)(\omega_i^2-\nu^2)}\nn\\
&=&-\frac{2\g\Gamma^2\nu^2}{\pi (\nu^2+\Gamma^2)}
\sum_{i\ge 1}\frac{\Delta}{\omega_i^2+\Gamma^2}
+\frac{\g\Gamma^2\nu}{\pi (\nu^2+\Gamma^2)}
\sum_{i\ge 1}[\frac{\Delta}{\omega_i-\nu}-\frac{\Delta}{\omega_i+\nu}]
\EEA
where we inserted the definition (\ref{ci=}) of the $c_i$.
The first sum may be replaced by an integral, while the second
can be carried out exactly,
\BEQ
\sum_{i=1}^\infty
[\frac{\Delta}{\omega_i-\nu}-\frac{\Delta}{\omega_i+\nu}]
=\lim_{N\to\infty}\left[
\psi(N-\frac{\nu}{\Delta})-\psi(1-\frac{\nu}{\Delta})
-\psi(N+\frac{\nu}{\Delta})+\psi(1+\frac{\nu}{\Delta})\right]
=\frac{\Delta}{\nu}-\pi \cot\frac{\pi \nu}{\Delta}\EEQ
where $\psi(z)=\d\ln\Gamma(z)/\d z$ is the di-Gamma function and
we used
\BEQ \psi(z+1)=\psi(z)+\frac{1}{z},\qquad 
\psi(1-z)=\psi(z)+\pi \cot \pi z
\EEQ 
The eigenfrequencies $\nu_k$ of the coupled system thus follow as
the roots of
\BEQ \label{E7}
\cot\frac{\pi\nu}{\Delta}-\frac{\Delta}{\pi\nu}=
-\frac{(a/m-\nu^2)(\nu^2+\Gamma^2)+\g\Gamma\nu^2}
{\g\Gamma^2\nu}
\EEQ

The transcendental equation has 
no solution for $0\le \nu\le \Delta$.
For $\nu>\Delta$ there is one solution in each period of 
the cotangens , except for the period that contains the point 
$\nu=\omega_0\equiv\sqrt{a/m}$, where there occur either 
three solutions or one. One can then check that in the limit of 
vanishing coupling $\gamma\to 0$,
there occur the modes $\omega_i=i\Delta$ ($i=1,2,\cdots$),
and $\omega_0$.  Notice, however, that this behavior
only pertains in the regime of infinitesimal coupling
$\gamma<\gamma_c$ with $\gamma_c\sim m\Delta$. 
For $\gamma\ge \gamma_c$, however, the interval
containing $\omega_0$ has only one solution, so 
$\omega_0$ is lost as a separate mode, 
its influence being taken by a shift of neighboring modes.

For finite $\gamma$ the solution of (\ref{E7}) shows that
the bath modes $\omega_k\gg\Delta$ now get shifted to
\BEQ 
\label{phi=}
\nu_k=k\Delta-\frac{1}{\pi}\phi(k\Delta)\Delta
=\omega_k-\frac{1}{\pi}\phi(\omega_k)\Delta\EEQ 
where 
\BEQ \phi(\nu)= 
\arctan\frac{\gamma\Gamma^2\nu}{(a-m\nu^2)(\nu^2+\Gamma^2)
+\gamma\Gamma\nu^2}\EEQ
Here the definition of the arctan is such that $\phi$
goes monotonously from $\phi(0)=0$ to $\phi(\infty)=\pi$.
We shall need 
\BEA \label{sinphi=} \sin\phi  (\nu)&=&
\frac{\gamma  \Gamma^2\nu}
{\{[{(a-m\nu^2)(\nu^2+\Gamma^2)+\gamma\Gamma\nu^2}]^2
+(\gamma\Gamma^2\nu)^2\}^{1/2}} \nn
&\approx&\frac{\gamma \nu}
{\{(a-m\nu^2)^2
+\gamma^2\nu^2 \}^{1/2}}
\EEA
where the approximation holds for large $\Gamma$.

\subsection{The Gibbsian state of the particle and its bath}

The steps of last section
allow to calculate the Gibbs free energy of the total system,
\BEA \beta F_{\rm tot}(T,\gamma)&=&
\sum_k\ln 2\sinh\half\beta\hbar\nu_k \EEA
For small $\Delta$ one may use the identity
\BEQ \label{sumint} \sum_{k=1}^\infty\, A(\nu_k)\,=\frac{1}{\Delta}\,
\int_0^\infty\d \omega_k\, A(\nu_k)
=\frac{1}{\Delta}\,\int_0^\infty\d \nu_k\,
\frac{\d \omega_k}{\d\nu_k}\, A(\nu_k)
=\int_0^\infty\d \nu\,[\frac{1}{\Delta}+\frac{1}{\pi}\,
\frac{\d\phi(\nu)}{\d \nu}]\, A(\nu)+{\cal O}(\Delta) \EEQ
and one gets
\BEA
\beta F_{\rm tot}(T,\gamma) &=&\beta F_B(T,\gamma=0)
+\beta F_p(a,\gamma,\Gamma,m,T)
\EEA
where the first term is the free energy of the bath in absence
of the particle. Neglecting its divergent zero point energy one gets
\BEA \beta F_B(T,\gamma=0)=\frac{1}{\Delta}
 \int_0^\infty\d\omega\ln(1-e^{-\beta\hbar\omega})
= -\frac{\pi^2T}{6\hbar\Delta}\label{comrad}\EEA
It is of order $1/\Delta$, showing the extensivity of the bath, and
implies the energy
\BEQ \label{Ubath=}
U_B(T,\gamma=0)=\frac{\pi^2}{6\hbar\Delta}\,T^2\EEQ 
and the linear specific heat and entropy
\BEQ \label{CB=SB}
C_B(T,\gamma=0)=S_B(T,\gamma=0)=\frac{\pi^2T}{3\hbar\Delta}\EEQ 

The free energy shift due to the central particle, its coupling to 
the bath and the resulting disturbance of the bath, follows
from Eq. (\ref{sumint}) as
\BEA 
\label{Fpeps=}
\beta F_p&=&\frac{1}{\pi} \int_0^\infty \d\nu\,
\ln[2\sinh\,\half\beta\hbar\nu]\,\frac{\d\phi}{\d \nu}
\\ &=&
\frac{\gamma\Gamma^2}{\pi}\int_0^\infty \d\nu
\,\ln[2\sinh \half\beta\hbar\nu]\,
\frac{a\Gamma^2+(m\Gamma^2+\gamma\Gamma-a)\nu^2
+3m\nu^4}{[(a-m\nu^2)(\nu^2+\Gamma^2)
+\gamma\Gamma\nu^2]^2+\gamma^2\Gamma^4\nu^2}
\label{Fpint=}\EEA
A useful identity is 
\BEA \beta F_p&=&\ln\,2\sinh (\half\beta\hbar\nu_1)
+\half\beta\hbar\int_0^\infty \d\nu\,
\,[\theta( \nu-\nu_1)-\frac{1}{\pi}\phi(\nu)]{\rm cotanh}\,
 (\half\beta\hbar\nu)\EEA
where $\nu_1$ is arbitrary and $\theta$ is the Heaviside 
step function. 

\subsubsection{Intermezzo: the characteristic frequencies of 
the damped oscillator}

The present model for a damped harmonic oscillator has three
characteristic frequencies, that do not depend on temperature. 
They just follow from the linear equations of 
motion, and thus have the same value at high and low temperatures.

The denominator in (\ref{Fpint=}) is a fourth order polynomial in $\nu^2$.
It decomposes as 
\BEQ m^2(\nu^2+\Gamma^2)P_3(i\nu)P_3(-i\nu) \label{P3s=}\EEQ
where
\BEA \label{P3=}
P_3(s)&=&s^3-\Gamma s^2+(\omega_0^2+\frac{\gamma\Gamma}{m})s
-\omega_0^2\Gamma=(s-\Gamma)(s^2+\omega_0^2)+\frac{\gamma\Gamma}{m}\,s
\EEA

The roots $s=\omega_{1,2,3}$ of $P_3(s)$ satisfy the relations
\BEA
&& \omega_1+\omega_2+\omega_3=\Gamma \label{rel1}\\
&& \omega_1\omega_2+\omega_2\omega_3+\omega_3\omega_1
=\omega_0^2+\frac{\gamma\Gamma}{m} \label{rel2}\\
&& \omega_1\omega_2\omega_3=\Gamma\omega_0^2 
\label{brut0}
\EEA

Two different situations can arise: Either all three roots are real 
(this is the case in the overdamped regime), or, in the underdamped regime, 
two of them are complex conjugate: 
$\omega _1^*=\omega _2$, whereas $\omega _3$ is real. In both cases
one has ${\rm Re}~\omega _{1,2,3}>0$, which indicates that with time 
the particle relaxes toward a stationary state.

For small $\gamma $ these roots read:
\BEA
\label{brut1}
\omega _{1,2}= \pm i \omega _0 +\frac{\gamma}{m}~
\frac{\Gamma}{2(\Gamma \mp i\omega _0 )}+ 
\left (\frac{\gamma}{m}\right )^2
\frac{\Gamma ^2(\Gamma \pm i\omega _0)}{8\omega _0(\Gamma\mp i\omega _0 )^3}
\EEA
\BEA
\omega _{3}=\Gamma  - \frac{\gamma \Gamma ^2}{m(\Gamma ^2+\omega _0^2)}
-\left (\frac{\gamma}{m}\right )^2
\frac{\Gamma ^3(\Gamma ^2-\omega _0^2)}{(\Gamma ^2+\omega _0^2)^3}
\label{brut2}
\EEA

On the other hand,
for a large $\Gamma$ one gets 
\BEA
\label{mega1Gam}
&&\omega _{1,2}=\frac{\gamma}{2m}
\left (1\pm \sqrt{1-\frac{4am}{\gamma ^2}}\,\, \right )
+\frac{1}{2\Gamma}\left (\frac{\gamma }{m}\right )^2
\left[ 1\pm \frac{1-2am/\gamma ^2}{\sqrt{1-4am/\gamma ^2}}
\right ] \\
&& \omega _3=\Gamma -\frac{\gamma }{m}-\frac{1}{\Gamma}
\left (\frac{\gamma }{m}\right )^2
\label{mega2}
\EEA
We shall only need them to leading order in $1/\Gamma$ 
\BEA
\label{mega1}
&&\omega _{1}=\frac{\gamma(1-w)}{2m}=\frac{2a}{\gamma(1+w)},
\qquad 
\omega_2=\frac{\gamma(1+w)}{2m},\qquad \omega _3=\Gamma -\frac{\gamma }{m}
\EEA
where we denoted 
\BEQ \label{eps=w=}\eps=\frac{am}{\gamma^2},\qquad w=\sqrt{1-4\eps}\EEQ
Later on we shall need
\BEQ \label{dwda} a\frac{\d w}{\d a}=m\frac{\d w}{\d m}
=-\frac{1-w^2}{2w},\qquad
 a\frac{\d \omega_{1,2}}{\d a}=\mp\frac{\gamma(1-w^2)}{4mw},
\qquad  m\frac{\d \omega_{1,2}}{\d m}=\pm\frac{\gamma}{4mw}(1\mp w)^2
\EEQ

For overdamping $(\eps<\frac{1}{4})$ $w$ is real positive.
Our interest is in particular the strong damping regime $\gamma^2\gg am$,
where
\BEA
\label{mega12}
&&\omega _{1}=\frac{a}{\gamma},\qquad 
\omega_2=\frac{\gamma}{m}(1-\frac{am}{\gamma^2}),\qquad
\omega _{3}= \Gamma-\frac{\gamma}{m}\EEA
and the approximations hold to leading order in $\eps$.

Already in the classical regime our system has
 three characteristic relaxation times:  for the
coordinate, for the momentum and for the noise. For large $\Gamma$ and 
$\gamma$ they are well separated

\BEQ\label{tauxpe=} \tau_x=\frac{1}{\omega_1}\approx\frac{\gamma}{a}\,\gg\,
\tau_p=\frac{1}{\omega_2}\approx\frac{m}{\gamma}\,\,
\gg \,\tau_\eta=\frac{1}{\Gamma}
\EEQ
In the quantum regime the quantum time-scale
\BEQ \label{tquant=}\tau_\hbar=\frac{\hbar}{T}\EEQ
is comparable to or larger than $\tau_x$, inducing quantum coherence
effects of the noise and thus new physics.

In case of underdamping, $(\eps>\frac{1}{4})$ one has $w=i\bar w$,
with 
\BEQ \bar w=\sqrt{4\eps-1}, \qquad 
\omega_{1,2}=\frac{\gamma(1\mp i\bar w)}{2m}\EEQ
This leads to the renormalized oscillation time $\tau_0$ and 
the damping time $\tau_d$
\BEQ\label{tau0taud} 
\tau_0=\frac{1}{\sqrt{\omega_0^2-\gamma^2/4m^2}},
\qquad \tau_d=\frac{2m}{\gamma}\EEQ

Since $\tau_d$ differs from $\tau_p$ by a factor of order unity,
we may skip the latter and use $tau_p$ and $\tau_0$ as the relevant
timescales in the underdamped regime.

It is worth to mention that the weak-coupling limit commutes with 
the quasi-Ohmic limit, in the sense that taking large $\Gamma $
in Eqs.~(\ref{brut1}, \ref{brut2}) we get the same main term and 
at least the first correction as having taken small $\gamma $ limit
in Eqs.~(\ref{mega1}, \ref{mega2}). 

\subsubsection{Continuing the main argument for the Gibbsian state}

In order to calculate the free energy (\ref{Fpint=}),
we shall first determine the following integral
\begin{equation}
\label{c1}
I(a,A,B)=
\int_0^{\infty}\frac{\d \nu ~\nu ~{\rm coth}(\half a\nu) }
{(A^2+\nu ^2)(B^2+\nu ^2)},
\end{equation}
We can write
\begin{eqnarray}
\label{c2}
&&I(a,A,B)=
\int_0^{\infty}\frac{\d \nu \nu }
{(A^2+\nu ^2)(B^2+\nu ^2)}+
2\int_0^{\infty}\frac{\d \nu \nu }
{(e^{a\nu }-1)(A^2+\nu ^2)(B^2+\nu ^2)}\nonumber \\
&&=\frac{1}{B^2-A^2}\ln \frac{B}{A}+
\frac{2}{B^2-A^2}\left [
\int_0^{\infty}\frac{\d \nu \nu }
{(e^{a\nu }-1)(A^2+\nu ^2)}-
\int_0^{\infty}\frac{\d \nu \nu }
{(e^{a\nu }-1)(B^2+\nu ^2)}
\right ]\nonumber \\
&&=
\frac{1}{B^2-A^2}\left [
\psi\left (\frac{aB}{2\pi}\right)-\psi\left(\frac{aA}{2\pi} \right )
\right ]-\frac{\pi}{a}\frac{1}{AB(A+B)},
\end{eqnarray}
where we used the known formula
\begin{equation}
\label{c3}
2\int_0^{\infty}\frac{t\d t}{(\exp (2\pi t)-1)(t^2+z^2)}=\ln z -
\psi (z)-\frac{1}{2z}
\end{equation}
 By integration we obtain

\BEA \label{JaAB=}
J(a,A,B)&=&\int_0^{\infty}\frac{\d \nu}{\pi}
{ ~\ln(2\sinh\half a\nu) }\left(
\frac{1}{A^2+\nu ^2}-\frac{1}{B^2+\nu ^2}\right)\nn
&=&-\frac{1}{A}\ln \Gamma\left (\frac{aA}{2\pi} \right )
   -\frac{1}{2A}\ln\frac{aA}{4\pi^2}
   +\frac{1}{B}\ln \Gamma\left (\frac{aB}{2\pi} \right )
   +\frac{1}{2B}\ln \frac{aB}{4\pi^2}
\EEA

In terms of the roots $\omega_i$, we may write
Eq. (\ref{phi=}) as
\BEQ \phi(\nu)=
\arctan \frac{\nu}{\omega_1}+
\arctan \frac{\nu}{\omega_2}+
\arctan \frac{\nu}{\omega_3}-
\arctan \frac{\nu}{\Gamma}\EEQ
The derivation follows immediately after using (\ref{P3s=}) 
with $P_3(s)=(s-\omega_1)(s-\omega_2)(s-\omega_3)$ and  expressing the 
$\arctan$ in  logarithm's. The integral in Eq. (\ref{Fpint=})
can now be done by adding to $\phi(\nu)$ a term 
$[(\Gamma-\omega_1-\omega_2-\omega_3)/\omega_4]
\arctan({\nu}/{\omega_4})$, which vanishes for any $\omega_4$
on account of Eq. (\ref{rel1}), and then using Eq. (\ref{JaAB=})
with $a=\hbar\beta$.

This finally brings the shift of the free energy due the presence 
of the particle 
\BEA
\label{artush}
\beta F_p=\ln\Gamma\left(\frac{\beta\hbar\Gamma}{2\pi}\right)
-\ln\Gamma\left(\frac{\beta\hbar \omega_1}{2\pi}\right)
-\ln\Gamma\left(\frac{\beta\hbar \omega_2}{2\pi}\right)
-\ln\Gamma\left(\frac{\beta\hbar \omega_3}{2\pi}\right)
-\ln\frac{\beta\hbar\omega_0}{(2\pi)^2}
\EEA
This is just equal to $-\ln Z'$ with $Z'$ calculated in Eq. (4.20) of
Grabert et al.~\cite{Grabert}. These authors did not point 
at the physical role of their $Z'$. Here we see it is the
part of the partition sum of the total system related to the
central particle and its coupling to the bath with its linear 
unperturbed spectrum $\omega_k=k\,\Delta$.
We nevertheless expect that the statics and the dynamics
hold for more general bath spectra, as long as the interaction is bilinear,
and the spectra ensure relaxation.

The internal energy reads

\BEA
\label{Up=}
U_p=\frac{\hbar\Gamma}{2\pi}\psi\left(\frac{\beta\hbar\Gamma}{2\pi}\right)
-\frac{\hbar \omega_1}{2\pi}\psi\left(\frac{\beta\hbar \omega_1}{2\pi}\right)
-\frac{\hbar \omega_2}{2\pi}\psi\left(\frac{\beta\hbar \omega_2}{2\pi}\right)
-\frac{\hbar \omega_3}{2\pi}\psi\left(\frac{\beta\hbar \omega_3}{2\pi}\right)
-T
\EEA

\subsection{The effective temperatures}

We shall now study two objects,  $T_x=a\langle x^2\rangle$
and $ T_p=\langle p^2\rangle/2m$,  that would in classical
equilibrium be equal to $T$ and which we shall interpret below
as effective temperatures. 
As in the classical situation, it holds that~\cite{Grabert}
\BEA T_x&=&a\langle x^2\rangle=2a\frac{\partial F_p}{\partial a}\EEA
We find
\BEA \label{Txexact}
T_x=-T+\frac{\hbar a}{\pi m}\,\left\{
\frac{(\omega_1-\Gamma)\,\psi\left(\frac{\beta\hbar
\omega_1}{2\pi}\right)}
{(\omega_2-\omega_1)(\omega_3-\omega_1)}
+\frac{(\omega_2-\Gamma)\,\psi\left(\frac{\beta\hbar
\omega_2}{2\pi}\right)
}{(\omega_1-\omega_2)(\omega_3-\omega_2)}
+\frac{(\omega_3-\Gamma)\,\psi\left(\frac{\beta\hbar \omega_3}{2\pi}\right)
}{(\omega_1-\omega_3)(\omega_2-\omega_3)}
\right\},
\EEA
Likewise,
\BEA \label{Tpexact} T_p&=&\frac{\langle p^2\rangle}{m}
=-2m\frac{\partial F_p}{\partial m}=
T_x+\frac{\hbar\gamma\Gamma }{\pi m}\,\left\{
\frac{\omega_1\,\psi\left(\frac{\beta\hbar \omega_1}{2\pi}\right)
}{(\omega_2-\omega_1)(\omega_3-\omega_1)}
+\frac{\omega_2\,\psi\left(\frac{\beta\hbar \omega_2}{2\pi}\right)
}{(\omega_1-\omega_2)(\omega_3-\omega_2)}
+\frac{\omega_3\,\psi\left(\frac{\beta\hbar \omega_3}{2\pi}\right)
}{(\omega_1-\omega_3)(\omega_2-\omega_3)}
\right\}
\EEA
To find the Gibbsian values for $\gamma\to 0$
one has to notice that
\BEA
\psi \left (\frac{i\omega _0}{2\pi T}\right )-
\psi \left (-\frac{i\omega _0}{2\pi T}\right ) = 
\frac{2\pi i}{\hbar\omega _0}~T
+i\pi ~ {\rm coth}~\frac{\hbar\omega _0}{2T}
\EEA
and this yields the standard weak-coupling result known from
all the books, 
\BEQ \label{TxpGibbs}
U=\half T_x+\half T_p=T_x=T_p=\frac{\hbar \omega _0}{2}
\coth\frac{\beta\hbar\omega _0}{2}=\half\hbar \omega _0+
\frac{\hbar\omega_0}{e^{\beta\hbar\omega _0}-1},\qquad(\gamma\to 0).\EEQ

\subsubsection{Thermodynamics and effective temperatures  at high $T$}

Using that for small $z$
\BEQ \ln \Gamma(z)=-\ln z-\gamma_Ez+\frac{\pi^2}{12}z^2\EEQ
where $\gamma_E=0.5772156$ is Euler's constant, one gets the free energy
\BEQ \label{FplargeT} 
F_p=T\ln\beta\hbar\omega_0+\frac{\beta\hbar^2}{48}
[\Gamma^2-\omega_1^2-\omega_2^2-\omega_3^2]\approx
T\ln\beta\hbar\omega_0+\frac{\beta\hbar^2(a+\gamma\Gamma)}{24m}
\EEQ
where $\omega_0=\sqrt{a/m}$ and $\Gamma$ has been taken large in the second 
identity.  The internal energy and entropy become
\BEQ \label{UplargeT} U_p=T+\frac{\beta\hbar^2(a+\gamma\Gamma)}{12m},\qquad
  S_p=\ln\frac{T}{\hbar\omega_0}+1+\frac{\beta\hbar^2(a+\gamma\Gamma)}{6m}
\EEQ

From eqs. (\ref{Txexact}), (\ref{Tpexact}) 
we obtain at large $T$
\BEQ\label{TxTplargeT}
T_x=T+\frac{\beta\hbar^2a}{12m}-
\frac{\beta^3\hbar^4a(a+\gamma\Gamma)}{720m^2};
\EEQ and 
\BEQ
T_p=T+\frac{\beta\hbar^2(a+\gamma\Gamma)}{12m}
-\frac{\beta^2\hbar^3\gamma\Gamma^2}{4\pi^3m}\zeta(3)+
\frac{\beta^3\hbar^4[\gamma\Gamma^3m-(a+\gamma\Gamma)^2]}{720m^2}
\EEQ

\subsubsection{Thermodynamics and effective temperatures at low $T$}

Further results can be obtained with the improved Stirling formula
\BEA
\label{samurai}
\ln \Gamma (z)= (z-\frac{1}{2})\ln z-z +\frac{1}{2}\ln (2\pi )+\frac{1}{12z}
-\frac{1}{360z^3}
\EEA
One gets for arbitrary and for large $\Gamma$ 
\BEA
\label{comrad1}
F_p&=& \frac{\hbar}{2\pi}[\Gamma\ln \Gamma
-\sum_{k=1}^3\omega _k\ln\omega _k]-\frac{\pi \gamma}{6\hbar a}\,T^2\nn
&=&\frac{\hbar\gamma}{2\pi m}[\ln \frac{2m\Gamma}{\gamma}\,+1]
-\frac{\hbar\gamma}{4\pi m}[(1+w)\ln(1+w)+(1-w)\ln(1-w)]
-\frac{\pi \gamma}{6\hbar a}\,T^2,
\\ \label{comrad2}
U_p&=& \frac{\hbar}{2\pi}[\Gamma\ln \Gamma
-\sum_{k=1}^3\omega _k\ln\omega _k]+\frac{\pi \gamma}{6\hbar a}\,T^2\nn
&=&\frac{\hbar\gamma}{2\pi m}[\ln \frac{2m\Gamma}{\gamma}\,+1]
-\frac{\hbar\gamma}{4\pi m}[(1+w)\ln(1+w)+(1-w)\ln(1-w)]
+\frac{\pi \gamma}{6\hbar a}\,T^2,
\\ \label{comrad3}
S_p&=&\beta (U_p-F_p)=\frac{\pi \gamma}{3\hbar a}\,T,
\EEA
Notice that $S_p$, the shift in total von Neumann entropy due to the 
presence of the Brownian particle, differs strongly from the von 
Neumann entropy of the particle itself, which remains finite at $T=0$,
as we shall show in Eqs. (\ref{v=}) and (\ref{SvN=}).
The non-additivity of entropies encountered here is a deep aspect of
quantum physics, where a subsystem can have a larger von Neumann
entropy than the full system.

Using (\ref{dwda}) one finds
at low temperatures and arbitrary  $am/\gamma ^2$, $w=\sqrt{1-4am/\gamma^2}$:
\begin{eqnarray}
\label{TT0g}
&&T_p=\frac{\hbar \gamma }{\pi m}\ln\frac{2\Gamma m}{\gamma}+
\frac{\hbar \gamma}{4\pi m\,w}[(1-w)^2\ln(1-w)-
(1+w)^2\ln(1+w)]
+{\cal O}(T^4), \\ 
\label{TTT0g}
&&T_x=\frac{\hbar a}{\pi\gamma}\,\frac{1}{w}\ln\frac{1+w}{1-w}
+\frac{\pi\gamma}{3\hbar a}T^2 +{\cal O}(T^4),
\end{eqnarray}

The above expressions simplify in the limit of strong damping,
see Eq.~(\ref{mega1}):
\BEA
F_p(T)&=&\frac{\hbar \gamma}{2\pi m}
(\ln\frac{\Gamma m}{\gamma} +1)+
\frac{\hbar a}{2\pi \gamma}\ln\frac{\gamma^2}{a m}-
\frac{\pi \gamma}{6\hbar a}\,T^2,\nn
U_p(T)&=&\frac{\hbar \gamma}{2\pi m}
(\ln\frac{\Gamma m}{\gamma} +1)+
\frac{\hbar a}{2\pi \gamma}\ln\frac{\gamma^2}{a m}+
\frac{\pi \gamma}{6\hbar a}\,T^2,
\label{mega3}
\EEA
and
\begin{eqnarray}
\label{TT0}
&&T_p=\frac{\hbar \gamma }{\pi m}\ln\frac{\Gamma m}{\gamma}+
\frac{\hbar a}{\pi \gamma}+{\cal O}(T^4), \\ 
\label{TTT0}
&&T_x=\frac{\hbar a}{\pi\gamma}\ln\frac{\gamma^2}{am}
+\frac{\pi\gamma}{3\hbar a}T^2 +{\cal O}(T^4),
\end{eqnarray}

\renewcommand{\thesection}{\arabic{section}}
\section{Thermodynamic aspects of adiabatic changes}
\setcounter{equation}{0}\setcounter{figure}{0} 
\renewcommand{\thesection}{\arabic{section}.}

\subsection{Generalized thermodynamic formulation}

We now make clear that the relation with standard thermodynamics 
can be continued much further by introducing 
the two {\it effective temperatures}
\begin{eqnarray}
\label{TT}
&&T_p=\frac{\langle p^2\rangle}{m}, \qquad 
\label{TTT}T_x=a\langle x^2\rangle,
\end{eqnarray}
One reason to do this is that
the stationary state for the harmonic potential has 
a quasi-Gibbsian expression for the Wigner function
\BEQ \label{Wpx=}
W(p,x)=W_p(p)W_x(x)=\frac{e^{-{\cal K}(p)/T_p}}{\sqrt{2\pi mT_p}}\,\,
\frac{e^{-{\cal V}(x)/T_x}}{\sqrt{2\pi T_x/a}}\EEQ
with 
${\cal K}(p)=p^2/2m$ the kinetic energy and ${\cal V}(x)=\half a x^2$ the potential 
energy. This expression is quasi-Gibbsian, since there occur
two different temperature-like variables.
(Notice that the normalization is $\int\d x\d p W=1$).  
There occur the Boltzmann entropies of
momenta and coordinate,
\BEA 
\label{kow3}
&&S_p=-\int \d p W(p)\ln[W(p)\,\sqrt{\hbar}\,]=\half \ln \frac{mT_p}{\hbar}
+\half,\\
\label{kow4}
&&S_x=-\int \d x W(x)\ln[W(x)\,\sqrt{\hbar}\,]=\half \ln \frac{T_x}{\hbar a}
+\half, \EEA
(in $S_p$ and $S_x$ we skipped terms $\ln\,2\pi)$.
The complete ``Boltzmann'' entropy is
\BEA
\label{kow2}
&&S_B=S_p+S_x=-\int \d p\d x W(p,x)\ln[W(p,x)\hbar]=
\half \ln \frac{mT_pT_x}{\hbar^2 a}+1
\label{STxp=}
\EEA

\subsubsection{Internal energy and interaction energy}

The energy of the central particle reads
\BEQ \label{kow1}
U=\frac{\langle p^2\rangle}{2m}+\half a \langle x^2\rangle=
\half T_p+\half T_x \EEQ 
For a discussion of this identification in systems with out
a self-interaction term, see section ~\ref{secthamiltonian}.

The interaction energy, i.e. the energy of the cloud of bath modes 
that surround the particle, is defined as  
\BEQ U_{\rm int}=U_{\rm tot}-U_B(\gamma=0)-U=U_p-U=U_p-\half T_p-\half T_x \EEQ 

At high temperatures one gets from Eqs. (\ref{UplargeT}) and
(\ref{TxTplargeT})

\BEQ U=T+\frac{\beta\hbar^2}{24m}(2a+\gamma\Gamma),\qquad
U_{\rm int}=\frac{\beta\hbar^2}{24m}(6a+7\gamma\Gamma)\EEQ
Since the energy of the cloud is involves $\hbar$, 
the non-triviality of the cloud is a quantum effect.

At low temperatures one gets the internal energy
\BEQ\label{UlowT} 
U=\frac{\hbar\gamma}{2\pi m}\ln\frac{2m\Gamma}{\gamma}-\frac{\hbar\gamma}
{2\pi m}[(1+w)\ln(1+w)+(1-w)\ln(1-w)]+\frac{\pi\gamma}{6\hbar a}T^2\EEQ
For large damping this reduces to
\BEQ U=\frac{\hbar\gamma}{\pi m}\ln\frac{m\Gamma}{\gamma}+\frac{\hbar a}
{2\pi \gamma}[\ln\frac{\gamma^2}{am}+1]+\frac{\pi\gamma}{6\hbar a}T^2.\EEQ
The interaction energy of the cloud is now independent of $w$,
\BEQ \label{UclowT}
U_{\rm int}=\frac{\hbar\gamma}{2\pi m}+{\cal O}(T^4),\EEQ
provided that $\Gamma$ is large.

\subsubsection{Generalized free energy and the first and second law}

The definition of the effective temperatures 
admits a clear thermodynamical interpretation.
For studying the role of an adiabatically slow
variation of an arbitrary parameter, such as $a$ or $m$, that we 
shortly denote by $\alpha$, the free energy $F$ is  defined as 
\begin{equation}
\label{dEEd}
F=-T_p\ln Z_p-T_x\ln Z_x,
\end{equation}
The definitions $Z_p=\int\d p \exp[-{\cal K}(p)/T_p]$, $Z_x=\int\d x \exp[-{\cal V}(x)/T_x]$ bring
\begin{equation}
\label{dEEd2}
F=-\half T_p\ln m T_p-\half T_x\ln\frac{T_x}{a}
\end{equation}
For considering changes in system parameters one needs 
\begin{eqnarray}
\label{asala1}
&&\d [-T_p\ln Z_p]=-\ln Z_p \d T_p-\frac{T_p}{Z_p}\d Z_p=
\frac{1}{Z_p}\int \d p e^{-\beta _p{\cal K}(p)}\d {\cal K}(p)-S_p\d T_p,\nonumber\\
&&\d [-T_x\ln Z_x]=-\ln Z_x \d T_x-\frac{T_x}{Z_x}\d Z_x=
\frac{1}{Z_x}\int \d x e^{-\beta _x{\cal V}(x)}\d {\cal V}(x)-S_x\d T_x \EEA
Eq. (\ref{dEEd}) then yields
\BEA \d F=- S_x\d T_x- S_p\d T_p+\dbarrm {\cal W}_{\rm rev}.
\end{eqnarray}
with, in agreement with the derivations (\ref{dEE}) and (\ref{dEEm})
below, the work added to the system
\BEQ\label{dwork} \dbarrm {\cal W}_{\rm rev}
=-T_p\frac{\d m}{2m}+T_x\frac{\d a}{2a} \EEQ

These relations are valid in spite of the fact that
both $T_x$ and $T_p$ are functions of $a$ and $m$.

Because of Eqs. (\ref{kow1}), (\ref{kow3}) and (\ref{kow4})
the definition (\ref{dEEd}) is compatible with the standard 
identification~\cite{1,Nhammer} 
\begin{equation}
\label{asala11}
F=U-T_pS_p-T_xS_x
\end{equation}
that one would write down immediately for a two-temperature
system. From this relation 
one will indeed reproduce the standard formulation for the first law
for situations with two temperatures,
\begin{eqnarray}
\label{asala2}
&&\d U=\dbarrm {\cal Q}_{\rm rev}+\dbarrm {\cal W}_{\rm rev}, \\
\label{asala3}
&&\dbarrm {\cal Q}_{\rm rev}= T_p\d S_p+T_x\d S_x,
\end{eqnarray}
where $\dbarrm {\cal Q}_{\rm rev}$ is the heat reversibly added to the
particle. A detailed discussion concerning the general definitions of
the work and heat is given below, in section \ref{work&heat}.

The generalized thermodynamical relations (\ref{asala1}-\ref{asala2})
are in close analogy with those proposed recently for 
nonequilibrium glassy systems \cite{1,Nhammer,4}.
Analogous to that situation, $F$ pertains to the particle alone,
and, except at high $T$, it differs from the $F_p$ of Eq. (\ref{artush}) 
in previous section, which relates to the whole system, to be more precise, 
to the particle and the cloud of bath modes around it. 

Let us recall that $F_p$ satisfies Gibbsian thermodynamics, while
$F$ does not. There are many physical systems, such as a Josephson
junction strongly coupled to the electromagnetic field, 
where the natural object to study is nevertheless $F$, since it 
relates for that case to properties of the junction only.

It is standard wisdom that  energy is dispersed if the variations are 
non-adiabatic changes. This is confirmed by Eq. (\ref{Picycle}),
which holds provided the whole time-domain where $m$ and $a$ vary is 
accounted for. This leads to the general result
\BEA \label{asala4}
&&\dbarrm \Q \le T_p\d S_p+T_x\d S_x
\end{eqnarray}
that is also known from the study of glasses and, more generally,
from two-temperature systems. 

\subsection{Violation of the Clausius inequality}
\label{CC}
\subsubsection{The Clausius inequality at small $T$.}

Let us now consider two concrete examples, and study the Clausius
inequality $\dbarrm {\cal Q}\le T\d S_{vN}$, which is one of possible
formulations of the second law. 

For a very slow variation of the spring constant $a$ 
one gets 
\begin{eqnarray}
\label{dEE}
&&\dbarrm {\cal W}_{\rm rev}
=\int \d x\d p W(p,x)\frac{\partial H}{\partial a}\d a
=\int \d x W(x)\half x^2\d a
=T_x\frac{\d a}{2a},
\EEA in agreement with Eq. (\ref{dwork}).
The first law implies for the heat added adiabatically to the particle
at low $T$
\BEA \label{dEE1}
&&\dbarrm {\cal Q}_{\rm rev}=\d U-\dbarrm\, {\cal W}_{\rm rev}=
(\frac{\p T_p}{\p a}+\frac{\p T_x}{\p a}
-\frac{T_x}{a})\frac{\d a}{2}
=-\frac{\pi\gamma }{3\hbar a^2}\,T^2\,\d a\,\,+{\cal O}(T^4\,\d a)
\label{dEE2}
\end{eqnarray}

It is seen that
$\dbarrm {\cal Q}_{\rm rev}=0$ at $T=0$ for all $\gamma$.
Using Eqs.~(\ref{TT}, \ref{TTT}) we derive for large $\gamma$
and very large $\Gamma$, 
\BEQ \d S_B=- \frac{\d a}{2a}\left[\frac{1}{\ln(\gamma^2/am)}
-\frac{ma}{\gamma^2\ln(\Gamma m/\gamma)}
+\frac{\pi^2\gamma^2T^2}{3\hbar^2a^2\ln(\gamma^2/am)}
(2-\frac{1}{\ln(\gamma^2/am)})+{\cal O}(T^4)\right] \EEQ

At $T=0$ the Clausius inequality says that no heat can be taken from the bath, 
at best heat can go from the central system (here: the Brownian particle)
 to the 
bath. In our situation $\dbarrm {\cal Q}$ is of order $T^2$, while $T\d S_{vN}$
 is of order $T$. Since we only do powercounting in $T$ and 
both expressions are non-trivial, we may replace here  $S_{vN}$ by $S_B$.
Thus for the case $\d a >0$, where
an amount of work $\dbarrm\, {\cal W}_{\rm rev}>0$ is done on the system, 
the Clausius relation is violated at low but non-zero $T$.

In the same way one can consider the variation of the (effective) mass $m$.
Here one has \begin{eqnarray}
\label{dEEm}
&&\dbarrm {\cal W}_{\rm rev}
=\int \d x\d p W(p,x)\frac{\partial H}{\partial m}\d m
=-\int \d p W(p)\frac{ p^2}{2m^2}\d m
=-T_p\frac{\d m}{2m},
\EEA
again in concordance with Eq. (\ref{dwork}). This implies
\BEA \label{dEE1m}
&&\dbarrm {\cal Q}_{\rm rev}
=(T_p\frac{\p S_p}{\p m}+T_x\frac{\p S_x}{\p m})\d m
=(\frac{\p T_p}{\p m}+\frac{\p T_x}{\p m}
+\frac{T_p}{m})\frac{\d m}{2}
=\frac{\hbar \gamma }{2\pi m^2}\d m+{\cal O}(T^2)
\end{eqnarray}
In contrast to the previous case there is a transfer of heat even for
if the bath temperature is zero. Thus, violation of the Clausius inequality 
is even stronger in this case, since for $\d m >0$ one has 
$\dbarrm {\cal Q}> 0$, even though $T\d S_{vN}=0$ (for $T\to 0$). 
This situation with $\dbarrm\, {\cal W}_{\rm rev}<0$ corresponds to work 
performed by the system on the environment. To underline that the heat 
comes from the cloud of bath modes, we notice that the general 
relations
\BEQ 
\dbarrm \W_{\rm rev}=\d F_p,\qquad \d U=\d U_p-\d U_{\rm int}=
\dbarrm\Q_{\rm rev}+\dbarrm \W_{\rm rev}\EEQ
imply
\BEQ \label{dQdUclT}\dbarrm\Q_{\rm rev}=T\d S_p-\d U_{\rm int} \EEQ
For changing $m$ in the $T=0$ situation it indeed holds that
\BEQ \dbarrm {\cal Q}_{\rm rev}=-\d U_{\rm int} \EEQ
for all values of $w$, even when $\Gamma$ is not very large.
For a change in $a$ it holds that $\d U_{\rm int}=\O(T^4)$, but
 Eq. (\ref{dQdUclT}) nevertheless reproduces 
Eq. (\ref{dEE1}), because of relation (\ref{comrad3}).

Let us briefly discuss consequences drawn from the
violation of the Clausius inequality in the quantum regime. First of
all it seen that it occurs in the overall Gibbsian state, so that
globally (i.e., when applied to the overall closed system)
thermodynamics is valid by definition. 
In particular, since the overall system does not
absorb heat during any variation of a parameter, 
and $\dbarrm\Q=0$ is consistent with the $T\to 0$ case of the
Clausius inequality (later we will see that this is also the case
at finite temperatures, where $\dbarrm\Q$ is still zero). 
Nevertheless, the local state of the particle is
not Gibbsian and does allow violations as we have seen.
We stress that this violation 
arises due to quantum entanglement, which leads to non-gibbsian effective
temperatures for the stationary state of the brownian particle.
If the effective temperatures for $T\to 0$ would equal to their gibbsian
values $\hbar\omega_0/2$, the state of the particle would be pure,
which is impossible since it does interact with the bath.

When later discussing the Thomson's formulation of the second law, we
will see that it is perfectly valid for the overall Gibbsian state, so
that the above violation of the Clausius inequality provides us with
an explicit example showing that at low temperatures the very
equivalence between different formulations of the second law is lost. 

A further aspect of this matter is the squeezing of phase space and
entropy, relevant for computing in the quantum regime. 
In a separate paper we have shown that the so-called Landauer bound for
the erasure of one bit of information, that arises from the Clausius 
inequality, is violated in a similar manner~\cite{ANLand}.

\begin{figure}[bhb]
\vspace{0.1cm}\hspace{-1.5cm}
\vbox{\hfil\epsfig{figure= 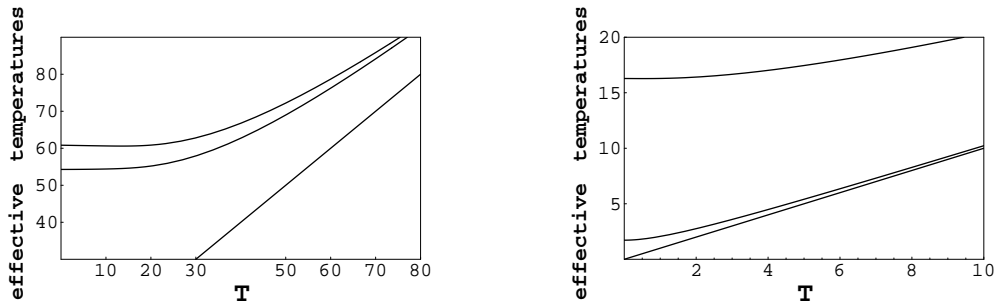,width=15cm,angle=0}\hfil}  
\vspace{0.75cm}
\caption{ The effective temperatures $T_p$, $T_x$ versus the bath temperature
$T$ for two values of the dimensionless damping. For the involved
parameters we take the following values: 
$\hbar\gamma/(4\pi m) =1$, $\hbar\Gamma/(2\pi) =100$.
Left figure: $am/\gamma^2=80 $ (underdamping); from the 
top to the bottom: $T_p$, $T_x$, $T$. Right figure: the same, but with
$am/\gamma^2=0.2$ (moderate overdamping).}
\label{fig-1}
\end{figure}

Notice again that the effective temperatures 
remain finite in the limit $T\to 0$
(see Eqs.~(\ref{TT0}, \ref{TTT0}) and Fig.~\ref{fig-1}), 
and both are larger than the bath temperature $T$.
The fact that they are non-equal is due to a mixed state of the particle.
Indeed, a quantum system non-weakly interacting with its environment, 
will be in a mixed state even if the whole closed system 
(the particle and environment together) is in a pure state 
(e.g., the vacuum state). 

The existence of different temperatures $T_x$, $T_p$ and $T$ 
for the subsystem and bath seems to contradict the zeroth law, which
states that systems interacting for a long time 
are in equilibrium, and share common temperature. Notice especially that
the above difference between temperatures is not a consequence of any
metastability and/or incomplete equilibration, so that our effective 
temperatures do not depend on the dynamics of the particle and 
have somewhat more definite status compared with
those defined, e.g., for glassy systems \cite{1,Nhammer,Nbh}. 
Typical derivations of the zeroth law (see e.g. Ref.~\cite{landau} for
one of the most clear examples) essentially use the assumptions
that the interaction with the bath is very weak, 
and that the total entropy can be considered as the sum of entropies
of the subsystem and the bath. Evidently, this last condition is not 
satisfied in our case, except for the limit $\gamma\to 0$, 
where $T_x$ and $T_p$ tend to their Gibbsian value, see
 Eq.~(\ref{TxpGibbs}), of the harmonic oscillator coupled very weakly 
to its bath, the situation treated in all textbooks.

Let us notice that in literature some other ways were proposed to establish
effective characteristics for non-underdamped Brownian oscillator.
In \cite{weiss} it is shown that there is a mapping to the Gibbsian
(underdamped)
oscillator through the definition of an effective mass and frequency.
For the description of He$_3$ Prokof'ev studies a related model with a gap
in $J(\omega)$ at small $\omega$~\cite{prkf}. In this approach he
makes a different identification for the effective temperature. Surely,
the choice which quantity to take as ``effective temperature'' is to some 
extent a matter of taste, that can only be justified by the induced
simplification of the physical results. For our thermodynamic 
approach other definitions of effective temperatures 
will not be very helpful. Our $T_p$ and $T_x$, however,
allow to formulate the generalized Clausius inequality and they 
also occur in the Maxwell-Boltzmann-like form (\ref{Wpx=}) 
of the Wigner function. Last but not least, these effective
temperatures enter in the same way as in glasses and other 
two-temperature systems, such a black holes.

\subsubsection{Von Neumann entropy}

In next part we shall discuss the von Neumann entropy of the central
particle. To investigate it one needs the density
matrix corresponding to the Wigner function (\ref{Wpx=}). 
For the harmonic oscillator this can be worked out explicitly.
One approach is to introduce an effective mass and an effective
frequency \cite{weiss}, and insert  these results in the expression 
for the entropy of the effective harmonic oscillator. 
We found it more insightful to redo the derivation.
The standard relation
\BEA
\langle x+\frac{u}{2}|\rho | x-\frac{u}{2} \rangle =
\int \d p~e^{-ipu/\hbar}W(p,x)
\EEA
connects the density matrix in coordinate representation
with the Wigner function. From this relation 
one gets the following formula~\cite{Grabert}
\begin{eqnarray}
\label{den2}
\langle x|\rho | x' \rangle &&=
\frac{1}{\sqrt{2\pi \langle x^2\rangle }}
\exp{[-\frac{(x+x')^2}{8\langle x^2\rangle}
-\frac{(x-x')^2}{2\hbar ^2/\langle p^2\rangle} ]}
\end{eqnarray}
The physical meaning of Eq.~(\ref{den2}) is clear: The diagonal
elements ($x=x'$) are distributed at the scale
$\sqrt{\langle x^2\rangle}$,
while the maximally off-diagonal elements
($x=-x'$), which characterize
coherence, are distributed with the characteristic scale
$\hbar/\sqrt{\langle p^2\rangle}$.

We have to find eigenfunctions and eigenvectors
of this density matrix
\BEA
\int \d x' \langle x|\rho | x' \rangle f_n(x')
=p_n f_n(x)
\EEA
The solution of this problem uses some tabulated 
formulas for Hermite polynoms, and results in
\begin{eqnarray}
&&p_n = \frac{1}{v+\half}\left [ 
\frac{v-\half}{v+\half}
\right ]^n, \\
&&f_n(x) = c\,H_n(c\,x)e^{-c^2x^2/2}, \\
&&c=
\left(\frac{\langle p^2\rangle}{\hbar ^2\langle x^2\rangle}\right)^{1/4},
\qquad
v=\frac{\Delta p\,\Delta x}{\hbar}=
\sqrt{\frac{\langle p^2\rangle\langle x^2\rangle}{\hbar ^2}}=
\sqrt{\frac{mT_pT_x}{\hbar ^2a}} \label{v=}
\end{eqnarray}
where $H_n$ are Hermite polynomials, and it holds that $v\ge \half$
due to the Heisenberg uncertainty relation. 
The result for the von Neumann entropy now reads~\cite{weiss}
\BEA\label{SvN=}
S_{vN} =-\sum_np_n\ln p_n=(v+\half)\ln(v+\half) -(v-\half)\ln(v-\half)
\EEA
The first terms in its large $v$-expansion are 
\BEA
\label{katmandu}&&
S_{vN} 
=\ln v+1-\frac{1}{24v^2}-\frac{1}{320v^4}-\frac{1}{2688v^6}
\EEA

From (\ref{kow3}), (\ref{kow4}) and (\ref{kow2}) we notice that
the same quantity $v$ governs behavior of the Boltzmann entropy 
\BEA
S_B=\half\ln\frac{\langle x^2\rangle\langle p^2\rangle}{\hbar^2}+1
=\ln{v}+1
\EEA
This appears to coincide with the leading terms of (\ref{katmandu}).
It is known to be larger than the Von Neumann entropy, and this is 
obvious from the sign of the correction terms.

If some parameter ($a$ or $m$) is varied, then the derivative
of $S_{vN}$ with respect to it reads:
\begin{eqnarray}
\d S_{vN}= \ln\frac{v+\half}{v-\half}\,\d v
\label{mustafa}
\end{eqnarray}
In other words, the sign of the change in $S_{vN}$ is determined by the sign
of the change in $v$. This holds as well for the change in $S_B$, so
qualitatively they carry the same information, and this already
was used above to simplify one point of the discussion.

Let us stress that von Neumann entropy $S_{vN}$ is the 
unique quantum measure of localization, whereas the entropies
$S_p$ and $S_x$ characterize localizations of momentum and coordinate
separately. The differences between $S_B$ and $S_{vN}$ are due to the
fact that in quantum theory momentum and coordinate cannot be measured 
simultaneously; in this sense $S_p$ and $S_x$ characterize two different 
measurement setups. Nevertheless, for the harmonic particle if
$S_{vN}$ increases (decreases), then $S_p+S_x$ increases (decreases)
as well. Notice that the real importance of $S_p$ and $S_x$ becomes
clear when they have to be used to generalize the Clausius inequality.
The von Neumann entropy cannot be used for this purpose
whenever $T_x\not =T_p$.

\subsubsection{Clausius inequality at large $T$.}

At low $T$ only power counting in $T$ was needed for showing the violation 
of the Clausius inequality. The precise definition of entropy, and the 
quantitative difference between the Boltzmann entropy 
and the von Neumann entropy were not essential, since then $T\d S\to 0$
anyhow. 
Here we wish to show that the same violation already happens at 
arbitrarily large temperature. To do this we have 
to use the von Neumann entropy of the subsystem.

In this section we consider very large temperatures, $T\gg
\hbar\Gamma$.  
Using (\ref{TxTplargeT}) we find from (\ref{dEE1m}) for a change in 
$m$ that
\BEQ 
\dbarrm {\cal Q}=[1-\frac{\beta^2\hbar^2a}{12m}]\frac{T}{2m}\,\d m,
\qquad
\d S_{vN}=[1-\frac{\beta^2\hbar^2(a+\gamma\Gamma)}{12m}]
\frac{1}{2m}\,\d m,
\EEQ
So, for a change $\d m>0$ it is seen that at arbitrarily 
large temperature the Clausius inequality is violated. 
The relative violation $(\dbarrm {\cal Q}-T\d S_{vN})/
\dbarrm {\cal Q}$ is of order $\hbar^2\gamma\Gamma/(mT^2)$.

For a change in $a$ we find
\BEQ
{\dbarrm {\cal Q}}
=\left\{-\frac{1}{2a}+\frac{\hbar^2\beta^2}{24m}
-\frac{\beta^4\hbar^4[a+(2/3)\gamma\Gamma]}{480m^2}\right\}\,T\,\d a
\EEQ
\BEQ\label{SvnTda}
T\d S_{vN}=\left\{-\frac{1}{2a}+\frac{\hbar^2\beta^2}{24m}
-\frac{\beta^4\hbar^4[a+\gamma\Gamma]}{480m^2}\right\}\,T\,\d a
\EEQ
These expressions differ at relative order
$\beta^4\hbar^4 \gamma\Gamma/m^2$, and the Clausius inequality is
violated for $\d a>0$.

The important conclusion of this section is that the violation
of the Clausius inequality already occurs at arbitrarily high
temperatures.
Later we point out that a similar conclusion can be drawn
about the violation of the zeroth law at large temperatures.

We stress that the Clausius inequality is violated for any
finite coupling, and the violating terms only disappear in the weak 
coupling limit $\gamma\to 0$, or in the classical limit $\hbar\to 0$,
equivalent to the high-temperature limit.

\subsubsection{Clausius inequality for comparing two systems.}
\label{Claustwosys}
For non-equilibrium systems the question has not been settled 
whether the von Neumann entropy is the true physical entropy. 
As we are inclined to believe that it is, we have discussed that
entropy above.

Let us, however, now consider cases where there is no doubt: 
For systems in true Gibbsian equilibrium the proper entropy of the subsystem
is surely its von Neumann entropy.  We can now compare two such 
equilibrium systems, having slightly different system parameters.
In standard thermodynamics such a comparison does not yield a new
insight, as the equilibrium 
state of the system is independent of its history.
We should point out that in the thermodynamics of glasses it 
is customary to compare cooling experiments at different 
but fixed pressures.
A related comparison was also made for black holes: 
it could be shown that comparing the situation of a single black 
hole before and after a small amount of matter was added,
is analogous to comparing two different black holes with 
slightly different masses~\cite{BCH}. 
This universality pointed at a thermodynamic behavior of 
black holes, and the physical 
framework could indeed be provided by one of us, by
drawing an analogy with the thermodynamics of glasses~\cite{Nbh}.

For our present case we can compare two equilibrium systems at 
slightly different temperatures. 
This has the benefit that the work term is absent, thus 
needing no interpretation, and it implies 
$\dbarrm{\cal Q}=\d U$. Using the fully exact expressions for 
the energy and the von Neumann entropy, it is then
straightforward to show that at large $T$

\BEQ \label{dQTdSvNlT}
\dbarrm {\cal Q}-T\d S_{vN}=\frac{\beta^2\hbar^2\gamma\Gamma}{24m}\,\d T
\EEQ

The standard {\it expectation} that this should vanish is again seen
to be invalid, and the Clausius inequality is violated for
$\d T>0$.  As before, the terms in the right hand side vanish only
in the weak coupling limit $\gamma\to 0$, the classical limit $\hbar\to 0$,
or the infinite temperature limit $\beta\to 0$.

\subsubsection{On our identification of the energy of the subsystem}
\label{procontra}

In section \ref{secthamiltonian} we have considered two physical situations.
In the first case the Hamiltonian contains a self-interaction term
$\sim x^2$. For that case the above results on the Clausius inequality
apply unambiguously. In the second case there is no such 
self-interaction, but the potential energy $\half b x^2$ is split
as $\half ax^2+\half\gamma\Gamma x^2$, and the first part
is counted in the energy of the subsystem, while the
last part is counted with the interaction energy. Let us now shortly
look at what happens when this is not done, and $\tilde U=\langle
\tilde H\rangle$ is considered as energy of the subsystem. 
At large $T$ one will have
$\tilde U=U+\half(\gamma\Gamma/a) T=T+\half(\gamma\Gamma/a) T$. 
Since the work is not modified by this identification, one will have
a shift in the change of heat, 
$\d\tilde Q=\d Q+\half\gamma\Gamma[\d T/a-T\d a/a^2]$.
From (\ref{SvnTda}) and (\ref{dQTdSvNlT}) it is seen that then
even at very large temperatures $\d\tilde Q-T\d S_{vN}$ will not vanish
whenever $\gamma$ is non-zero. Thus, when there is no self-interaction
our identification of $\H$ as the Hamiltonian of the subsystem
is already mandatory for having a proper classical limit.
The underlying reason is that the Wigner function has the Maxwell-Boltzmann 
form $\exp(-[\frac{1}{2m} p^2+\half ax^2]/T)$, involving $a$ and not 
$b=a+\gamma\Gamma$.
This fixes the entropy, and leaves one consistent choice for the energy.

\renewcommand{\thesection}{\arabic{section}}
\section{Exact dynamical solution}
\renewcommand{\thesection}{\arabic{section}.}
\setcounter{equation}{0}

We now consider the situation where our closed system starts at time $t=0^-$
from a Gibbsian initial distribution. It could arise if long before
the total system was coupled to a `superbath', that allowed relaxation to
equilibrium, after which the connection was cut~\cite{superbath}. 
A more realistic situation occurs when the bath has small non-linearities, that 
drive the whole system to its global Gibbsian state. 

\subsection{The case when the initial state is a modified Gibbsian}
\label{casimir}

We assume that for $t<0$ the system is in a Gibbsian state at 
temperature $T$ with certain parameters $a=a_0$, $m=m_0$, 
$\gamma=\gamma_0$.
At $t=0$ these parameters are instantaneously
changed to $a$, $m$ and $\gamma$, and the system will relax
to a  steady state. This new setup generalizes previous studies 
with for times $t\le 0$ particle and bath are uncoupled, 
described by $\gamma_0=0$. An important benefit is that
in the strong damping limit
the present initial state can be close to the final state,
which is, of course, impossible if $\gamma_0=0$ but $\gamma$ is large. 
When making the change $\gamma_0\to \gamma$ at $t=0$, a  amount of work
$\half(\gamma-\gamma_0)\Gamma\langle x^2\rangle_0$ has to be supplied
to the system. This was truly large in our Letter ~\cite{ANQBMprl}, 
where we took $\gamma_0=0$, but $\Gamma$ and $\gamma$ large.
In the present setup we can choose 
$\gamma_0=\gamma$, but $a_0\neq a$, implying that the work need not be 
not large, even when the Debye frequency $\Gamma$ is large.

\subsubsection{The eigenmodes of the initial state}

In the most general case the Hamiltonian has for $t<0$
the parameters $a_0$, $\gamma_0$ and $m_0$. It reads

\BEQ \H=\sum_{i\ge 0}\frac{p_i^2}{2m_i}+\half\sum_{i,j\ge 0}
\sqrt{m_im_j}\,x_iA_{ij}x_j \EEQ
with 
\BEA A_{00}=\frac{a_0+\gamma_0\Gamma}{m_0}; \qquad 
A_{0i}=A_{i\,0}=-\frac{c_{i}^{(0)}}{\sqrt{m_0m_i}},\qquad
A_{ij}=\omega_i^2\delta_{ij}\EEA
where $c_{i}^{(0)}$ is given by Eq. (\ref{ci=}) with
$\gamma_0$ replacing $\gamma$,
\BEQ \label{ci0=}
c_i^{(0)}=\sqrt{\frac{2\gamma_0 m_i\omega_i^2\Delta}{\pi}\,
\frac{\Gamma^2}{\omega _i^2+\Gamma^2}}
\EEQ 
Let us denote the eigenvalues of $A$ by $\nu_k^2$. From a 
previous section, Eqs. (\ref{phi=},\ref{sinphi=}), we infer that
the eigenfrequencies are shifted, 
\BEQ \nu_k=\omega_k-\frac{1}{\pi}\phi_0(\omega_k)\,\Delta
\EEQ
where $\phi_0$, satisfying $0\le\phi_0\le \pi$, is given by
\BEQ \label{phi0=}
\phi_0(\nu)=\arctan
\frac{\gamma_0\Gamma^2\nu}{(a_0-m_0\nu^2)(\nu^2+\Gamma^2)
+\gamma_0\Gamma\nu^2}\EEQ
In later sections we only need that for large $\Gamma$
\BEQ \label{sinphi0=}
\sin\phi_0  (\nu)=\frac{\gamma_0 \nu}
{\{(a_0-m_0\nu^2)^2+\gamma_0^2\nu^2\}^{1/2}}
\EEQ

The eigenvectors are 
\BEQ e_0^k=\alpha_k,\qquad e_i^k=\frac{c_i^{(0)}\,\alpha_k}
{\sqrt{m_0m_i}(\omega_i^2-\nu_k^2)}\EEQ
with normalization factor
\BEQ \frac{1}{\alpha_k^2}=1+\sum_{i\ge 1}
\frac{[c_i^{(0)}]^{\,2}}{m_0m_i(\omega_i^2-\nu_k^2)^2} \EEQ
The following normalization conditions hold:
\BEQ \sum_{k}e_i^ke_j^k=\delta_{ij},\qquad
\sum_{i\ge 0}e_i^ke_i^l=\delta_{kl}\EEQ

For small $\Delta$ one may use 
\BEQ \sum_{i=-\infty}^\infty\frac{1}{[(i-k)\pi
+\phi_0(\omega_{k})]^2}=\frac{1}{\sin^2\phi_0(\omega_k)}\EEQ 
to find 
\BEQ 
\alpha_k=\sqrt{\frac{2\Delta m_0(\Gamma^2+\omega_k^2)}
{\pi\gamma_0\Gamma^2}}
\,\sin\phi_0^k,\qquad
e_i^k=\sqrt{\frac{\Gamma^2+\omega_k^2}{\Gamma^2+\omega_i^2}}\,
\frac{2\Delta\omega_i\sin\phi_0^{k}}{\pi(\omega_i^2-\nu_k^2)}
\EEQ
where $\phi_0^k=\phi_0(\omega_k)$.
In the zero coupling limit $\gamma_0\to 0$ one has $\phi_0^{k}\to 0$, 
so that $\nu_k\to\omega_k$ and indeed $e_i^k\to\delta_{ik}$. 
The latter setup occurs in the standard treatments where bath 
and subsystem are initially uncoupled.

\subsubsection{The noise}

We consider Gibbsian equilibrium for $t\le 0$.  
Let us now introduce the creation and annihilation operators
$b_k^\dagger$, $b_k$, satisfying $[b_k,b^\dagger_l]=\delta_{kl}$ by
\BEQ \label{xi=pi=}
x_i=\sum_k\sqrt{\frac{\hbar}{2m_i\nu_k}}\,e_i^k
(b^\dagger_ke^{i\nu_kt}+b_ke^{-i\nu_kt}),\qquad
p_j=i\sum_k\sqrt{\frac{\hbar m_j\nu_k}{2}}\,e_j^k
(b^\dagger_ke^{i\nu_kt}-b_le^{-i\nu_kt})
\EEQ
\BEQ \label{xbk=}
x=\sum_k\sqrt{\frac{\hbar\Delta(\Gamma^2+\nu_k^2)}
{\pi\gamma_0\Gamma^2\nu_k}}\sin\phi_0^k\, 
(b^\dagger_ke^{i\nu_kt}+b_ke^{-i\nu_kt}),\quad
p=\sum_k\sqrt{\frac{\hbar\Delta(\Gamma^2+\nu_k^2)}
{\pi\gamma_0\Gamma^2\nu_k}}\,\sin\phi_0^k\, m_0\nu_k
(i\,b^\dagger_le^{i\nu_kt}-i\,b_le^{-i\nu_kt})
\EEQ
They indeed satisfy $[x_i,p_j]=i\hbar\delta_{ij}$ due to the normalization
condition, as well as $[x,p]=i\hbar$, $[x,p_i]=[x_i,p]=0$.
 For $t<0$ the Hamiltonian then reads
\BEQ \H=\half\sum_k\hbar\nu_k(b^\dagger_kb_k+b_kb^\dagger_k)
=\sum_k\hbar\nu_k(b^\dagger_k b_k+\half)\EEQ

In the Gibbsian state that describes our closed system for $t<0$, 
the density matrix is 
\BEQ \rho=\frac{1}{Z}\,e^{-\beta \H}\EEQ
It has the Bose occupation numbers 
\BEQ \label{boseocc}
\langle b^\dagger_kb_k+b_kb^\dagger_k\rangle=
1+\frac{2}{e^{\beta\hbar\nu_k}-1}=
{\rm coth}(\half\beta\hbar\nu_k)
 \EEQ

Combining (\ref{etat=}) and (\ref{xi=pi=}) we now have for the noise
\BEQ 
\eta(t)=\sum_{i\ge 1}\sum_k\sqrt{\frac{\hbar}{2m_i\nu_k}}\,\,
c_i\,e_i^k\left[(b^\dagger_k+b_k)\cos\omega_it
+i\frac{\nu_k}{\omega_i}(b^\dagger_k-b_k)\sin\omega_it\right]\EEQ

To carry our the $i$-sum, we have to evaluate
\BEA\label{etasumik} && \sum_{i\ge 1}\frac{c_ie_i^k\cos{\omega_it}}{\sqrt{m_i}}
=\sqrt{\frac{2\gamma\Gamma^2\Delta}{\pi(\Gamma^2+\nu_k^2)}}\, 
\frac{2\Delta\sin\phi_0^k}{\pi}
\sum_{i\ge 1}(\frac{\nu_k^2}{\omega_i^2-\nu_k^2}+
\frac{\Gamma^2}{\omega_i^2+\Gamma^2})\cos\omega_it\EEA
Gradstein \& Rhyzhik ~\cite{GradsteinRhyzik}
present on page 40, Eq. 1.445.2 the equality
\BEQ \sum_{k=1}^\infty \frac{\cos kx}{k^2+\alpha^2}=
-\frac{1}{2\alpha^2}+\frac{\pi}{2\alpha}\,\frac{\cosh(\pi\alpha-\alpha x)}
{\sinh\pi\alpha}\EEQ
According to them it holds for $0\le x\le 2\pi$, but it actually only
holds for $0\le x\le \pi$, while further it is symmetric and periodic.
We have to apply this with $x=t\Delta$, which is surely between $0$ and 
$\pi$, for the cases $\alpha =i\nu/\Delta$ and $\alpha=\Gamma/\Delta$
\BEA \sum_{i\ge 1}\frac{c_ie_i^k\cos{\omega_it}}{\sqrt{m_i}}
&=&\sqrt{\frac{2\gamma\Gamma^2\Delta}{\pi(\Gamma^2+\nu_k^2)}}\,\sin\phi_0^k
\left[\Gamma \frac{\cosh({\pi\Gamma}/{\Delta}-\Gamma t)}
{\sinh({\pi\Gamma}/{\Delta})}-
\frac{\nu_k\cos({\pi\nu_k}/{\Delta}-\nu_k t)}
{\sin({\pi\nu_k}/{\Delta})}\right]\nn\\
&=&\sqrt{\frac{2\gamma\Gamma^2\Delta}{\pi(\Gamma^2+\nu_k^2)}}\,\left[
\Gamma \sin\phi_0^k\,e^{-\Gamma t}+\nu_k\cos(\phi_0^k+\nu_k t)\right]
\EEA
In the last step we have neglected terms of order $\exp(-2\Gamma/\Delta)$,
which are extremely small.
The primitive of this relation yields 
\BEA \sum_{i\ge 1}\frac{c_ie_i^k\sin{\omega_it}}{\sqrt{m_i}\,\omega_i}
&=&\sqrt{\frac{2\gamma\Gamma^2\Delta}{\pi(\Gamma^2+\nu_k^2)}}\,\left[
-\sin\phi_0^k\,e^{-\Gamma t}+\sin(\phi_0^k+\nu_k t)\right]
\EEA
When we insert this in expression (\ref{etasumik}) for 
the noise we have the explicit result
\BEQ \eta(t)=\sum_k
\sqrt{\frac{\gamma\hbar\nu_k\Gamma^2\Delta}{\pi(\Gamma^2+\nu_k^2)}}
\left[(\frac{\Gamma-i\nu_k}{\nu_k}b_k^\dagger 
+\frac{\Gamma+i\nu_k}{\nu_k}b_k)\sin\phi_0^k e^{-\Gamma t}+
e^{i\phi_0^k+i\nu_kt}b_k^\dagger +
e^{-i\phi_0^k-i\nu_kt}b_k\right]\EEQ
The memory of the initial state (i.e. the dependence on $a_0$, $m_0$
and $\gamma_0$, coded in $\phi_0^k$) is washed out after a time 
$\tau_\eta=1/\Gamma$, apart from a harmless phase factor.
 Notice that the time-dependencies are $\exp(i\nu_kt)$,
as one would have expected from Eq. (\ref{xi=pi=}). Also note that the
$\exp(-\Gamma t)$ terms underline the special role of $t=0$, just as
it does elsewhere in Eq. (\ref{pdot=}). We shall later verify that
this passes a consistency check.

\subsubsection{The noise correlator}

The noise correlator now decomposes in two parts
\BEQ 
K(s,t)=\half\,\langle\,{\eta(t)\eta(s)+\eta(s)\eta(t)}\,\rangle
=K_0(s-t)+K_1(s,t)\EEQ
The first term is the stationary noise know from the situation where
system and bath were initially uncoupled, 
\BEQ \label{K0t=}
K_0(s-t)=\frac{1}{\pi}\int_0^\infty\d\omega\,\bar{K}_0(\omega)
\cos\omega(t-s) \EEQ with spectrum
\BEQ\label{K0om=}
 \bar{K}_0(\omega)
=\frac{\gamma\hbar\omega}{\tanh\half\beta \hbar \omega}\,
\frac{\Gamma^2}{\Gamma^2+\omega^2}\EEQ
It indeed does not involve parameters of the initial state.
The connection between properties of the noise and the friction kernel
is the consequence of quantum fluctuation-dissipation theorem 
\cite{landau,klim-rev,weiss}.

As shown in Ref.~\cite{aslangul},
the quantum noise has correlation $K_0(t)=-\ln (\Gamma t)>0$ at 
small times, $t\ll 1/\Gamma$. At $T=0$ there occurs for large times
the celebrated powerlaw, anti-correlated decay
\BEQ K_0(t)=-\frac{\hbar\gamma}{\pi t^2} \EEQ
This is cut off  at times larger than
the universal quantum coherence time $ \tau_\hbar={\hbar}/{T}$, where
\begin{equation}
\label{5}
K_0(t)=-\frac{\pi \gamma T^2}{\hbar}\left [
\sinh \left (\frac{\pi t}{\beta\hbar}\right )
\right ]^{-2},
\end{equation}
The divergence of this expression at $t=0$ shows that a regulator 
like $\Gamma$ is needed.

Let us briefly explain qualitative reasons for the above
structure of the quantum noise. As seen from Eq.~(\ref{etat=}), the
quantum noise is just a weighted sum of the unperturbed coordinates of the
baths oscillators. For $T\to 0$ the unperturbed bath appears in its
lowest energy level, and since energy and coordinate do not commute
(just because coordinate and momentum do not commute), the quantum
noise fluctuates even for $T\to 0$, and brings a non-trivial structure
to $K(t)$ in contrast to the classical case, where the noise is just
absent for zero temperatures. On the other hand, the total intensivity of the quantum
noise is zero for $T\to 0$: as seen from (\ref{K0om=}), $\int \d t
K(t)=2\gamma T$. For the total integral to be zero, the correlator $K(t)$
should change its sign at some intermediate time $t$. 
For longer times the quantum noise is anticorrelated.
The correlator displays a power-law behavior, since the correlation 
time-scale $\hbar/T$ is now infinite. 
A colored noise generated by the low-temperature quantum thermal bath 
will be the main cause of our effects. The classical white noise situation 
$K_0(t)= 2\gamma T\delta(t)$ is recovered by
taking the high-temperature limit ($T\gg \hbar\Gamma$).

The second term is due to the initial correlation of particle and bath,
\BEA K_1(s,t)=K_{11}e^{-\Gamma(s+t)}+K_{12}(s)e^{-\Gamma t}
+K_{12}(t)e^{-\Gamma s}\EEA
\BEA K_{11}=\frac{1}{\pi}\int_0^\infty\d\omega\,\bar{K}_0(\omega)
\frac{\Gamma^2+\omega^2}{\omega^2}\sin^2\phi_0(\omega)\EEA
\BEA K_{12}(t)=\frac{1}{\pi}\int_0^\infty\d\omega\,\bar{K}_0(\omega)
[\frac{\Gamma}{\omega}\cos(\phi_0(\omega)+\omega t)-
\sin(\phi_0(\omega)+\omega t)]\sin\phi_0(\omega) \EEA
where $\phi_0$ is defined by (\ref{phi0=}).

The standard case of initially uncoupled Brownian particle and bath
is recovered for $\gamma_0 \to 0$, where $\phi_0\to 0$. 
Then $K_{12}$ and $K_{11}$ vanish, making the noise correlator 
time-translation invariant. In the general case,
the initial correlation only affects the very
short time regime $t\le \tau_\eta=1/\Gamma$ (remember that we assume that
$\Gamma$ is larger than other characteristic frequencies
of the damped Brownian particle).

\subsubsection{Variances and covariance arising from the initial state}

The Gibbsian initial state leads to three coupled Gaussian random variables: 
the random initial conditions, $z_1=p_0$ and $z_2=x_0$, 
and the noise $z_3(t)=\eta(t)$. More precisely,
when we discretize the time-axis in points $t_i$, the function
$z_3(t)$ becomes a set of variables $z_{3,i}$.  
Their correlations  and cross correlations are

\BEA \label{autoc}
&&\langle p_0^2\rangle=m_0T_p(a_0,m_0,\gamma_0)
,\qquad \langle x_0^2\rangle=\frac{T_x(a_0,m_0,\gamma_0)}{a_0}
,\qquad 
K(s,t)=\half\langle\eta(s)\eta(t)+\eta(t)\eta(s)\rangle,\\
&&\label{crossc}\langle p_0x_0\rangle=0, \qquad
S_1(t)=\half\langle \eta(t)p_0+p_0\eta(t)\rangle,
\qquad S_2(t)=\half\langle\eta(t)x_0+x_0\eta(t)\rangle\EEA
The new terms read
\BEA \label{S1t=} 
S_1(t)&=&\frac{\hbar m_0\sqrt{\gamma}}{\pi\sqrt{\gamma_0}}
\int_0^\infty\d\nu\,\nu \sin\phi_0(\nu)\,\sin(\phi_0(\nu)+\nu t)\,
\coth\half\beta\hbar \nu
\nn \\ \label{S2t=} 
S_2(t)&=&\frac{\hbar\sqrt{\gamma}}{\pi\sqrt{\gamma_0}}
\int_0^\infty\d\nu \sin\phi_0(\nu)\,\cos(\phi_0(\nu)+\nu t)\,
\coth\half\beta\hbar \nu
\EEA

\subsection{Exact solution of the Langevin equation}

Next we consider the exact solution of 
the quantum Langevin equation:
\BEA
\label{e1}
\dot{x}= \frac{p}{m},
\qquad
\dot{p}= - ax -\gamma(t)x(0)-
\int _{0}^{t}\d t' \gamma  (t-t')\dot{x}(t')+\eta (t),
\EEA

The general solution of Eq.~(\ref{e1})
is obtained with help of Laplace transformation. The reader may recall
the following standard relations between functions $A(t)$, $B(t)$ and 
their Laplace-transforms 
${\cal L}\{A\}=\hat{A}(s)=\int_0^\infty\d t\,e^{-st}A(t)$,
\BEA
{\cal L}\left \{
\int _0^t\d t' A(t-t')B(t')
\right \}=\hat{A}(s)\hat{B}(s),
\qquad
{\cal L}\left \{\dot{A}\right \}=-A(0)+s\hat{A}(s),
\EEA

One gets

\BEQ ms\hat x(s)-mx(0)=\hat p(s),\qquad 
s\hat p(s)-p(0)=-a\hat x(s)-\hat \gamma(s)s\hat x(s)+\hat\eta(s)
\EEQ
Together this yields 
\BEQ \hat x(s)=\frac{1}{m}[mx(0)s+p(0)+\hat\eta(s)]\,\hat f(s)\EEQ
where
\BEQ \hat{f}(s)=\frac{m}{ms^2+a+{s}\hat{\gamma}(s)}
\EEQ
Thus, the solution of Eq.~(\ref{e1}) reads
\BEA \label{xt=}
&&x(t)=x(0)\dot f(t)
+\frac{1}{m}p(0)f(t)+\frac{1}{m}\int_0^{t}\d t'
f(t-t')\eta (t'),
\\ \label{pt=}
&&p(t)=p(0)\dot{f}(t)+mx(0)\ddot f(t)+\int_0^{t}\d t'
\dot{f}(t-t')\eta (t')
\EEA
where 
$\omega _0 =\sqrt{a/m}$;
$\hat{f}(s)$ and $\hat{\gamma}(s)$ are the Laplace transforms of 
$f(t)$, $\gamma (t)$.
Expanding $\hat f(s)$ for large $s$, one finds that $f(0)=\ddot f(0)=0$, 
$\dot f(0)=1$. Now we turn to our standard case of the Drude-Ullersma spectrum:
\BEA
\gamma (t) = \gamma \Gamma e^{-\Gamma |t|},\qquad
\hat{\gamma } (s) = \frac{\gamma \Gamma}{\Gamma +s}
\EEA

For $\hat{f}(s)$ one has
\BEA \label{hatf=}
\hat{f}(s)=\frac{m(\Gamma+s)}
{(s+\Gamma )(ms^2+a)+s\gamma \Gamma}
=\frac{\Gamma+s }
{(s+\omega _1)(s+\omega _2)(s+\omega _3)}=\frac{\Gamma+s}{P_3(-s)}\EEA
where $P_3(s)$ was defined in (\ref{P3=}), where also its roots
$\omega _{1,2,3}$ are discussed.

Likewise one has for the initial Gibbsian states
\BEA \label{hatf0=}
\hat{f}_0(s)=\frac{m_0(\Gamma+s)}
{(s+\Gamma )(m_0s^2+a_0)+s\gamma_0 \Gamma}
\EEA

One may write
\BEQ \label{hatfexp} \hat f(s)=\sum_{i=1}^3\frac{f_i}{s+\omega_i},\qquad f(t)=
\sum_{i=1}^3{f_i}e^{-\omega_it},\qquad f_i=\frac{\Gamma-\omega_i}
{(\omega_{i+1}-\omega_i)(\omega_{i-1}-\omega_i)}
\EEQ
where, in this connection, $\omega_0=\omega_3$,
not to be confused with the definition $\omega_0=\sqrt{a/m}$ elsewhere
in the work.

For large $\Gamma$ one has 
\BEQ \label{f12=}f_1=-f_2=\frac{m}{\gamma w},\qquad f_3=\frac{1}{\Gamma}\EEQ
with $w$ defined in (\ref{eps=w=}). 
 
Let us now set, in analogy with (\ref{xbk=}),
\BEQ \label{xbeta=}
x(t)=\sum_k \sqrt{\frac{\hbar\Delta(\Gamma^2+\nu_k^2)}
{\pi\gamma\Gamma^2\nu_k}}\,\,[\beta_k(t)b^\dagger_k+\beta_k^\ast(t) b_k],
\EEQ This implies 
\BEQ p(t)=m\sum_k \sqrt{\frac{\hbar\Delta(\Gamma^2+\nu_k^2)}
{\pi\gamma\Gamma^2\nu_k}}\,\,[\dot\beta_k(t)b^\dagger_k+\dot
\beta_k^\ast(t) b_k]
\EEQ
One has from (\ref{xt=}), (\ref{hatf=}) and (\ref{hatfexp}):
$\beta_k=\beta(\nu_k)$ with
\BEQ \beta(\nu)=\sqrt{\frac{\gamma}{\gamma_0}}\,\sin\phi_0(\nu)
\sum_{i=1}^3 f_i(-\omega_i+i\nu )e^{-\omega_it}
+\frac{\gamma \Gamma^2\nu }{m(\Gamma^2+\nu ^2)}
\sum_{i=1}^3 f_i\left[\sin\phi_0(\nu)\frac{\Gamma-i\nu }{\nu }
\,\frac{e^{-\omega_it}-e^{-\Gamma t}}{\Gamma-\omega_i}
+e^{i\phi_0(\nu)\,}\frac{e^{i\nu  t}-e^{-\omega_it}}{\omega_i+i\nu }\right]
\EEQ
The $e^{-\Gamma t}$ terms cancel since, due to (\ref{hatf=}), 
\BEQ \sum_i\frac{f_i}{\omega_i-\Gamma}=\hat f(-\Gamma)=0\EEQ
Next one can check that
\BEQ\label{phifrel}
e^{i\phi(\nu)\,}=
\frac{m(\Gamma^2+\nu ^2)\sin\phi(\nu)}{\gamma\Gamma^2
\nu \hat f(i\nu )},\qquad e^{i\phi_0(\nu)}=
\frac{m_0(\Gamma^2+\nu ^2)\sin\phi_0(\nu)}{\gamma_0\Gamma^2
\nu \hat f_0(i\nu )} \EEQ
which brings
\BEQ \frac{\gamma \Gamma^2\nu }{m(\Gamma^2+\nu ^2)}\times
e^{i\phi_0(\nu)\,}\times\sum_{i=1}^3  \frac{f_i}{\omega_i+i\nu }
=\frac{\gamma m_0\hat f(i\nu )}
{\gamma_0 m\hat f_0(i\nu )}\sin\phi_0(\nu)
\EEQ
This leads to the exact result
\BEA \label{betanu=}
\beta(\nu,t)&=&\beta_0e^{i\nu t}
+\sum_{i=1}^3 \beta_i(\nu)e^{-\omega_it}\nn
\beta_0(\nu)&=&\sin\phi(\nu)\,e^{i\phi_0(\nu)-i\phi(\nu)}\nn
\beta_i(\nu)&=&\,\sin\phi_0(\nu)\, f_i\,
\left[
\sqrt{\frac{\gamma}{\gamma_0}}(-\omega_i+i\nu)
+\frac{\gamma \Gamma^2}{m(\Gamma+i\nu)(\Gamma-\omega_i)}
-\frac{\gamma m_0}{\gamma_0m\,\hat f_0(i\nu)(\omega_i+i\nu)}\right]
\EEA

For large times only the first term remains, and the initial condition
only enters through its phase factor $\exp(i\phi_0-i\phi)$, which has no 
physical effect, thus showing that the central particle relaxes to its 
equilibrium state independent of its initial condition.
The expression for $\beta_i$ can be simplified by writing it as the ratio of
two polynomials, and using 
the fact that $\omega_i$ is a zero of $1/\hat f(-\omega)$, 
allowing to eliminate the $\omega_i^3$ term of the numerator. This brings
\BEQ\label{betai=}
\beta_i(\nu)=
\,\sin\phi_0(\nu)\,\,\frac{a-a_0+(m_0-m)\nu^2}{m}\,\, 
\,\frac{f_i}{\omega_i+i\nu}
\EEQ
which is still exact. 
Using (\ref{phifrel}) we can also express the result as
\BEQ \label{betanut=} \beta(\nu,t)=
\frac{\gamma\Gamma^2\nu\,\, e^{i\phi_0(\nu)}}{m(\Gamma^2+\nu^2)}
\left\{\hat f(i\nu)e^{i\nu t}+[a-a_0-(m-m_0)\nu^2\,]\hat f_0(i\nu)
\sum_{i=1}^3 \frac{f_i}{\omega_i+i\nu}e^{-\omega_it}
\right\}\EEQ
It is trivial to check that, when no change is made 
at $t=0$ ($\gamma_0=\gamma$, $m_0=m$ and $a_0=a$), the 
result $\beta(\nu)=\sin\phi\,\exp(i\nu t)$ extends the
negative time behavior (\ref{xbk=}) to all positive times, 
even though the noise and the damping payed a
special (but in that case unphysical) role to $t=0$.

In the rest of this work we shall be mainly interested 
in the situation $\gamma_0=\gamma$,
$m_0=m$ while $a_0$ is different from but close to $a$. 
One gets in the regime
of large $\Gamma$ and $t\gg 1/\Gamma$ to linear order in $a-a_0$
\BEQ \label{bnu0=}
\beta(\nu)=\sin\phi(\nu)\,e^{i\phi_0(\nu)-i\phi(\nu)+i\nu t}\,\,
\left[1+\frac{a-a_0}{\gamma w}
\left(\frac{e^{-\omega_1t-i\nu t}}{\omega_1+i\nu}
-\frac{e^{-\omega_2t-i\nu t}}{\omega_2+i\nu}\right)\right]\EEQ
where $w$ is defined  in (\ref{eps=w=}) and $\omega_{1,2}$ in (\ref{mega1}).

\renewcommand{\thesection}{\arabic{section}}
\section{Energy oscillation and negative entropy production}
\renewcommand{\thesection}{\arabic{section}.}
\setcounter{equation}{0}
 
We consider the dynamical evolution of a system initially in 
equilibrium characterized by a spring constant $a_0$, which
at $t=0$ is instantaneously changed to $a_1=a$. 
These parameters are connected as
\BEQ a_0=(1-\alpha_0)a \EEQ
We  shall assume that $|\alpha_0|\ll 1$. We also assume 
a large Debye frequency $\Gamma$, but this does not lead to
principal changes.

\subsection{Non-monotonous relaxation of the energy at low $T$}

Let us now consider how the system relaxes to its steady state.
From (\ref{xbeta=}) and (\ref{boseocc}) one has 
\BEA \langle x^2\rangle=\int_0^\infty\d\nu\,\
\frac{\hbar(\Gamma^2+\nu^2)}
{\pi\gamma\Gamma^2\nu}\,\beta^\ast\beta\coth\half\beta\hbar\nu,
\qquad \langle p^2\rangle=m^2\int_0^\infty\d\nu\,\
\frac{\hbar(\Gamma^2+\nu^2)}
{\pi\gamma\Gamma^2\nu}\,\dot\beta^\ast\dot\beta\coth\half\beta\hbar\nu
\EEA
The infinite time values, discussed already in Eqs. 
(\ref{Txexact},\ref{Tpexact}), can be checked from these expressions. 
For the evolution from the initial state to these values,
we shall consider times $t\gg 1/\Gamma$, and we can just
take the limit $\Gamma\to\infty$ since no divergencies occur,
except for the $\ln\Gamma$ term of the static part $\langle p^2\rangle$ 
at $T=0$.  Inserting (\ref{bnu0=}) we get to linear order in $\alpha_0$

\BEA\label{Txaa0=} V(t)&=& \half a \langle x^2\rangle=\half T_x+
\alpha_0\frac{\hbar a}{2\pi\gamma}\,C_x(\frac{\gamma t}{2m}),\\
K(t)&=&\frac{\langle p^2\rangle}{2m}=\half T_p+
\alpha_0\frac{\hbar a}{2\pi\gamma}\,C_p(\frac{\gamma t}{2m}),\\
\label{Ualp=}
U(t)&=&\half T_p+\half T_x+\alpha_0\frac{\hbar a}{2\pi\gamma}
\,C_E(\frac{\gamma t}{2m})
\EEA
with the relaxation functions of coordinate, momentum and energy
\BEA \label{Cxtau=} 
C_x(\tau)&=&\frac{2(1-w^2)}{w}\int_\minfty^\infty
\frac{ \d y \,y\coth (\beta \hbar \gamma y/4m)}
{[(1+w)^2+y^2][(1-w)^2+y^2]}\left[
\frac{e^{-(1-w+iy)\tau}}{1-w+iy}-\frac{e^{-(1+w+iy)\tau}}{1+w+iy}
\right] \\
C_p(\tau)&=&\frac{2}{w}\int_\minfty^\infty
\frac{ \d y \,y\coth (\beta \hbar \gamma y/4m)}
{[(1+w)^2+y^2][(1-w)^2+y^2]}\left[
\frac{iy(1-w)e^{-(1-w+iy)\tau}}{1-w+iy}-
\frac{iy(1+w)e^{-(1+w+iy)\tau}}{1+w+iy}\right]\\
C_E(\tau)&=&\frac{2}{w}\int_\minfty^\infty
\frac{ \d y \,y\coth (\beta \hbar \gamma y/4m)}
{[(1-iy)^2-w^2]}\left[
\frac{(1-w)e^{-(1-w+iy)\tau}}{(1-w+iy)^2}-
\frac{(1+w)e^{-(1+w+iy)\tau}}{(1+w+iy)^2}\right]
\EEA
Of course, one just has $C_E=C_p+C_x$. 
The integration variable is $y=2m\nu/\gamma$.
The appearance of the dimensionless timescale $\tau=\gamma t/2m$ is 
natural, since in the underdamped regime, where $w$ is imaginary,
 the damping time is
just $\tau_d=2m/\gamma$, see Eq. (\ref{tau0taud}). In the overdamped
regime the timescales $\tau_x$ and $\tau_p$ from Eq. (\ref{tauxpe=})
are coded in the terms $(1\mp w)\tau$ in the exponentials, respectively. 
In particular,
for strong overdamping one has $t/\tau_x=(1-w)\gamma t/2m\to at/\gamma$.

\subsubsection{Classical regime}
At large $T$ the tanh linearizes. One can do contour integration
to find for overdamping, i.e. for $\eps=am/\gamma^2 <\frac{1}{4}$
and $w=\sqrt{1-4\eps}>0$, the exact results
\BEA \label{CxpElargeT}
C_x(\tau)&=&\frac{\pi\gamma T}{\hbar a}\,e^{-2\tau}\left[\cosh(w\tau)
+\frac{\sinh (w\tau)}{w}\right]^2\\
C_p(\tau)&=&\frac{\pi\gamma T}{\hbar a}\,e^{-2\tau}(1-w^2)
\left[\frac{\sinh(w\tau)}{w}\right]^2\\
\label{CEclass}
C_E(\tau)&=&
\frac{\pi\gamma T}{\hbar a}\,e^{-2\tau}\left\{\left[\cosh(w\tau)
+\frac{\sinh (w\tau)}{w}\right]^2+
(1-w^2)\left[\frac{\sinh(w\tau)}{w}\right]^2\right\}
\EEA
For overdamping ($0< w<1$) these functions are strictly positive.
For underdamping one has to replace $w\to i\bar w$, 
implying $\cosh w\tau\to\cos\bar w\tau$
and $\sinh (w\tau)/w\to\sin(\bar w\tau)/\bar w$. Then $C_x$ and $C_p$
get zeroes, but remain non-negative, while $C_E$ remains strictly positive.

For the relaxation of the energy this implies
\BEQ U=T+\half\alpha_0T\,\,\left\{\left[\cosh(w\tau)
+\frac{\sinh (w\tau)}{w}\right]^2+(1-w^2)
\left[\frac{\sinh(w\tau)}{w}\right]^2\right\}\,e^{-2\tau}
\EEQ
For strong overdamping ($\eps\to 0$, $w\approx 1-2\eps$)
this becomes a simple exponential decay,
\BEQ\label{sipleexp}
 U=T+\half\alpha_0T\, e^{-4\eps\tau}
\EEQ
In case of underdamping, $\eps>\frac{1}{4}$, 
one has $w=i\bar w$ with $\bar w=\sqrt{4\eps-1}$. 
This yields by analytic continuation
\BEQ U=T+\half\alpha_0T\,\left\{\left[\cos(\bar w\tau)
+\frac{\sin (\bar w\tau)}{\bar w}\right]^2+
\frac{1+\wb^2}{\wb^2}\sin^2(\bar w\tau)
\right\}\,e^{-2\tau}
\EEQ
The term multiplying $\alpha_0$ is an oscillating function, 
and it is strictly positive.  Its derivative,
\BEQ \label{dotUlT}
\dot U=-\frac{\gamma\alpha_0T}{m}\,\frac{1+\bar w^2}{\bar w^2}\,
\sin^2 (\bar w\tau)\,e^{-2\tau}
\EEQ
 has zeros but does not change sign.
Physically this means: depending on the sign of $\alpha_0$, both the
kinetic and the potential energy oscillate either above or below 
their final value, and the total energy flow is unidirected; 
it goes towards the bath when $a>a_0$, i.e. $\alpha_0>0$
and from the bath to the particle in the opposite case.

In the limit of weak damping ($\wb\gg1$) one gets
\BEQ \label{UdotU=}
U=T+\half\alpha_0T\,
\left\{1+\frac{\sin (2\omega_0t)}{\bar w}
\right\}\,e^{-\gamma t/m},\qquad
\dot U=-\frac{\gamma\alpha_0T}{m}\,
\sin^2 (\omega_0t)\,e^{-\gamma t/m}
\EEQ
Notice that the small but oscillating term in $U$ 
has become of leading order for $\dot U$.

\begin{figure}[bhb]
\vspace{0.1cm}\hspace{-1.5cm}
\vbox{\hfil\epsfig{figure= 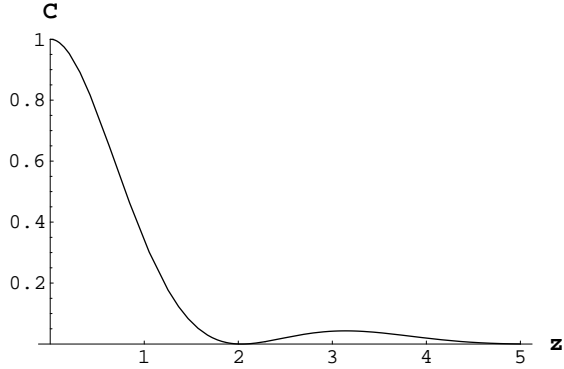,width=8cm,angle=0}\hfil}  
\vspace{0.75cm}
\caption{ The underdamped situation. $C_x$
as a function of the rescaled dimensionless time $z=\tau\delta=\tau/|w|$, 
normalized to unity at $z=0$, for large $T$ and $\delta=0.5$}
\label{espan4}
\end{figure} 

\begin{figure}[bhb]
\vspace{0.1cm}\hspace{-1.5cm}
\vbox{\hfil\epsfig{figure= 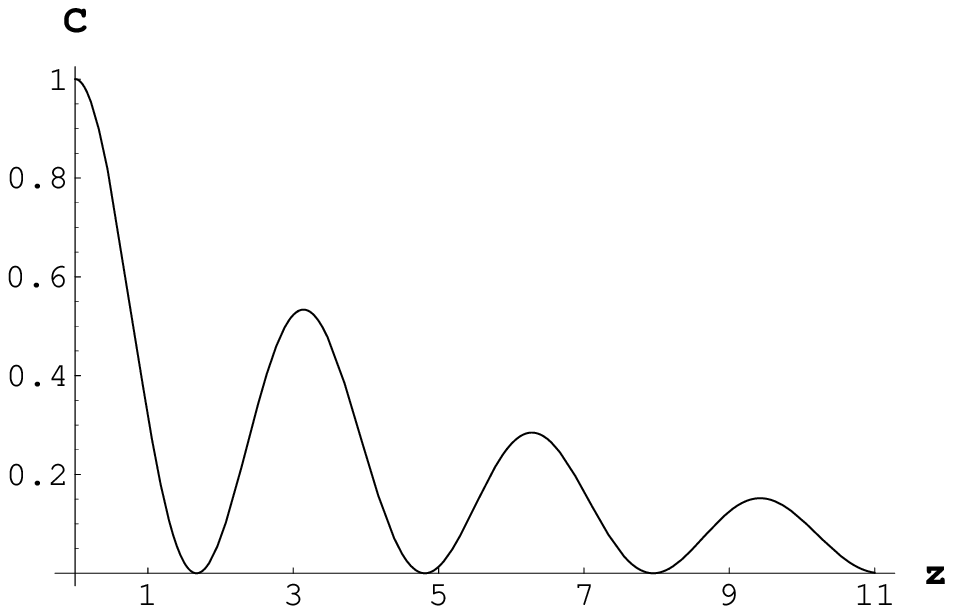,width=8cm,angle=0}\hfil}  
\vspace{0.75cm}
\caption{ The underdamped situation. $C_x$
as a function of the rescaled dimensionless time $z=\tau\delta=\tau/|w|$, 
normalized to unity at $z=0$, for large $T$ and $\delta=0.1$}
\label{espan5}
\end{figure} 

\subsubsection{Weak damping regime}

In the weak damping limit it holds that 
\BEQ \wb=-iw \approx\frac{2\sqrt{am}}{\gamma} \gg 1 \EEQ
For performing the integrals, 
we write $C_x(\tau)$ first as an integral from $0$ to
infinity, and make the shift $y\to \wb+u$, yielding up to order $1/\wb^2$
\BEA \label{Cxint2}
C_x(\tau)=e^{-\tau}\,\int_{-\wb}^\infty\d u\frac{\coth(b+bu/\wb)}{u^2+1}
\left(-\frac{u\cos u\tau+\sin u\tau}
{1+u^2}+\frac{\cos(2\omega_0t+u\tau)}{2\wb}\right)
\EEA
where a correction factor $1+u/\wb$ in numerator and denominator 
have canceled, and we denoted
\BEQ \label{b=} b=\half\beta\hbar\omega_0\EEQ
Evaluating this to leading order in $1/\wb$ we get
\BEA \label{Cxwd} C_x(\tau)=\frac{\pi\gamma }{4\sqrt{am}}\,
\left(\frac{\half\beta\hbar\omega_0}{\sinh^2\half\beta\hbar\omega_0}+
\coth\half\beta\hbar\omega_0\cos 2\omega_0t\,\right)\,e^{-2\tau}
\EEA

For large $T$ this agrees with Eq. (\ref{CxpElargeT}), of which 
the last factor now becomes $\cos^2\omega_0t$.
For $T=0$ and $\tau=0$ it agrees with Eq. (\ref{kuchma}) below.
Likewise
\BEA C_p(\tau)=\frac{\pi\gamma }{4\sqrt{am}}\,
\left(\frac{\half\beta\hbar\omega_0}{\sinh^2\half\beta\hbar\omega_0}-
\coth\half\beta\hbar\omega_0\cos 2\omega_0t\,\right)\,e^{-2\tau}
\EEA
As in the classical limit, $C_E'=C_x'+C_p'$ picks up a contribution
of $C_E$ that is subleading but oscillating; it is
most easily obtained by evaluating $C_E'$ in a manner similar to
(\ref{Cxint2}),
\BEA \label{CEpr} C_E'(\tau)=-\frac{\pi\gamma }{\sqrt{am}}
\left(\frac{\half\beta\hbar\omega_0}{\sinh^2\half\beta\hbar\omega_0}-
\coth\half\beta\hbar\omega_0 \cos2\omega_0t\right)\,e^{-2\tau}
\EEA

When inserting $C_p+C_x$ in Eq. (\ref{Ualp=}) we have
for the leading decay of the energy
\BEQ U=\half\hbar\omega_0\coth \half\beta\hbar\omega_0
+\half\alpha_0T  \frac{(\half\beta\hbar\omega_0)^2}
{\sinh^2\half\beta\hbar\omega_0}\,e^{-\gamma t/m}
\EEQ
showing  that to leading order in $\gamma$ in the weak damping limit 
the energy does not oscillate,
but monotonously leaks into the bath (when $\alpha_0>0$) or is taken
from the bath (when $\alpha_0<0$). At low temperature this happens
with an exponentially small rate. But the rate of energy transfer,
determined by (\ref{CEpr}), 
\BEQ\label{dotU=} \dot U=-\frac{\gamma \alpha_0T}{2m}\left( 
 \frac{(\half\beta\hbar\omega_0)^2}{\sinh^2\half\beta\hbar\omega_0}-
\frac{\half\beta\hbar\omega_0 }{\tanh \half\beta\hbar\omega_0}
\cos2\omega_0t
\right)\,e^{-\gamma t/m}
\EEQ
is an oscillating function that changes sign in each period
whenever $T$ is not infinite.
Thus the rate of energy transfer is not uni-directed except 
for the classical limit. When averaged over one period, the 
cosine is subleading and a unidirected flow emerges.

\subsubsection{Quantum regime for non-weak damping}
At $T=0$ one has $y\coth (\beta \hbar \gamma y/4m)=|y|$. 
For time $\tau=0$ one finds by direct integration
\BEA 
\label{Cx(0)=}
C_x(0)=\frac{1}{w^2}-\frac{1-w^2}{2w^3}\ln\frac{1+w}{1-w},
\EEA
and we define the short hand
\BEQ \label{lambda=} \lambda(w)=\frac{1}{2w}\ln\frac{1+w}{1-w}\EEQ
It further holds that
\BEA 
\label{kuchma}
C_p(0)=-C_x(0),
\qquad C_E(0)=0\EEA
These results can be verified using the relations
\BEA a\langle x^2(t=0^+)\rangle-T_x(a)&=&\frac{a}{a_0}T_x(a_0)-T_x(a)=
-\alpha_0a^2 \frac{\d (T_x/a)}{\d a}=\alpha_0\frac{\hbar a}{\pi\gamma}
C_x(0)\nn
 \frac{1}{m}\langle p^2(t=0^+)\rangle-T_p(a)&=&
-\alpha_0a \frac{\d T_p}{\d a}=\alpha_0\frac{\hbar a}{\pi\gamma}C_p(0)
\EEA
So after the instantaneous change of the spring constant ($t\gg1/\Gamma$),
the deviation of the potential energy from its final value is,
to leading order in $\alpha_0$, just opposite to the one of the kinetic 
energy. Consequently, the particle has already the proper energy,
but this will not remain so; for $\alpha_0>0$ first a flow from the bath
will occur and then a reversed flow, after which the equilibrium will
be reached by a second energy flow from the bath to the particle.

At large times ($\tau\gg 1$) and still $T=0$ 
one gets in case of overdamping ($w>0$)
\BEA 
\label{app}
C_x(\tau)=&&-\frac{1}{2\eps^2\tau^2}\,e^{-\tau}
[\cosh w\tau+\frac{\sinh w\tau}{w}]\nn
C_p(\tau)=&&-\frac{1}{\eps^2\tau^3}\,e^{-\tau}\,\frac{\sinh w\tau}{w}
\nn
C_E(\tau)=&&-\frac{1}{2\eps^2\tau^2}\,e^{-\tau}
[\cosh w\tau+\frac{\sinh w\tau}{w}]\EEA
and for underdamping ($\bar w>0$)
\BEA C_x(\tau)=&&-\frac{1}{2\eps^2\tau^2}\,e^{-\tau}
[\cos\bar w\tau+\frac{\sin\bar w\tau}{\bar w}]
\nn
C_p(\tau)=&&
-\frac{1}{\eps^2\tau^3}\,e^{-\tau}\,\frac{\sin\bar w\tau}{\bar w}\nn
C_E(\tau)=&&-\frac{1}{2\eps^2\tau^2}\,e^{-\tau}
[\cos\bar w\tau+\frac{\sin\bar w\tau}{\bar w}] \EEA
The latter expressions all exhibit an infinity of oscillations around $C=0$. 
For overdamping one has  $C_x(0)>0$, while it has a negative tail;
consequently there remains one oscillation even in the limit of strong 
damping. 
In that limit ($\gamma$ large), one may set
$\sigma=(1-w)\tau=at/\gamma$. For large, but fixed $\sigma$ one gets 
$C_x=C_E=-2\exp(-\sigma)/\sigma^2$, $C_p=-8\eps\exp(-\sigma)/\sigma^3 $.

\subsubsection{Strong damping at low $T$}

Let us now write $C_x(t)$ as
\BEA
&&C_x(t)=f_x(\tau,w)+f_x(\tau,-w),\\
&&f_x(\tau,w)=\frac{2(1-w^2)}{w}\,e^{-(1-w)\tau}\,\int_\minfty^\infty\d y\,
\frac{y\coth(\,b\,y/\sqrt{1-w^2}\,)[\,(1-w)\cos y\tau-y\sin y\tau\,]}
{[\,(1+w)^2+y^2\,]\,[\,(1-w)^2+y^2\,]^2},
\label{fff}
\EEA
where $b$ was defined in Eq. (\ref{b=}).

\begin{figure}[bhb]
\vspace{0.1cm}\hspace{-1.5cm}
\vbox{\hfil\epsfig{figure= 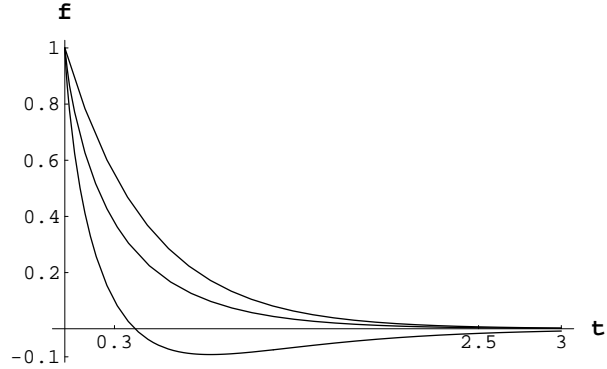,width=8cm,angle=0}\hfil}  
\vspace{0.75cm}
\caption{ The case of strong damping. $f_1(t)$
as function of dimensionless time $t$, normalized
to unity at $t=0$, for different values of the dimensionless
temperature $\theta=1/(b\sqrt{\eps})=2T\gamma/(\hbar a)$. 
Upper curve: $\theta\to\infty$ (the expression given by (\ref{bro}) taking
into the normalization). 
Middle curve: $\theta=1$ (see Eq.~(\ref{averoes})~).
Lower curve: $T=\theta=0$ (the expression given by (\ref{bora})~). 
In the latter case there is still an oscillation, despite the
strong damping.}
\label{fig1}
\end{figure}

We investigate in some greater detail  two particular cases:
$w\to 1$ (strong overdamping) and $w=0$ (the border between overdamping and
underdamping). For the first case one changes the integrating variable
$y\to y/(1-w)$ and arrives at:
\BEA
\label{averoes}
&&f_x(\tau,w)=f_1(2\eps\tau),\qquad f_1(\sigma)=e^{-\sigma}\,
\int_\minfty^\infty\d y\,
\frac{y\coth(\,b\,y\sqrt{\eps}\,)[\,\cos y\sigma-y\sin y\sigma\,]}
{[\,1+y^2\,]^2},\\
&&f_x(\tau,-w)=-\eps e^{-2\tau}\,
\int_\minfty^\infty\d y\,
\frac{y\coth(\,b\,y\sqrt{\eps}\,)\,\cos 2\eps y\tau}
{[\,1+y^2\,]^2}.
\EEA
Recall that $w=\sqrt{1-4\eps}$, and in the limit $\eps\to 0$ one has
$w=1-2\eps\to 1$.
It is seen that in this limit $f_x(\tau,-w)$ is small compared to
$f_x(\tau,w)$ due to an extra factors $\eps$ and, above all,
a quickly decaying exponential $e^{-2\tau}=e^{-\sigma/\eps}$. 
Thus, we will omit $f_x(\tau,-w)$. Then one has a scaling form: 
\BEA\label{bro1}
C_E(\tau,\eps)=
C_x(\tau,\eps)=f_1(2\eps\tau)=f_1(\frac{at}{\gamma})=f_1(\frac{t}{\tau_x}). 
\EEA
Notice also that for this function small and large
temperatures are determined by the dimensionless ratio:
$b\sqrt{\eps}=\half\beta \hbar a/\gamma=\half \hbar/(\tau_xT)$.
If this parameter is small (which is always achieved for large temperature
and also for fixed temperature and large damping), then 
we go back to the situation of eq. (\ref{sipleexp}),
\BEA
\label{bro}
f_1(2\eps\tau)=\frac{\pi T}{\hbar\omega_0\sqrt{\eps}}\,e^{-4\eps\tau}.
\EEA
In the zero-temperature limit one takes $y\coth
(b\,y\sqrt{\eps})=|y|$ and gets:
\BEA
\label{bora}
f_1(\sigma)=2e^{-\sigma}\,
\int_0^\infty\d y\,
\frac{y\,[\,\cos y\sigma-y\sin y\sigma\,]}
{[\,1+y^2\,]^2}.
\EEA
This function can be exactly expressed through Meijer functions, but
we will not write down this representation explicitly, since it is useful
only for numerical computations.  Notice that (\ref{kobalt}) can be
once more checked with help of (\ref{bora}).
The behavior of $f_1(\sigma)$ for different temperatures is presented in
Fig.~\ref{fig1}. 

It starts with $f(0)=1$, becomes negative at $\tau_0=0.407211889989$,
goes trough a minimum, and finally bends up to $0^-$ for  $\tau\to\infty$.
The minimum is characterized by
\BEQ  \label{mindata}
\sigma_{\rm min}=0.87908730804,\qquad f_1(\sigma_{\rm min})=-0.0918980496,
\qquad c_2\equiv f_1''(\sigma_{\rm min})=0.404842
\EEQ
In this limit $C_p$ has an interesting behavior. We discussed already that
$C_p(0)=-C_x(0)$. For small $\sigma$, $C_p$ quickly grows,
goes trough zero, and then becomes of order $\eps$, starting as
$\eps\ln 1/\sigma$ for small, but not too small $\sigma$. 
For finite $\sigma$ one thus has $C_E\approx C_x$, 
implying that now the total energy makes one oscillation, 
despite the strong damping.

For $\alpha_0>0$ it says that, after initially energy has been put on
the particle by the change of $a_0\to a$, this energy leaks away into 
the bath. However, at intermediate times more leaks away than 
in the final state, so a part has to come back at moderately late times.
This non-monotonous behavior (``bouncing'') 
is familiar of the noise correlator,
which is anti-correlated at large times in the quantum regime.

\begin{figure}[bhb]
\vspace{0.1cm}\hspace{-1.5cm}
\vbox{\hfil\epsfig{figure=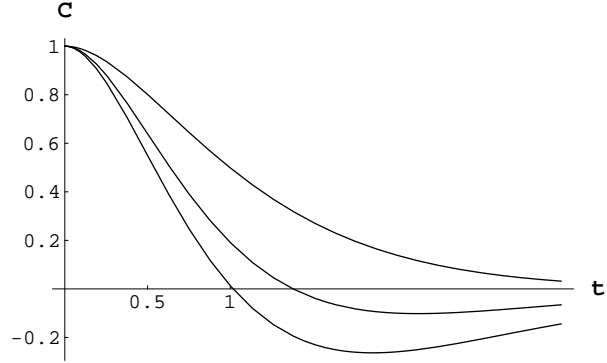,width=8cm,angle=0}\hfil}  
\vspace{0.75cm}
\caption{ The case with $w=0$ (the border between overdamping and
underdamping). $C_x(t)$
as a function of dimensionless time $t$, normalized
to unity at $t=0$, for different values of the dimensionless
temperature $\theta=1/b=T\sqrt{2m}/(\hbar \sqrt{a})$. 
Upper curve: $\theta=1$; 
middle curve: $\theta=0.2$;
lower curve: $T=\theta=0$. 
In the two latter cases there is an oscillation.}
\label{katon}
\end{figure}

Let us now turn to the behavior of $C_x$ for $w=0$:
\BEA
\label{tora}
C_x(\tau)=8\tau\,e^{-\tau}\,
\int_0^\infty\d y\,
\frac{y\,\coth (b\,y)\,[\,\cos y\tau-y\sin y\tau\,]}
{[\,1+y^2\,]^3}+8\,e^{-\tau}\,
\int_0^\infty\d y\,
\frac{y\,\coth (b\,y)\,[\,(1-y^2)\cos y\tau-2y\sin y\tau\,]}
{[\,1+y^2\,]^4}.
\EEA
The behavior of this function is depicted in Fig.~\ref{katon}. It is
seen that the cases $w=0$ and $w=1$ are qualitatively similar.
As expected, the negative tail of $C_x(t)$ is more pronounced for $w=0$.

\begin{figure}[bhb]
\vspace{0.1cm}\hspace{-1.5cm}
\vbox{\hfil\epsfig{figure= 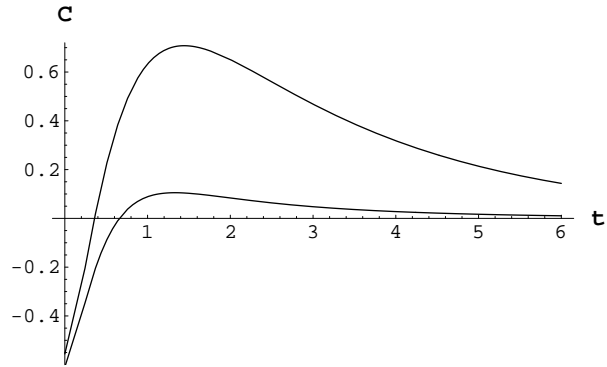,width=8cm,angle=0}\hfil}  
\vspace{0.75cm}
\caption{ The overdamped situation: $\epsilon=0.1$ ($w=0.7745$). $C_p(t)$
as a function of time, for different values of the dimensionless
temperature $\theta=1/(b\sqrt{\eps})=2T\gamma/(\hbar a)$. 
Upper curve: $\theta=33$; 
lower curve: $\theta=0$.}
\label{fig21}
\end{figure}

For $C_p(t)$ one has:
\BEA
\label{uruk1}
&&C_p(t)=f_p(\tau,w)+f_p(\tau,-w),\\
&&f_p(\tau,w)=\frac{2(1-w)}{w}\,e^{-(1-w)\tau}\,\int_\minfty^\infty\d y\,
\frac{y\coth(\,b\,y/\sqrt{1-w^2}\,)[\,y^2\cos y\tau+y(1-w)\sin y\tau\,]}
{[\,(1+w)^2+y^2\,]\,[\,(1-w)^2+y^2\,]^2}.
\label{uruk2}
\EEA
The behavior of $C_p(t)$ in the overdamped situation can be studied
along exactly the same lines as for $C_x(t)$,
\BEA
C_p(t)&&=2\epsilon\,
e^{-2\epsilon\tau}\,
\int_0^\infty\d y\,
\frac{y\coth(\,b\,y\sqrt{\eps}\,)[\,y^2\cos (2\,\epsilon\,\tau y)
+y\sin (2\,\epsilon\,\tau y) \,]}
{[\,1+y^2\,]^2\,(1+\epsilon^2y^2)}\nonumber\\
&&-2\,e^{-2\tau}\,
\int_0^\infty\d y\,
\frac{\coth(\,b\,y/\sqrt{\eps}\,)[\,y\cos (2\,\tau y)
+\sin (2\,\tau y) \,]}{[\,1+y^2\,]^2}.
\EEA
Due to the additional factor $\epsilon$, this is smaller than $C_x(t)$
for $\tau>1$, and this justifies Eq.~(\ref{bro1}).
Nevertheless, in the qualitative level $C_p(t)$ displays nearly the
same behavior as $C_{x}(t)$. This is demonstrated by Fig.~\ref{fig21}.

\begin{figure}[bhb]
\vspace{0.1cm}\hspace{-1.5cm}
\vbox{\hfil\epsfig{figure=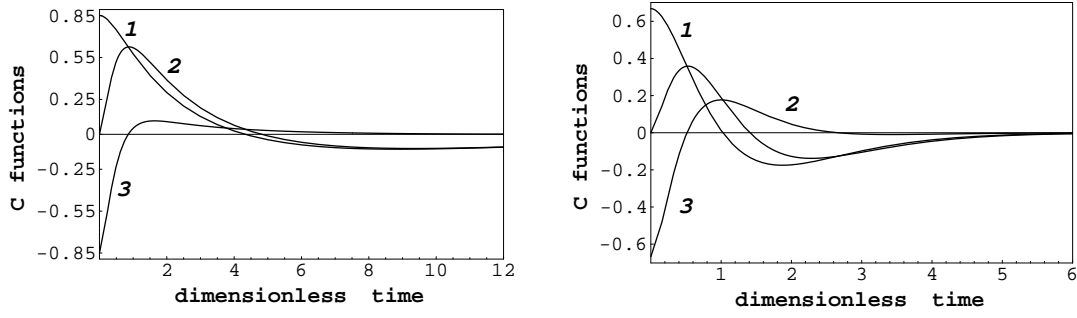,width=15cm,angle=0}\hfil}  
\vspace{0.75cm}
\caption{ The zero-temperature behavior of the $C$-functions versus
the dimensionless time. Left figure: $w=0.9$ (strong overdamping), 
{\it 1}: $C_x$, {\it 2}: 
$C_E=C_p+C_x$, {\it 3}: $C_p$. Right figure: the same but with $w=0.1$
(weak overdamping).
}\label{fig121}
\end{figure}

\begin{figure}[bhb]
\vspace{0.1cm}\hspace{-1.5cm}
\vbox{\hfil\epsfig{figure= 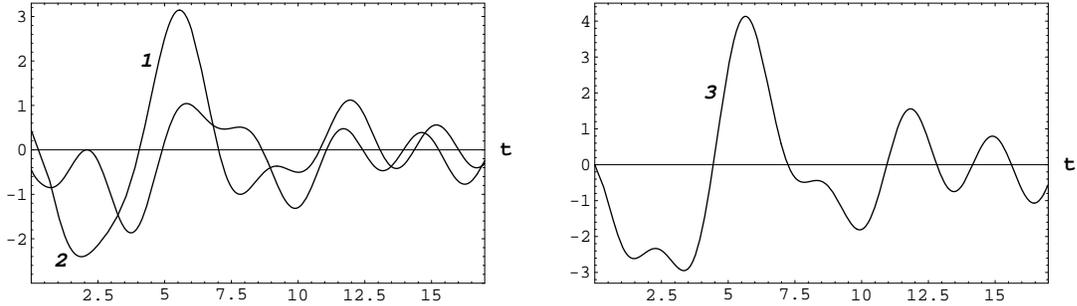,width=15cm,angle=0}\hfil}  
\vspace{0.75cm}
\caption{ The underdamped situation. ${\it 1}$: $C_x$
as a function of the rescaled dimensionless time $t=\tau/|\bar{w}|$, 
for $T=0$ and $\bar{w}=2$; ${\it 2}$: 
$C_p$ as a function of $t$;
${\it 3}$: 
$C_E$ as a function of $t$ both for the same values of the parameters. }
\label{espan1}
\end{figure}

\begin{figure}[bhb]
\vspace{0.1cm}\hspace{-1.5cm}
\vbox{\hfil\epsfig{figure= 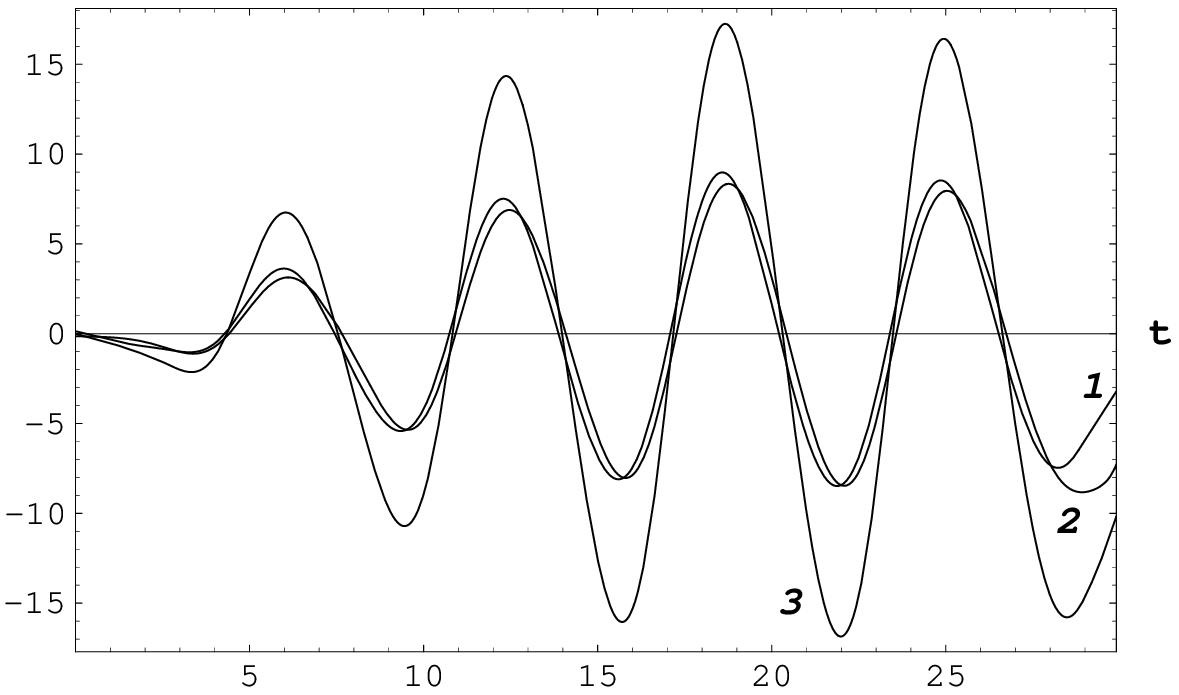,width=8cm,angle=0}\hfil}  
\vspace{0.75cm}
\caption{ The underdamped situation. ${\it 1}$: $C_x$, ${\it 2}$:
$C_p$, ${\it 3}$: $C_E$,
as functions of the rescaled dimensionless time
$t=\tau/|\bar{w}|$, for $T=0$ and $\bar{w}=10$. 
As compared with Fig.~(\ref{espan1}), the amplitude of oscillations is
much larger. It is seen also that $C_x$ and $C_p$ are quite close to
each other.} 
\label{espan2}
\end{figure}

Let us now investigate properties of $C_x(t)$ in the underdamped
limit, where $w=i\bar w=i\sqrt{4\eps-1}$, and $\bar w$ is real. In the additional
weak damping limit one has $\bar w\sim1/\gamma \to \infty$.
Using (\ref{fff}, \ref{averoes}) one gets:
\BEA
&&C_x(t)=\frac{8(\bar w^2+1)}{\bar w^4}\,e^{-\tau}\,{\rm Im}\left[\,e^{i\tau\bar w}
\int_0^{\infty}\d y\,\frac{y\coth \left(\frac{by\bar w}{\sqrt{\bar w^2+1}}\right)
\,[(1/\bar w-i)\cos \,(y\tau\bar w)
-y\sin \,(y\tau\bar w)\,]}{[(1/\bar w+i)^2+y^2]\,[(1/\bar
w-i)^2+y^2]^2}\right],\\
&&C_p(t)=\frac{8}{\bar w^2}\,e^{-\tau}\,
{\rm Im}\left[\,(1/\bar{w}-i)\,e^{i\tau\bar w}
\int_0^{\infty}\d y\,\frac{y\coth \left(\frac{by\bar w}{\sqrt{\bar w^2+1}}\right)
[\,y^2\cos \,(y\tau\bar w)
+(1/\bar w-i)y\sin \,(y\tau\bar w)\,]}{[(1/\bar w+i)^2+y^2]\,[(1/\bar
w-i)^2+y^2]^2}\right].
\EEA
The behavior of these functions, as well as $C_E(t)=C_x(t)+C_p(t)$, 
is depicted in Figs.~\ref{espan1},
\ref{espan2} for $T=0$. It is seen that for the initial time of order
$1/\bar w$, $C_x$ and $C_p$ oscillate with the amplitude
higher for larger $\bar{w}$. 

\subsubsection{Moments of the relaxation functions at low temperature}

We can also determine the integral
\BEQ
\int_0^\infty \d\tau C _x(\tau)=
8(1-w^2)\int_\minfty^\infty
\frac{ \d y \,y\coth (\beta \hbar \gamma y/4m)}
{[(1+w)^2+y^2]^3[(1-w)^2+y^2]^3}
[(1-w^2)^2-2y^2(1+w^2)-3y^4]
\EEQ

At $T=0$ it can be simply checked it vanishes at $w=0$ and near $w=1$:
\BEQ \int_0^\infty \d\tau C _x(\tau)=8[B(1,5)-2B(2,4)-3B(3,3)]=0;\quad
\int_0^\infty \d\tau C _x(\tau)
=(1-w)\int_\minfty^\infty
{\d y \,|y|}\frac{(1-w)^2-y^2}{[(1-w)^2+y^2]^3}=0\EEQ
where $B(w,z)=\Gamma(w)\Gamma(z)/\Gamma(w+z)$ is the Beta-function.
It actually holds that for all $w$  that 
\BEQ
\label{kobalt}
\int_0^\infty \d\tau C_x(\tau, T=0)=0
\EEQ
This surprising zero-temperature outcome will have important 
consequences when it comes to work extraction. It is similar to 
$\int_\minfty^\infty\d t\,K(t)=2\gamma T\to 0$ for $T\to 0$, 
where $K(t)$ is the autocorrelation function of the quantum noise. 
For small $T$ one gets
\BEQ
\label{kobaltT}\label{C0(T)=}
C_x^{(0)}(T)=\int_0^\infty \d\tau C_x(\tau)=\frac{64 \pi^2}{3(1-w^2)}\,
\left(\frac{m T}{\hbar \gamma(1-w^2)}\right)^2=
\frac{1}{12\eps}\,
\left(\frac{2\pi\gamma T}{\hbar a}\right)^2
\EEQ

It can also be verified that at zero $T$
\BEA \label{C1(T)=}
C_x^{(1)}&=&-\int_0^\infty\d\tau\,\tau C_x(\tau,T=0)=
\frac{1}{24\eps^2}\,\tilde C_x^{(1)}\nn
\tilde C_x^{(1)}&=&
\frac{(3+ w^2)(3w^2-1)+3(1-w^2)^3\lambda(w)}{8w^4}\,
\EEA
where $\lambda$ is defined in Eq. (\ref{lambda=}),  and
\BEQ \label{C2(T)=}
C_x^{(2)}=-\int_0^\infty\d\tau\,\tau^2C_x(\tau,T=0)=\frac{8}{3(1-w^2)^3}
=\frac{1}{24\eps^3} \EEQ
These coefficients are positive for all $w$. The minus signs in the integrals arise because the negative tail of $C_x(\tau,T=0)$ 
gets a larger weight than its positive center. 
These results follow from the Laplace-transform

\BEA\label{hatCx=}
&&\hat C_x(2u,T=0)=\int_0^\infty\d\tau \,C_x(\tau,T=0)\,e^{-2u\tau}\nn
&&=\frac{1}
{2uw^2} +\frac{1-w^2}{8u^2(1+u)}\left[
\frac{(1+2u-w)\ln(1+2u-w)}{(u-w)w(1+u-w)}
-\frac{(1+2u+w)\ln(1+2u+w)}{(u+w)w(1+u+w)}\right.
\\&& \left.\frac{2u^2(1+u)^2+2u(1+u)w+w^2-w^3}
{(u+w)w^3(1+u-w)}\ln(1-w)-\frac{2u^2(1+u)^2-2u(1+u)w+w^2+w^3}
{(u-w)w^3(1+u+w)}\ln(1+w)\right]\nonumber
\EEA

At $T=0$ all even moments of $C_p$ vanish. This implies
in particular that the integral of $C_E=C_p+C_x$ vanishes. 
The Laplace transform of $C_E$ reads in that case
 \BEA\label{hatCE=}
\hat C_E(2u,T=0)&=&\int_0^\infty\d\tau \,C_E(\tau,T=0)\,e^{-2u\tau}\nn
&=&
\frac{(1+w)(1+2u+w)}{4u^2w(1+u)(1+u+w)}   
 \ln\frac{1+2u+w}{1+w}-
\frac{(1-w)(1+2u-w)}{4u^2w(1+u)(1+u-w)}\ln\frac{1+2u-w}{1-w}
\EEA
The Laplace transform $\hat C_p(2u,T=0)$ follows 
as $\hat C_E(2u,T=0)-\hat C_x(2u,T=0)$.

For later use we introduce the coefficients
\BEA\label{CE012=} C_E^{(0)}
&=&\int_0^\infty\d\tau  C_E(\tau)=
\frac{1}{12\eps}\,\left(\frac{2\pi\gamma T}{\hbar a}\right)^2
+\O(T^4)\nn
C_E^{(1)}
&=&-\int_0^\infty\d\tau \tau C_E(\tau)=\frac{1}{24\eps^2}+\O(T^2)
\\ C_E^{(2)}
&=&-\int_0^\infty\d\tau^2 \tau C_E(\tau)=\frac{1}{24\eps^3}+\O(T^2)
\nonumber\EEA 
They differ from the $C_x^{(0,1,2)}$ only by the factor 
$\tilde C_x^{(1)}$, which goes to unity for large damping.

\subsection{Entropy production versus energy dispersion}

To derive the rate of entropy production we first need 
the Wigner function and its temporal evolution.

\subsubsection{Fokker-Planck equation for the Wigner function}

To derive the evolution equation for the Wigner function,
we shall write the Langevin eq. (\ref{pdot=}) in the form
\BEQ \dot p = -ax-\frac{\gamma}{m}p+\eta +\delta\dot p\EEQ
where $\delta\dot p$ is small as $1/\Gamma$. Indeed, from this definition and 
the exact dynamical solution (\ref{xt=},\ref{pt=}) one may derive 
\BEQ \delta\dot p(t)=mx_0\dot g(t)+p_0g(t)+\int_0^t\d t'\,g(t-t')\eta(t')\EEQ
where
\BEQ 
g(t)=\frac{\gamma}{m}\sum_{i=1}^3\frac{\omega_i^2}{\Gamma-\omega_i}
\,f_i\,e^{-\omega_it}
\EEQ 
is of order $1/\Gamma$ for large $\Gamma$. 
Now recall that for the harmonic situation the Wigner function is
given us
\begin{equation}
\label{wigner}
W(p,x,t)=
\int \d p_0~\d x_0 W(p_0,x_0,0)
\langle \delta (p(t)-p)\delta (x(t)-x)\rangle,
\end{equation} 
where the average is taken with respect to the noise,
$W(p,x,t)$ and $W(p_0,x_0,0)$ are final and initial 
Wigner functions,
while $p(t)$, $x(t)$ are the solutions of (\ref{pdot=})
for the corresponding initial conditions, and 
for a particular realization of the Gaussian noise.
Eq.~(\ref{wigner}) is not the most general definition of the Wigner
function, but it is exact for harmonic systems.

We now seek a closed equation for the Wigner function (\ref{wigner}).
Differentiating $W(y_1,y_2,t)$ we get 
\begin{equation}
\label{11}
\frac{\partial W(y_1,y_2,t)}{\partial t} =
-\sum_{k=1}^2\frac{\partial (v_k W)}{\partial y_k} - \frac{\partial }
{\partial y_1}
\langle \delta (p(t)-y_1)\delta (x(t)-y_2)[\eta (t)+\delta\dot p(t)]\rangle,
\end{equation}
where 
\begin{equation}
\label{11.1}
v_1=-ax-\frac{\gamma}{m}p,\quad
v_2=\frac{p}{m}.
\end{equation}
are the damped Newtonian acceleration and the velocity, respectively.
The term $\delta\dot p$ is a linear combination of $p_0$, $x_0$ and $\eta(t')$.
Due to the Gibbsian initial state, these are Gaussian random variables
and their cross-correlations were given in Eqs. (\ref{autoc})-(\ref{S2t=}). 
Let us denote these variables by the vector $z=\{p_0,x_0,\eta(t)\}$, 
and their correlations by the matrix $M_{ij}=\langle z_iz_j\rangle$. 
One then has for its joint distribution 
\BEQ P_0(z)=\frac{1}{\sqrt{{\rm det}(2\pi M)}} \,\exp(-\half z_iM^{-1}_{ij}z_j) 
\EEQ
Using  the relation 
\BEQ z_iP_0(z)=-\sum_jM_{ij}\frac{\delta}{\delta z_j}P_0(z) \EEQ
one can perform a partial integrations, which brings a closed equation 
for $W$.
The final result is that we obtain a diffusion-type equation 
(Fokker-Planck-Kramers-Klein equation) for $W$ itself:
\begin{equation}
\label{ko1}
\frac{\partial W(p,x,t)}{\partial t}={\cal L}W=
-\frac{p}{m}\frac{\partial W}{\partial x}+\frac{\partial}{\partial p}
\left [(\frac{\gamma}{m}p+ax)W \right]+[D_x(t)-D_p(t)]
\frac{\partial ^2}{\partial p\partial x}W
+\gamma D_{p}(t)
\frac{\partial ^2W}{\partial p^2},
\end{equation}
where the diffusion coefficients $D_{x}$ and ${D}_{p}$ are instantaneous 
functions $t$. (Notice that in ref. ~\cite{ANQBMprl}
we used the notation  $D_{pp}=D_p$, $D=D_x$, $D_{xp}=D_x-D_p$.)
The derivation along this road is somewhat lengthy.
A quicker way to derive the result is to use the solution of the 
Fokker-Planck equation, determined by the moments
\BEQ \langle p^2\rangle=2mK(t),\qquad \langle px\rangle=\frac{m}{a}\dot V(t),
\qquad \langle x^2\rangle=\frac{2}{a}V(t)\EEQ
where $K(t)=\langle {\cal K}(p)\rangle$ and $V(t)=\langle {\cal V}(x)\rangle$ 
are the expectation values of kinetic and potential energy, respectively. 
The time-dependent Wigner function thus reads
\BEQ W(p,x,t)=
\frac{a}{2\pi\sqrt{4amKV-m^2\dot V^2}}\,
\exp\left(-\frac{aKx^2-\dot V px+Vp^2/m}{4KV-m\dot V^2/a}\right) 
\EEQ
Inserting this in equation (\ref{ko1}), one finally gets
\BEQ D_p(t)=2K(t)+\frac{m}{\gamma}[\dot K(t)+\dot V(t)] \EEQ
\BEQ D_x(t)=2V(t)+\frac{m}{\gamma}[\dot K(t)+\dot V(t)] 
+\frac{\gamma}{a}\dot V(t)+\frac{m}{a}\ddot V(t) \EEQ

Let us also define the time-dependent but current-less state 
\begin{eqnarray}
\label{ko33}
W_{st}(p,x,t)=\frac{\sqrt{a}}{2\pi\sqrt{mD_p(t)D_x(t)}}
\exp \left(-\frac{p^2}{2mD_p(t)}-\frac{ax^2}{2D_x(t)}\right)
\end{eqnarray}
for which indeed the right hand side of the Fokker-Planck equation 
(\ref{ko1}) vanishes, though the left hand side does not. 
This is the locally-stationary distribution.
For sufficiently long times, that is when $D_x(t)$ and $D_p(t)$ are
changing with time slowly enough, $W_{st}$ becomes a solution of
the Fokker-Planck equation.

\subsubsection{H-function and entropy production}

The {\it $\H$-function} is defined as the information theoretical
distance between the actual Wigner function $W(x,p,t)$ and the
locally-stationary Wigner function $W_{st}(x,p,t)$:
\BEA
{\cal H}=\int \d x\,\d p\,
W(x,p,t)\ln\frac{W(x,p,t)}{W_{st}(x,p,t)}\ge 0. 
\EEA
The ${\cal H}$-function is non-negative due to the inequality
\BEA
&&\frac{W}{W_{st}}\ln \frac{W}{W_{st}}\ge \frac{W}{W_{st}}-1,\\
&&{\cal H} 
 \ge \int \d x \d p[W(x,p,t)-W_{st}(x,p,t)]=0.
\EEA
Thus, ${\cal H}$ is equal to zero only for $W(x,p,t)=W_{st}(x,p,t)$, i.e., 
in the stationary state. Since values of ${\cal H}$ at intermediate
times are higher than its final value, it is reasonable to look
at its rate of change.
In particular, ${\cal H}$ changes with time due to the time-dependence
of the reference Wigner distribution $W_{st}$, whereas the remaining
part of $\d {\cal H}/\d t$ appears to be induced solely by the bath
(see below). 
We define the {\it entropy production} $\d _iS/\d t$ by
\BEQ
-\frac{\d {\cal H}}{\d t}=\int\d x\,\d p\,
\dot{W}_{st}(x,p,t)\,\frac{W(x,p,t)}{W_{st}(x,p,t)}
+\frac{\d _iS}{\d t} . 
\EEQ
This leads to
\BEQ
\frac{\d_iS}{\d t}=-\int\d x\,\d p\,
\dot{W}(x,p,t)\,\ln \frac{W(x,p,t)}{W_{st}(x,p,t)}  
\EEQ
This definition has the following properties: 1) It is equal to zero
in the stationary state; 2) It is equal to zero if
the brownian particle does not couple with the bath; 
3) It is non-negative in the classical case, where $D_x=D_p=T$. The last
two properties are proved below.

Using (\ref{ko33}) and denoting
\BEA
R(x,p,t)=\frac{W(x,p,t)}{W_{st}(x,p,t)},
\EEA
one gets:
\BEA
\frac{\d {\cal H}}{\d t}=&&
\int\d x\,\d p\,[\,{\cal L}\,W(x,p,t)\,]\,\ln R(x,p,t)-
\int\d x\,\d p\,R(x,p,t)\,\dot{W}_{st}(x,p,t)\nonumber\\
=&&\int\d x\,\d p\,W(x,p,t)\,{\cal L}^{\dagger}\ln R(x,p,t)-
\int\d x\,\d p\,R(x,p,t)\,\dot{W}_{st}(x,p,t).
\EEA
where ${\cal L}$ is the Fokker-Planck operator of the 
right hand side of Eq. (\ref{ko1}). Noting that
\BEA
{\cal L}^{\dagger}\ln R=\frac{1}{R}\,{\cal L}^{\dagger}\,R
-\frac{1}{R^2}\left(
\gamma D_p(t)\left[\frac{\partial R}{\partial p}\right]^2+
[D_x(t)-D_p(t)]\frac{\partial R}{\partial x}\,\frac{\partial
R}{\partial p}
\right),
\EEA
and making once more integration by parts one ends up with
\BEA
\frac{\d _iS}{\d t}=\int\d x\,\d p\,
\frac{W(x,p,t)}{R^2(x,p,t)}\left(
\gamma D_p(t)\left[\frac{\partial R(x,p,t)}{\partial p}\right]^2+
[D_x(t)-D_p(t)]\frac{\partial R(x,p,t)}{\partial x}\,\frac{\partial
R(x,p,t)}{\partial p}\right).
\EEA
Now it is clear that in the classical white-noise limit, where
$D_x=D_p$, the entropy production is non-negative.
The positivity of $\d _iS/\d t$ just means that from the
global viewpoint the approach to the stationary state is monotonous. 
In contrast, in the quantum case the positivity of the entropy
production is endangered. It is also clear that for a free brownian 
particle ($\gamma=0$) the entropy production is zero. 

Finally we mention that the difference between $\d S_B/\d t$ and the
entropy production is just the entropy flux:
\BEA
\frac{\d _eS}{\d t}\equiv\frac{\d S_B}{\d t}-\frac{\d _iS}{\d t}=
-\int\d x\,\d p\,\dot{W}(x,p,t)\ln W_{st}(x,p,t).
\EEA
It takes the value
\BEQ
\frac{\d _eS}{\d t}=
\frac{\dot K(t)}{D_p(t)}+\frac{\dot V(t)}{D_x(t)} \EEQ
Let us recall that in the relaxation process no work is performed,
so a change in energy can only be due to a change of heat 
exchanged with the bath. Therefore the last relation can be written as
\BEQ \frac{\d _eS}{\d t}=\frac{\dot \Q_p }{D_p}+\frac{\dot \Q_x}{D_x}\EEQ
where $\dot Q_p=\dot K$ and $\dot Q_x=\dot V$ are the changes of
heat in the momentum and coordinate sector, respectively,
while $D_{p,x}$ are the corresponding diffusion coefficients in the 
Fokker-Planck operator ${\cal L}$ of Eq. (\ref{ko1}).  
Notice that this entropy flow deviates from the standard expression
$\d_eS/\d t=\dot\Q/T=\dot\Q_p/T+\dot\Q_x/T$, that could not make sense
since $\dot\Q$ does not scale with $T$ at low $T$. 

The Boltzmann entropy reads
\BEQ
S_B=-\int\d p\d x W(p,x,t)\ln \frac{1}{\hbar}W(p,x,t)=1+
\half\ln\left[\frac{m}{\hbar^2a}\left(4KV-\frac{m}{a}\dot V^2\right)\right]\EEQ
Its rate of change is 
\BEQ
\frac{\d S_B}{\d t}=
\frac{2a\dot KV+2aK\dot V-m\dot V\ddot V}{4aKV-m\dot V^2} \EEQ

The entropy production is the difference between them and appears 
to be quadratic in the deviation from the equilibrium state.
To second order in the small parameter $\alpha_0$ it becomes 
\begin{equation}
\label{Sprod}
\frac{\d _iS}{\d t}=\frac{\hbar^2 \gamma a\alpha_0^2}{16\pi^2 m^2}\left\{
\frac{C_x'^2}{T_x^2}+\eps (C_x'+C_p')(\frac{C_p'}{T_p^2}+\frac{C_x'}{T_x^2})
+\frac{T_p-T_x}{2T_pT_x^2}\,C_x'C_x^{\,\prime\prime}\right\}
\EEQ
where $C_{p,x}'$ denote the dimensionless derivatives 
$\d C_{p,x}(\tau)/\d\tau$.

\subsubsection{Classical limit}
In the classical limit, where $T_x=T_p=T$, the rate of entropy production
this becomes the sum of two squares, much alike the energy relaxation
function $C_E$ of eq. (\ref{CEclass}). 

\BEQ \frac{\d _iS}{\d t}=4\frac{a \alpha_0^2}{\gamma}\,e^{-4at/\gamma}
\,\frac{\sinh^2w\tau}{w^2}
\left\{\left(\cosh w\tau+\frac{\sinh w\tau}{w}\right)^2+(1-w^2)
\frac{\sinh^2w\tau}{w^2}\right\}\EEQ

The total entropy production is thus
\BEQ \Delta_i S=\int_0^\infty\d t\frac{\d _iS}{\d t}=\frac{\alpha_0^2}{4}\EEQ
This result holds for all $w$.

In Eq. (\ref{DelPiif}) we derive the general result for the energy 
dispersion. In the present setup we have $\alpha(t)=\alpha_0\delta(t)$,
yielding
\BEQ \label{dPia0}
\Delta\Pi=\frac{\hbar a\alpha_0^2}{4\pi\gamma}\,C_x(0)\EEQ
With help of Eq. (\ref{CxpElargeT}) we find
\BEQ \frac{\Delta\Pi}{T}=\frac{\alpha_0^2}{4}\EEQ
This just coincides with $ \Delta_i S$, explaining that both 
describe the same physics.

In the strong damping limit $w\to 1$ one a simple exponential decay, 
\BEQ \frac{\d _iS}{\d t}=\frac{a \alpha_0^2}{\gamma}\,e^{-4at/\gamma}\EEQ
In the weak damping limit, but still at high temperatures, the result 
oscillates, but is non-negative, 
\BEQ \label{entprodbTsg}
\frac{\d _iS}{\d t}=\frac{\gamma \alpha_0^2}{m}\,\sin^2\omega_0t\,
e^{-2\gamma t/m}\EEQ
Notice the similarity with the rate of energy decay (\ref{UdotU=}).

\subsubsection{Weak damping limit at moderate temperature}
\label{weakentr}

In the weak-damping limit $\gamma\to0$, where $T_p=T_x=\half \hbar\omega_0
{\rm coth}(\half\beta\hbar\omega_0)$, the entropy production follows
from (\ref{Sprod}), (\ref{Cxwd}) and (\ref{CEpr}) to leading order as
\BEA\label{Sprodwd} &&\frac{\d _iS}{\d t}=\frac{\gamma\alpha_0^2}{4m}
\left(\sin^22\omega_0t+(\cos2\omega_0t-
\frac{\beta\hbar\omega_0}{\sinh\beta\hbar\omega_0})^2+
\frac{2}{\pi}\Delta \psi\,\tanh \half\beta\hbar\omega_0\sin4\omega_0t
\right) e^{-2\gamma t/m}
\EEA
The term $\Delta\psi$  comes from the 
difference $T_p-T_x$, given in (\ref{Tpexact}). It reads 
\BEQ \Delta\psi=\psi(\frac{\beta\hbar\Gamma}{2\pi})-
\Re\psi(i\frac{\beta\hbar\omega_0}{2\pi})
\EEQ
At high $T$ the last contribution of (\ref{Sprodwd}) vanishes 
(at least, it is of order $1/\Gamma$, which we discard everywhere 
in this work), so Eq. (\ref{entprodbTsg}) is recovered.

 The term with $\Delta\psi$ is responsible for the
occurrence of both positive and negative values of the rate of entropy 
production. At low $T$ one has $\Delta\psi\to \ln{\Gamma}/{\omega_0}$, 
which is moderately large. Therefore, below some specific temperature 
$T^\ast\sim \hbar\omega_0/\ln(\Gamma/\omega_0)$  the rate of entropy 
production can be negative, a surprising result. This finding goes
against the formulation of the second law in the form of positivity of 
the entropy production. In our system the negative rates are not totally
unexpected since oscillatory behavior is also exhibited already in the 
rate of energy decay (\ref{dotU=}). 

The integrated entropy production is, to leading order in $\gamma$, 
insensitive to the oscillations. When averaged over one period, the 
cosine and sine are subleading and a positive rate emerges.
The full integral reads 

\BEQ \Delta_i S=\frac{\alpha_0^2}{8}\left(
1+\frac{(\beta\hbar\omega_0)^2}{\sinh^2\beta\hbar\omega_0}\right)
\EEQ
while the energy dispersion is
\BEQ \frac{\Delta\Pi}{T_x}=\frac{\alpha_0^2}{8}\left
(1+\frac{\beta\hbar\omega_0}{\sinh \beta\hbar\omega_0}\right) \EEQ
Both expressions have the same order of magnitude, and coincide
at large and small $T$.

Consequently, in the Gibbsian limit the rate of entropy production
 oscillates in case of underdamping, as does the rate of internal 
energy. After averaging over one period the oscillations are wahsed out.
This justifies our identifications of entropy flux and production.

\subsubsection{Entropy production at zero temperature}
\label{strongentr}

Also at zero temperature the entropy production can be negative. 
Let us consider
the case of  strong damping, where $\eps\ll 1$ and $T_x\sim\eps T_p$, 
implying
\begin{equation}
\label{o223}
\frac{\d _iS_x}{\d t}\approx \frac{\hbar^2 \gamma \alpha_0^2}{16\pi^2 m^2
T_x^2}\left\{
(1+\eps)C_x'^2+\half C_x'C_x''\right\}=\frac{a(1+\eps)}{4\gamma\lambda^2}
f'(\sigma)[f_1'(\sigma)+\eps f_1''(\sigma)]+\O(\eps^2)
\EEQ
Now we know that $f_1$ has a negative minimum at $\sigma_{\rm min}$.
Let us expand, using the numerical constants from Eq. (\ref{mindata}),
 \BEQ f_1(\sigma)=f_1(\taum)+\half c_2(\sigma-\taum)^2 \EEQ
Then we get
\begin{equation}
\label{dSidt}
\frac{\d _iS_x}{\d t}=\frac{a(1+\eps)\alpha_0^2}{4\gamma\lambda^2}
c_2^2(\sigma-\taum)(\sigma-\taum +\eps)
\EEQ
This is negative for $\taum-\eps<\sigma<\taum$. The minimum is of order
$-\eps^2$, and the area of the negative part is of order $\eps^3$.
Notice that negative value holds over a time window $\delta\sigma=\eps$,
corresponding to $\Delta t=\tau_p=m/\gamma$. This is much less than the
free oscillation period $\tau_0=\sqrt{m/a}$, so after averaging over
one period it disappears. However, in the (strongly) overdamped regime
there are no oscillations, so there is no compelling reason to average
over one period.

For a numerical investigation of the entropy production at $T=0$ 
we will first of all introduce a new parametrization for the
effective temperatures:
\BEA
&&T_p=\frac{\hbar\gamma}{\pi m}\,\theta_p,\qquad
\theta_p=\frac{1}{4w}\left[
(1+w)^2\ln\left(\frac{\Lambda}{1+w}\right)-
(1-w)^2\ln\left(\frac{\Lambda}{1-w}\right)
\right],\\
&&T_x=\frac{\hbar\gamma}{\pi m}\,\varepsilon\,\theta_x,\qquad
\theta_x=\frac{1}{w}\ln\frac{1+w}{1-w},\\
&&\theta=\frac{\theta_x}{\theta_p},
\EEA
where $\Lambda=2m\Gamma/\gamma$ is a large dimensionless parameter.
Then Eq.~(\ref{Sprod}) can be presented in a more convenient form:
\BEA
\frac{\d _iS}{\d
t}=\frac{a\alpha^2_0}{16\gamma}\,\frac{1}{\varepsilon^2\theta^2_x} 
\left(
C'^2_x+\varepsilon (C'_x+C'_p)(C'_x+\theta^2\varepsilon^2C'_p)
+\frac{1-\theta}{2}\,C'_xC''_x
\right).
\EEA
As for the functions involved in this expression, one notice:
\BEA
&&\frac{\d f_x(\tau,w)}{\d \tau}
=-\frac{4(1-w^2)}{w}\,e^{-(1-w)\tau}\,\int_0^\infty\d y\,
\frac{y\coth\left(\,b\,y\sqrt{\frac{1-w}{1+w}}\,\right)\,\cos [\,y(1-w)\tau\,]\,}
{[\,(1+w)^2+(1-w)^2y^2\,]\,[\,1+y^2\,]},\\
&&\frac{\d ^2f_x(\tau,w)}{\d \tau^2}
=\frac{4(1-w^2)(1-w)}{w}\,e^{-(1-w)\tau}\,\int_0^\infty\d y\,
\frac{y\coth\left(\,b\,y\sqrt{\frac{1-w}{1+w}}\,\right)\,(\cos\,[
y(1-w)\tau\,]+y\sin [\,y(1-w)\tau\,]\,)}
{[\,(1+w)^2+(1-w)^2y^2\,]\,[\,1+y^2\,]},\\
&&\frac{\d f_p(\tau,w)}{\d \tau}
=-\frac{4(1-w)^2}{w}\,e^{-(1-w)\tau}\,\int_0^\infty\d y\,
\frac{y^2\coth\left(\,b\,y\sqrt{\frac{1-w}{1+w}}\,\right)\,\sin [\,y(1-w)\tau\,]\,}
{[\,(1+w)^2+(1-w)^2y^2\,]\,[\,1+y^2\,]},
\label{fff+}
\EEA
Recall that $C_{p,x}(\tau)=f_{p,x}(\tau,w)+f_{p,x}(\tau,-w)$.

The behavior of the entropy production is depicted in Fig.~\ref{espan17}.
It is seen that there is a small
region, where the curves are negative. For $w=0.1$ (weak overdamping) 
the negative region is yet noticeable, but already for $w=0.7$ (moderate
overdamping) this region is almost indistinguishable. This is in
agreement with the above analytical analysis in the limit $\eps\to 0$.

\begin{figure}[bhb]
\vspace{0.1cm}\hspace{-1.5cm}
\vbox{\hfil\epsfig{figure= 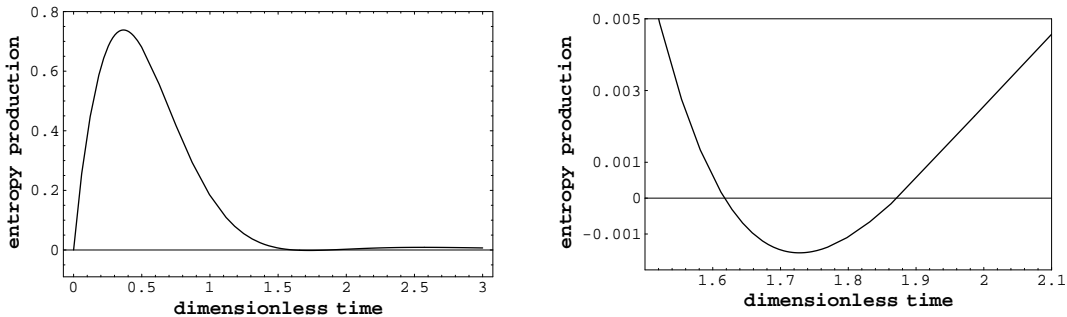,width=15cm,angle=0}\hfil}  
\vspace{0.75cm}
\caption{Rate of entropy production (up to a factor
$a\alpha^2_0/(16\gamma\varepsilon^2\theta^2_x)$ ) versus
dimensionless time $\tau$. Left figure: $T=0$, $w=0.1$.
Right figure: magnification of the region around $\tau=1.75$.}
\label{espan17}
\end{figure}

\renewcommand{\thesection}{\arabic{section}}
\section{Work and heat}
\renewcommand{\thesection}{\arabic{section}.}
\setcounter{equation}{0}

So far we have discussed the system's relaxation from a non-equilibrium
initial state. Since the total system is isolated, 
in this process energy is transferred from the subsystem to the bath,
or vice versa. This energy is related to the unobservable bath modes,
so it is identified as  heat.
In particular, in no way work was added to or extracted
from the system, except for the initial moment, where the
strength of the central spring was modified. 

We shall now consider the possibility of additional changes in the
spring constant and its implications for work extracted from the system.

\subsection{General definition of work and heat}
\label{work&heat}

The behavior of a statistical system under interaction with external
macroscopic sources is the standard area of applications for any 
thermodynamical theory. As known well, in this setup one neglects 
the influence of the statistical system to the dynamics of the source.
Therefore, one can keep the parameters of the system as given functions
of time, and solve the corresponding equations for the system's dynamics.
We start with general remarks about the energy budget of any variation.

Let us consider the change of a system parameter $\alpha $.
It is assumed to be intrinsic, that is to say, to characterize the Brownian
particle but not the bath or the interaction between the particle and
bath. In the situation discussed in the body of this work, 
$\alpha$ can stand for the spring constant $a$ of the harmonic
potential, the effective width of the anharmonic potential, or
the mass $m$ (for electrical circuits and junctions mass is connected
with inductivity and can be subjected to variations; this also
makes sense in systems where $m$ is an effective mass,
that can be modified by changing other system parameters, such as pressure).

First one has to identify the Hamiltonian of the subsystem. In Eq. 
(\ref{hamiltonian}) we have choosen $\H=\half p^2/m+\half ax^2$, 
as in absence of the bath.
It should be stressed that we did not include the self-coupling 
$\half \gamma\Gamma x^2$ or (part of) the interaction
energy in $\H$. Our choice is the natural one
in the sense that the limits of large Debye frequency $\Gamma$ and subsequent
large damping $\gamma$ lead to moderately large values of the energy
of system and bath, and  not to large terms of opposite sign
that cancel in the total energy, as would occur, e.g. if 
$\half \gamma\Gamma x^2$ were counted for the subsystem. 
See also the discussion in section \ref{procontra}.

A change with time of the mean energy is considered due to a variation 
of a parameter $\alpha $ according to the prescribed trajectory $\alpha (t)$:
\begin{equation}
\label{dE}
\d U = \d \int\d x\d p\, W(p,x,t)\, \H(p,x) =
\int \d x \d p \,\H(p,x)\, \d W(p,x,t)+\int \d x \d p\, W(p,x,t)\, \d \H(p,x),
\end{equation}
where $W(p,x)$ is the Wigner function of the Brownian particle.
The last term results from the change in the Hamiltonian,
so it is a mechanical, non-statistical object.
Following other authors\cite{landau,keizer,keizerbook,balian,klim}, 
we shall associate 
it with the work $\dbarrm {\cal W}$ produced by external sources,
in close relation with the definition of work in classical mechanics
and standard thermodynamics. 
The first term in the right hand side represents the
variation due to the statistical redistribution of the 
phase space. We shall identify it with the change in heat $\dbarrm {\cal Q}$. 
Eq. (\ref{dE}) can then be written as the usual first law: 
\begin{equation}
\label{1law}
\d U =\dbarrm {\cal Q}+\dbarrm {\cal W} 
\end{equation}

The work, as defined in Eq.~(\ref{dE}) can be shown to be the change of the
total closed system's (the particle plus bath) energy due to 
the variation of the parameter $\alpha$. First one notices
that for closed systems with a unitary evolution any change of energy is 
determined solely by work. This fact is due to the conservation of energy, and
can be easily illustrated using the von Neumann equation of motion for the 
density matrix $\rho _{tot}$ of the total system.
Indeed, since
$$
\frac{\d \rho _{tot}}{\d t}=-\frac{i}{\hbar}[\rho_{tot},\H_{tot}],
$$
one has 
\begin{eqnarray}
\label{sorrento}
\d U_{tot}&&=\d ~{\rm tr}(\rho _{tot}\H_{tot})=
{\rm tr}(\rho _{tot}~\d \H_{tot})+{\rm tr}(\H_{tot}~\d \rho _{tot})
\nonumber \\
&&={\rm tr}(\rho _{tot}~\d \H_{tot})
-\frac{i}{\hbar} ~\d t ~ {\rm tr}(\H_{tot}~[\rho _{tot},\H_{tot} ])=
{\rm tr}(\rho _{tot}~\d \H_{tot})
\end{eqnarray}
due to the cyclic character of the trace.
If now $W_{tot}(p,x,p_1,x_1,...)$ is the Wigner function of the whole system,
then  this implies the identity
\begin{equation}
\label{russa1}
\d U_{\it tot}
=\int \d p~\d x\prod _{k}\d p_k\d x_k\,
W_{tot}(p,x,p_1,x_1,...)\,\d \H_{tot}(p,x,p_1,x_1,...)=
\int \d x ~\d p\, W(p,x,t)\, \d \H(p,x),
\end{equation}
since we only consider cases where forces are attached to the central 
particle, so $\d \H_{tot}(p,x,p_1,x_1,...)=\d \H(p,x)$, implying that
the $p_k$, $x_k$ integrals over the full Wigner function just bring the
Wigner function of the subsystem.
Taking into account that no heat was added to the total system,
$\dbarrm {\cal Q}_{\it tot}=0$, we may conclude that the work 
$\dbarrm {\cal W}$ extracted from the subsystem 
equals the work subtracted from the total system,
$\dbarrm {\cal W}_{\it tot}=\d U_{\it tot}=\dbarrm {\cal W}$.

We again consider the situation where at $t=0$  the system was Gibbsian,
with spring constant $a_0$, and it is instantaneously changed to a
new value $a$. For achieving this an amount of work $\W_0$,
given in (\ref{W0add}) has to be added to the system. For $a>a_0$ this is just 
the work needed to make the spring attached to the particle
stiffer. When $a_0<a$ this work is negative;
energy is extracted, since the spring is weakened.
For work extraction from the thus created non-equilibrium state
 we shall make additional changes in the spring constant.

\subsection{Maximally extractable work}

Now the total system consisting of central particle and bath is out of 
equilibrium, some work can be extracted from it. 
Before the interaction between the bath and the particle has been switched 
on, the total energy was 
\BEQ U_{\rm tot}(0^-)=U_{B}(T)+U_p(T,a_0)\EEQ
where $U_{B}(T)=\pi^2T^2/(6\hbar\Delta)$ and  
$U_p(T)$ are the initial energies of the unperturbed bath  
and the perturbation due to the Brownian particle, defined in 
(\ref{Ubath=}) and by $U_p=\p[\beta F_p]/\p \beta$ with
$F_p$ taken from (\ref{artush}),
respectively. After the switching of the interaction 
has been  completed, the energy has become  
\BEQ U_{\rm tot}(0^+)=\W_0+ U_{B}(T)+U_p(T,a_0)
\EEQ

Let us now consider what is the maximal amount of work that can be extracted 
from the overall isolated system in the considered non-equilibrium state.
First of all, we notice that we are interested in the work done due to
the non-equilibrium character of this state, and not by a work which might be
done due a change of the Hamiltonian. Therefore, during extraction processes
the parameters of the Hamiltonian $\H_{\rm tot}$ will
be either fixed or vary cyclically, such that after the process has
been completed, the system has the same Hamiltonian as initially.
To determine the maximal amount of extracted work we will employ the
following formulations of the second law, which are undoubtedly valid
for the considered thermally isolated system \cite{landau,balian,Lenard}. 
{\it (i)} No work can be extracted from a system in its equilibrium state.
(Let us recall that thermal isolation 
means that no external supply of heat is given; the allowed transformations
are variation of parameters by external sources). 
{\it (ii)} The converse is true as well under certain general conditions
\cite{Lenard}: If no work can be extracted by any means 
from a system in a given state, then this state is equilibrium.

As follows from {\it (i)} some work
can be extracted from non-equilibrium states. In the same way 
{\it (ii)} implies that if work has been extracted in all possible ways,
the system is left in an equilibrium state at some temperature $T_{\rm fin}$.
Since the overall system is thermally isolated, the extracted work 
is in magnitude equal to the complete change of energy
(this is the statement of the first law): 
$|{\cal W}_{\rm max}|= U_{\rm tot}(0^+)-U_{\rm fin}(T_{\rm fin})$,
where $U_{\rm fin}(T_{\rm fin})=U_{\rm tot}(T_{\rm fin},a)$ is 
the energy of the final equilibrium state.
Because the entropy is conserved during variations of parameters,
the optimal final equilibrium state will have a density matrix 
\BEA
\rho _{\rm fin} = \frac{\exp [-\beta _{\rm fin} \H_{\rm tot}]}{Z},
\EEA  
involving the temperature $T_{\rm fin}=1/\beta _{\rm fin}$,
which is determined by constancy of the von Neuman entropy of the 
total system:
\BEQ S_{\rm tot}(T)=S_B(T,\gamma=0)+S_p(T,a_0)=
S_{\rm tot}(T_{\rm fin})=S_B(T_{\rm fin},\gamma=0)
+S_p(T_{\rm fin},a)\EEQ
with $S_B(T,\gamma=0)=\pi^2T/3\hbar\Delta$ from (\ref{CB=SB}).
Remembering that the level splitting
$\Delta$ of the bath modes is very small, one can solve
\BEQ T_{\rm fin}=T+\frac{3\hbar\Delta}{\pi^2}[S_p(T,a_0)-S_p(T,a)]\EEQ

This yields 
\BEA \label{Wmax}
|{\cal W}_{\rm max}|&=& \W_0+U_{\rm tot}(T,a_0)-U_{\rm tot}(T,a)
-TS_p(T,a_0)+T_p(T,a)\nn&=&
\W_0+F_{\rm tot}(T,a_0)-F_{\rm tot}(T,a)
=\W_0+F_p(T,a_0)-F_p(T,a)
\EEA
In the last step we canceled the contributions of the unperturbed bath.
Not unexpectedly, the result depends on the free energies of the total system.

Notice that for a cycle consisting of the changes $a_0\to a$ and (much)
later $a\to a_0$, the maximally extractable work becomes the sum of the
amounts of work $\W_0(a_0\to a)+\W_0(a\to a_0)$, so for cycles in principle
all work can be recovered.

\subsubsection{Values at high and low $T$}

In the classical limit the free energy is given in (\ref{FplargeT}),
just the value for a harmonic oscillator, whether or not coupled 
to other harmonic oscillators. The maximally extractable work is
\BEQ \label{Wsurpl}
|{\cal W}_{\rm max}|=
\half [\frac{\alpha_0}{1-\alpha_0}+\ln(1-\alpha_0)]T
=\frac{1}{4}\alpha_0^2T+\O(\alpha_0^3)\EEQ

At low $T$ the difference in free energy of the total system
between the equilibrium states at 
the initial and final value of the spring constant is
\BEQ F_{\rm tot}(a)-F_{\rm tot}(a_0)= F_p(a)-F_p(a_0)
=\frac{\hbar a}{2\pi\gamma}\left
[\frac{\alpha_0}{w}\ln\frac{1+w}{1-w}+\frac{\alpha_0^2}{2w^2}
(1-\frac{1-w^2}{2w}\ln\frac{1+w}{1-w})\right]+(\alpha_0+\alpha_0^2)
\frac{\pi\gamma T^2}{6\hbar a}
\EEQ

The energy added at $t=0$ is
\BEQ \label{W0add}
\W_0=\half(\frac{a}{a_0}-1)T_x(a_0)=\frac{\hbar a}{2\pi\gamma}
\left
[\frac{\alpha_0}{w}\ln\frac{1+w}{1-w}+\frac{\alpha_0^2}{w^2}
(1-\frac{1-w^2}{2w}\ln\frac{1+w}{1-w})\right]+(\alpha_0+2\alpha_0^2)
\frac{\pi\gamma T^2}{6\hbar a}\EEQ

So by making the instantaneous change in $a$ the maximally 
extractable work (\ref{Wmax}) reads
\BEQ \label{W0surplT=0}
|{\cal W}_{\rm max}|=\alpha_0^2\left[\frac{\hbar a}{4\pi\gamma}
\,\frac{1}{w^2}(1-\frac{1-w^2}{2w}\ln\frac{1+w}{1-w})
+\frac{\pi\gamma T^2}{6\hbar a}\right]
\EEQ
If we let the system relax, this will run away in the bath on a timescale
$\tau_x$. By making clever subtraction schemes, we
may recover some of it, and in principle all of it.

For the case $T=T_{\rm fin}=0$ Eq. (\ref{Wmax}) merely says that that 
all energy exceeding the groundstate energy of the new system can, 
in principle, be extracted.

\subsection{Work extraction by further sudden changes}
\label{Suddenwork}

Here we present the formalism of work-extraction via sudden
changes of a parameter. Besides presenting the general setup, we will
display the validity of the Thomson's equilibrium formulation of the
second law within the present situation.

Let there be a closed system with a Hamiltonian $\H$ in a state
$\rho (t_1)$ at the moment $t_1$. Certain parameters of the
Hamiltonian are varied in a very fast way such that for $t_1+$ its
Hamiltonian becomes $\H_1$, but the state remains $\rho (t_1)$ due to the
sudden character of the variation. The work done by an external source
reads:
\BEA
\W_1={\rm tr}[\rho (t_1) (\H_1 -\H)]
\EEA
In the second step the system is allowed to involve according to the
new Hamiltonian $\H_1$. At the moment $t_2$ when the system reaches
the state:
\BEA
\rho (t_2)=e^{-i(t_2-t_1)\H_1/\hbar}\rho(t_1)e^{i(t_2-t_1)\H_1/\hbar}
\label{o1}
\EEA
its parameters are suddenly returned to their original value. The work
done in this step reads:
\BEA
\W_2={\rm tr}[\rho (t_2)(\H-\H_1)].
\EEA
The total work done by the source for this cyclic variation of the
parameter reads:
\BEA
\W=\W_1+\W_2={\rm tr}\left[\rho(t_1)\left(
e^{i(t_2-t_1)\H_1/\hbar}\,\H\,e^{-i(t_2-t_1)\H_1/\hbar}
-\H\right)\right],
\label{o2}
\EEA
where we have used Eq.~(\ref{o1}). Notice that we consider the closed
overall system, and only due to this fact the evolution of the system
for times between $t_1$ and $t_2$ is given by the Hamiltonian $\H_1$.

It is not difficult to see from Eq.~(\ref{o2})
that the second law is satisfied for the present setup. 
Let us first assume that at the moment $t$ the system was in the ground
state of $\H$: $\rho (t_1)=|0\rangle\langle 0|$. Then one has:
\BEA
\W=\langle 0|\,
e^{i(t_2-t_1)\H_1/\hbar}\,\H\,e^{-i(t_2-t_1)\H_1/\hbar}\,
|0\rangle -\langle 0|\,\H\,|0\rangle
\ge 0,
\EEA
just by the definition of the ground state. 
The same statement, namely $W\ge 0$, holds when $\rho (t)$
is the Gibbs distribution of the initial state at positive temperature 
$T=1/\beta$: $\rho (t_1)=\exp (-\beta \H)/Z$, $Z={\rm tr} \exp(-\beta \H)$
\cite{Lenard}. 
 
\label{Suddenwork+}
Our work extraction mechanism involves a second 
change of the spring constant, which is cyclic: at time
$t_2$ we impose a jump  $a\to a_2=a(1-\alpha_2)$ 
and it keeps that value, until at $t_3$ it is put back to $a$.
The work involved in this cyclic two-step process is
\BEQ \Delta \W=U_{\rm tot}(t_2^+)-U_{\rm tot}(t_2^-)+
U_{\rm tot}(t_3^+)-U_{\rm tot}(t_3^-)=
\half (a_2-a)[\langle x^2\rangle_{t_2}-\langle x^2\rangle_{t_3}]\EEQ

The change in particle energy between $t_2^-$ and $t_3^+$ is
\BEQ 
\Delta U=\frac{1}{2m} [\langle p^2\rangle_{t_3}-\langle p^2\rangle_{t_2}]
+\half a[\langle x^2\rangle_{t_3}-\langle x^2\rangle_{t_2}]
\EEQ
Thus the change in heat during the work extraction process is 
\BEQ 
\Delta \Q=\frac{1}{2m} [\langle p^2\rangle_{t_3}-\langle p^2\rangle_{t_2}]
+\half a_2[\langle x^2\rangle_{t_3}-\langle x^2\rangle_{t_2}]
\EEQ

The values of $\langle x^2\rangle$ and $\langle p^2\rangle$ at time $t_2$
are set by the spring constants $a_0$ and $a$ solely, and can be taken from 
previous section.  When we take $t_3$ large, 
we can take for that situation the limiting values for a system with 
spring constant $a_2$. We then find
\BEA \Delta \W&=&(a_2-a)[\frac{T_x(a)}{2a}
+\frac{\alpha_0\hbar}{2\pi \gamma}\,C_x(\frac{\gamma t_2}{2m})
-\frac{T_x(a_2)}{2a_2}]
\label{DelW=}\\
\Delta U&=&\half T_p(a_2)-\half T_p(a)
+\frac{a}{2a_2}T_x(a_2)-\half T_x(a)-
\frac{\alpha_0\hbar a}{2\pi \gamma}\,C_E(\frac{\gamma t_2}{2m})
\\ \Delta \Q&=&
\half T_p(a_2)-\half T_p(a)
+\frac{1}{2}T_x(a_2)-\frac{a_2}{2a}T_x(a)
-\frac{\alpha_0\hbar a}{2\pi \gamma}\,C_p(\frac{\gamma t_2}{2m})
-\frac{\alpha_0\hbar a_2}{2\pi \gamma}\,C_x(\frac{\gamma t_2}{2m})
\EEA

\subsubsection{ Classical regime}

In the classical case with strong damping we use (\ref{sipleexp}) and obtain
\BEQ \Delta \W=\half\alpha_2^2T-\half\alpha_0\alpha_2Te^{-2at_2/\gamma}
\EEQ

This has a minimum 
\BEQ\label{DWmin}
\alpha_2^\ast =\half\alpha_0e^{-2at_2/\gamma},\qquad 
\Delta \W_{\rm min}=-\frac{1}{8}\alpha_0^2T\,\,e^{-4at_2/\gamma}
\EEQ
Work can only be extracted under proper conditions, that is to say,
when  $\alpha_2$ is between $0$ and $\alpha_0\exp(-2\omega_2t_2)$.
Else the cycle $a\to a_2\to a$ disperses energy. 

The heat absorbed by the subsystem is at linear order in $\alpha_0$ 
insensitive to the work extraction,
\BEQ \Delta \Q=\Delta U=
-\frac{1}{4}\alpha_0Te^{-4at_2/\gamma}\EEQ

There are two cases:

$\bullet$ 
$\alpha_0>0$: The central spring is stiffened, $a>a_0$, and energy is 
supplied at $t=0$. This energy leaks away, mostly as heat into the bath,
$(\Delta \Q<0)$ and partly as work extracted from the
total system $(\Delta W<0)$, more precisely, from the particle.

The ratio of extracted work to maximally extractable energy is, 
in the regime where $\alpha_0$ is small but finite and 
$T\gg\hbar\omega_0$,
\BEQ \eta= \frac{|\Delta \W |}{|\W_{\rm max}|}
=\half\,e^{-4\omega_2 t_2}
\EEQ
So our mechanism extracts maximally 50$\%$ of maximum; 
for doing this it must start immediately ($t_2=0$) and 
last as long as possible ($t_3=\infty$).

$\bullet$
 $\alpha_0<0$: the central spring is weakened at $t=0$. Energy
is taken out from the system. The amount, $\half T(|\alpha_0|-\alpha_0^2)$
is less than the amount that could have been extracted by an adiabatic change, 
$\half T(|\alpha_0|+\half\alpha_0^2)$. After that has been done, heat flows
from the bath to the particle $(\Delta \Q>0)$, as if the particle were at a
lower temperature. In the course of this process work can be extracted,
maximally the absolute value of Eq. (\ref{DWmin}), 
as is usual for two-temperature systems. The basic issue to extract work
is to have a mechanism that, given the initial change in $a$,
is the  closer to adiabaticity.

These conclusions also hold for moderate and weak damping.

\subsubsection{ Low temperature regime}

At $T=0$ we get from Eq. (\ref{DelW=})
\BEQ \Delta \W
=\frac{\hbar a}{2\pi\gamma} \left[\alpha_2^2C_x(0)
-\alpha_0\alpha_2\,C_x(\frac{\gamma t_2}{2m})\,\right] 
\EEQ
where $C_x(0)$ is defined in (\ref{Cx(0)=}) and $C_x(\tau)$ in
(\ref{Cxtau=}). The minimum occurs for
\BEQ\label{a1a0min}
 \alpha_2^\ast=\frac{\alpha_0}{2C_x(0)}\,C_x(\frac{\gamma t_2}{2m})
,\qquad
 \Delta \W_{\rm min}=-\frac{\hbar\omega_2\alpha_0^2}{8\pi C_x(0)}
\,C_x^2(\frac{\gamma t_2}{2m})
\EEQ
The change in heat and internal energy is
\BEQ \Delta \Q=\Delta U=-\alpha_0\frac{\hbar a}{2\pi\gamma}
C_E(\frac{\gamma t_2}{2m})
\EEQ
We consider again the separate  cases:

$\bullet$ $\alpha_0>0$. The spring is stiffened at $t=0$ and energy $\W_0$ 
is supplied. At $t=0^+$ the energy is, to linear order in $\alpha_0$,
equal to its final value, since $C_E(0)=0$. This changes since energy comes
 from the bath, on a timescale $\tau_p$ which is short for strong damping, where $\tau_p=m/\gamma$.
In the initial time regime $\gamma t_2/{2m}<\tau_0$ 
this mainly leaks away to the bath, and a small part can be extracted as work.
In the regime $ \tau_0<\gamma t_2/{2m}<\sigma_{\rm min}$ this also happens,
but the energy of the subsystem goes below its final value, so the particle
becomes ``too cold''. In the final
regime $\gamma t_2/{2m}>\sigma_{\rm min}$ energy flows back to the particle
and again a small part can be extracted as work. This is then work extracted
from the non-equilibrium bath, and the surprise is that this can be done
although initially energy was put on the particle. This recovery of 
energy stored in the bath is a quantum effect.

$\Delta W<0$ means that work is extracted from the total system.
Eq. (\ref{a1a0min}) is the maximally extractable amount of work 
with the present mechanism.
As efficiency factor way may normalize with respect to 
the maximally extractable energy from eq.  (\ref{Wsurpl}), 
the energy that would
otherwise leak away into the bath.
\BEQ\label{etaCt} \eta=\frac{-\Delta \W_{\rm min}}{|\W_{\rm max}|}=\half
C^2(\omega_2 t_2)\EEQ
When one starts the extraction quickly after the initial change 
$(\tau_1\approx 0)$ one can still get half of the work back in this way,
the same rate as in the classical regime.
Even more is obtained when one still starts at $t_2=0$ but
stops at the moment that the energy current goes no longer towards 
the particle, but away from it, i.e.  at $\omega_2 t_3=\sigma_{\rm min}$.
One then has 
\BEQ \Delta \W=-\half\alpha_2a[\langle x^2\rangle_{t=0;\,a_0}
-\langle x^2\rangle_{t_3;\,a_2}]
=\frac{\hbar a}{2\pi\gamma}
\{\alpha_2^2-\alpha_0\alpha_2[1-C_x(\omega_2t_3)]\}
\ge -\frac{\hbar a}{8\pi\gamma}\alpha_0^2[1-C_x(\omega_2t_3)]^2
\EEQ
The efficiency is

\BEQ
\eta=\frac{-\Delta\W}{|\W_{max}|}=\half[1-C_x(\omega_2 t_3)]^2
\EEQ
which indeed has the proper behavior for $t_3\to 0$ and $t_3\to\infty$.
The maximum is, due to (\ref{bro1}) and (\ref{mindata}),  
\BEQ \eta_{\rm max}=\half[1-C_x(\sigma_{\rm min})]^2= 0.5961\EEQ
the maximum exceeds the classical efficiency $\eta=1/2$.
So the quantum statistical excess energy flow from the particle into 
the bath indeed allows a more optimal recovery of energy initially put 
on the particle.

The most interesting feature is that with the present mechanism 
it is also possible to extract work solely from the bath,
a mechanism forbidden by the original Thomson formulation of
the second law. Indeed, after $\omega_2t=\sigma_{\rm min}$ energy 
will flow back from the bath to the particle. By starting the
extraction mechanism at $\omega_2t_2=\sigma_{\rm min}$ and 
exploiting all times after this for the work extraction,
the maximal efficiency (\ref{etaCt}) is

\BEQ \eta_{\rm max}=\half C^2(\sigma_{\rm min})= 0.00422262576 \EEQ

In contrast to the classical case, it goes to a finite limit when 
$\alpha_0\to 0$. This occurs because in the quantum case the
energy $\sim\alpha_0^2$, with respect to which the extracted work 
has been normalized,  is one order of magnitude smaller than the 
initially supplied energy $\W_0\sim\alpha_0$.

$\bullet$ $\alpha_0<0$: The spring is weakened and energy $\W_0$ is
extracted. Some energy can be extracted. 
For $\omega_2t_2<\sigma_{\rm min}$ it comes  from the bath, but in 
the regime $\omega_2t_2>\sigma_{\rm min}$ the particle has an excess 
energy, which then supplies the work.

\subsection{Work extraction by smooth changes of the spring constant}

Let us now consider the case where the spring constant 
$a(t)=[1-\alpha(t)]a$, is slightly changed ($|\alpha(t)|\ll 1$) in a 
smooth manner, starting from the equilibrium state $a(-\infty)=a$.
In appendix A we derive for the rate of work added to the system
\BEQ \frac{\d \W}{\d t}=\frac{\d \W_{\rm rev}}{\d t}+\frac{\d \Pi}{\d t}
\EEQ
where
\BEQ \frac{\d \W_{\rm rev}}{\d t}=
-\frac{\gamma}{2m}\, \frac{\d \alpha(\tau)}{\d \tau}
\left[\half T_x+\frac{\hbar a}{2\pi\gamma }\alpha(\tau)C_x(0)\right]
\EEQ
is the adiabatic (`reversible' or `recoverable') rate of work and
\BEQ \label{dPIdtau2=} \frac{\d \Pi}{\d t}=
\frac{\hbar a}{4\pi m}\,\frac{\d \alpha(\tau)}{\d \tau}\,
\int_0^\infty\d \tau\,\alpha'(\frac{\gamma t}{2m}-\tau) C_x(\tau)
\EEQ
is the rate of energy dispersion. 

\subsubsection{Completed changes}

Integrating over the full change
one has 
\BEQ 
\Del \W_{\rm rev}=-\half(\alpha_f-\alpha_i) T_x-
\frac{\hbar a}{4\pi\gamma }C_x(0)[\alpha_f^2-\alpha_i^2]\EEQ
where $\alpha_i=\alpha(-\infty)$ and  $\alpha_f=\alpha(\infty)$, and
\BEQ\label{DelPiif} \Del \Pi=\frac{\hbar a}{4\pi\gamma }
\int_\minfty^\infty\d\tau_1\,\alpha^\prime(\tau_1)\int_\minfty^\infty
\d\tau_2\,\alpha^\prime(\tau_2)
C_x(\,|\tau_1-\tau_2|\,)\EEQ
For a full process (covering the whole region where $\alpha'\neq 0$) 
$\Pi$ is non-negative, since it is an integral over a non-negative function,
\BEQ \label{Picycle}
\Del \Pi=\frac{\hbar a}{2\gamma}
\int_\minfty^\infty \d \zeta |A(\zeta)|^2 \Re\hat C_x(i\zeta)
\EEQ
where the Laplace transform $\hat C_x$ was defined in Eq. (\ref{hatCx=});
it can be verified that $ \Re\hat C_x(i\zeta)$ is positive for all real
$\zeta$. Furthermore
\BEQ A(\zeta)=\int\frac{\d\tau}{2\pi}\,
\alpha^\prime(\tau)\,e^{i\zeta\tau}\EEQ 

The positive energy dispersion for a completed, non-adiabatic cyclic
change of system parameters is the Thomson formulation of the second law.
We see that a positive dispersion also holds for non-cyclic but completed
changes, as is known to occur on general grounds.

We can check previous case 
$\alpha(t)=\alpha_0\theta(-t)$, 
$\alpha'(\tau)=-\alpha_0\delta(\tau)$, for which Eq. (\ref{W0add})
is at $T=0$ equivalent to 
\BEQ \W_0=\W_{\rm rev}+\Pi=\half\alpha_0T_x+
\frac{\hbar a}{2\pi\gamma}\alpha_0^2C_x^{(0)}\EEQ

\subsubsection{Incomplete changes}
\label{incompletecyc}

Let us now consider the temporal build up of this result in the
regime of strong damping.
Let $\alpha$ have the form
\BEQ \alpha(\tau)=\am h(\Omega t)\EEQ
where $\alpha_m$ is a small amplitude and $h$ is a bounded function 
($|h|\le 1$), with $h(\minfty)=0$, $h'(\infty)=0$. 
If $h(\infty)\neq 0$ it could be an error function; if $h(\infty)=0$ a Gaussian. 
$1/\Omega$ is the typical duration time of the change. 
Using Eqs. (\ref{C0(T)=},\ref{C1(T)=},\ref{C2(T)=})
we get from (\ref{dPIdtau=})
\BEA\label{dPIdt=}\frac{\d \Pi}{\d t}&=&
\frac{\hbar a}{4\pi m}\,\alpha'\,[\alpha'C_x^{(0)}+\alpha''C_x^{(1)} 
-\half \alpha'''C_x^{(2)} ]
\nn
&=&\frac{\hbar \Omega^2\amsq}{12\pi}\,h'(\Omega t)
\left[\left(\frac{2\pi\gamma T}{\hbar a}\right)^2h'(\Omega t)+
\frac{\gamma\Omega  \tilde C_x^{(1)}}{a}h''(\Omega t) -
\frac{\gamma^2\Omega^2}{a^2}h'''(\Omega t)\right]\EEA
where $\tilde C_x^{(1)}$ was defined in (\ref{C1(T)=}); 
for weak damping it equals $3\pi\sqrt{am}/8\gamma$, while for 
large damping it becomes unity. The integrated effect is 
\BEQ\label{DelPi=}
\Del \Pi(t)=\frac{\hbar \Omega\amsq}{12\pi}\left\{
\left(\frac{2\pi\gamma T}{\hbar a}\right)^2
\int_\minfty^{\Omega t} \d \xi[h'(\xi)]^2+\frac{\gamma\Omega 
\tilde C_x^{(1)}}{2 a }[h'(\Omega t)]^2
+\frac{\gamma^2\Omega^2}{a^2}\left[-h'(\Omega t)h''(\Omega t)+
\int_\minfty^{\Omega t}\d \xi[h''(\xi)]^2\right]\right\}\EEQ
For a completed change the second and third term vanish, leading 
to a positive energy dispersion, in concordance with (\ref{Picycle}).
It is seen that then the  standard behavior $\Del\Pi\sim \Omega$ applies
when $T$ is large enough or when $\Omega$ is small enough. However, in the
quantum regime 
where the duration $1/\Omega$ is smaller than the quantum 
timescale $\tau_\hbar=\hbar/T$, the last term in Eq. (\ref{DelPi=}) 
dominates, with a new  behavior $\Del\Pi\sim \Omega^3$. 

Another new quantum effect is that at low $T$ and for typical times
the second term in (\ref{dPIdt=}) 
is larger in magnitude than the other ones. This too
occurs since the integral of $C_x(\tau)$ vanishes at $T=0$,
and leads to new possibilities, that we discuss now.

\subsubsection{Work extraction from a smooth cyclic change}

One definition for a perpetuum mobile of the second kind is 
that there is a machine that performs a cycle 
in which it receives heat from a bath and converts this fully into work
done on the surrounding. Additional requirements can occur; 
we shall discuss them in section ~\ref{Sectperpmob}. Here we analyze 
whether such a full energy conversion can be realized in our setup.
The aim is thus to have a cyclic change of a
system parameter with the properties 
\BEQ \Delta U=0,\qquad  \Delta \Q=-\Delta\W>0\EEQ

Including in (\ref{Vtcont}) also the contribution of the momentum,
we have to linear order in $\alpha$
\BEQ U(t)=\half T_p(a)+\half T_x(a)+\frac{\hbar a}{2\pi\gamma }
[\alpha(\frac{\gamma t}{2m})C_E(0)-
\int_0^\infty\d \tau \,\alpha'(\frac{\gamma t}{2m}-\tau)
C_E(\tau)\,]\EEQ
Let us choose for $\alpha(\tau)$ a curve in the shape of a double bell, and 
consider the system at some time $t_1$ after the first peak, and compare 
it to the situation at a  later time $t_2$, after the second peak. 
We require that the spring constant has the same values at these instants, 
$\alpha(\tau_1) =\alpha(\tau_2)$, where $\tau_{1,2}=\gamma t_{1,2}/2m$. 
This implies for the internal energy
\BEQ \Delta U=U(t_2)-U(t_1)=\frac{\hbar a}{2\pi\gamma }
\int_0^\infty\d\tau[\alpha'(\tau_1-\tau)-\alpha'(\tau_2-\tau)]
C_E(\tau) \EEQ
For slow changes this can be expanded
\BEQ \Delta U=\frac{\hbar a}{2\pi\gamma }
\{[\alpha'(\tau_1)-\alpha'(\tau_2)]C_E^{(0)}+
[\alpha''(\tau_1)-\alpha''(\tau_2)]C_E^{(1)} 
-\half[\alpha'''(\tau_1)-\alpha'''(\tau_2)]C_E^{(2)} 
\}
\EEQ
where the coefficients are given in (\ref{CE012=}).

Let us assume that we have two consecutive changes, characterized
by a common bell-shaped function $\hc(x)$ with $|\hc(x)|\le 1$, but involving
different rates of change $\Omega_{1,2}$,
\BEA \label{alshape}
h(\tau)=\hc(\Omega_1 t)+\hc(\Omega_2(t-t_2^c))\EEA 
where the parameters $\Omega_1>\Omega_2>0$ and $t_2^c>0$ are such that the 
profiles have negligible overlap.
Choosing the times $t_{1,2}$ as
\BEQ t_1=\frac{\bar x}{\Omega_1},\qquad 
t_2=t_2^c+\frac{\bar x}{\Omega_2}\EEQ
we indeed satisfy the cyclic condition $\alpha(\tau_1)=\alpha(\tau_2)=
\am\hc(\bar x)$. The difference in energy is
\BEQ \label{DeltaU=}
\Delta U=\frac{\am\hbar (\Omega_1-\Omega_2)}{12\pi }
\left\{\left(\frac{2\pi\gamma T}{\hbar a}\right)^2\hc'(\bar x)+
\frac{\gamma(\Omega_1+\Omega_2)}{a}\hc''(\bar x)
-\frac{\gamma^2(\Omega_1^2+\Omega_1\Omega_2+\Omega_2^2)}{a^2}
\hc'''(\bar x)
\right\}
\EEQ
Let us assume that $\bar x$ is fixed but such that $\hc'<0$ while 
$\hc''>0$;
this is possible because $\hc$ has both convex and concave parts,
 implying that there is an interval with such behavior.
In the example $\hc(x)=\exp(-x^2/2)$, there is an inflection point $x_{\rm if}=1$
and one needs $\bar x>1$. 
Let us also assume that $T\ll \hbar a/\gamma$. 
 Then for
\BEQ \Omega_1+\Omega_2= \frac{4\pi^2\gamma T^2}
{\hbar^2 a}\, \frac{|\hc'(\bar x)|}{\hc''(\bar x)}
\EEQ 
the first two terms can cancel,
while the exposed correction term and the higher ones are small.
Thus under these conditions it is possible to have a cyclic
process with $\Delta U=0$ to order 
$\am$; if $\amsq$ corrections are taken into account,
the condition for cancellation is shifted an amount of order $\am$
and can again be met. Thus it is possible to start from the equilibrium state,
make a first cyclic change of $a$ and then a second, which process
contains itself a cyclic change of $a$  with $\Delta U=0$. 
The work during this cycle comes solely from the energy dispersion. 
Using Eq. (\ref{DelPi=}) we obtain the leading terms
\BEA\label{DPicycle}
\Delta \W&=&\frac{\amsq\hbar}
{12\pi}\left\{\left(\frac{2\pi\gamma T}{\hbar a}\right)^2
\left(\Omega_1\int_{\bar x}^\infty \d \xi[\hc'(\xi)]^2+
\Omega_2\int_\minfty^{\bar x}\d \xi[\hc'(\xi)]^2\right)+
\frac{\gamma(\Omega_2^2-\Omega_1^2)\tilde C_x^{(1)}}{2 a}
[\hc'(\bar x)]^2\right\}\nn
\label{DPicycle2}
&=&\frac{\hbar\amsq}
{12\pi}\left(\frac{2\pi\gamma T}{\hbar a}\right)^2
\left\{\Omega_1\int_{\bar x}^\infty \d \xi[\hc'(\xi)]^2+
\Omega_2\int_\minfty^{\bar x}\d \xi[\hc'(\xi)]^2+\half
(\Omega_2-\Omega_1)\tilde C_x^{(1)}\frac{|\hc'(\bar x)|^3}{\hc''(\bar x)}\right\}
\EEA 

The higher order terms are small for the same reason as above.
If $\Delta W<0$ this amounts to work  exerted by the system on the 
environment.  One can always have $\hc''(\bar x)$ small enough
(by choosing $\bar x$ close  to the inflection point) to make
 the combination of the $\Omega_1$ terms negative, and choose
$\Omega_2$ so small that the whole expression remains negative. 
So it is indeed possible to have a cycle where the extracted work 
$|\Delta \W|$ comes solely from the bath. 

We should stress that these work cycles are realizable only because
in the first part of the process, for $t<t_1$, energy was lost
($\W(t_1)-\W(-\infty)>0$); 
a part of this is recovered. If, on the other hand,
 all the work is counted, then no work extraction
is possible ($\W(\infty)-\W(-\infty)>0$), c.f. Eq. (\ref{DelPiif}).

Alternatively one may conclude that there are non-equilibrium initial
conditions (for instance, the state of the total system at time $t_1$)
 which allow cycles which 
fully transform heat obtained from the bath into work done on the environment.

From the analysis it is clear that under less strict conditions it is even 
possible to make a cycle that extracts work that comes partly from the
 bath and partly from the system (``efficiency larger than 100$\%$'').

\subsubsection{Perpetuum mobile with many work extraction cycles}
\label{perpmob}

One can make several of these cycles. Even
though the previous finding that complete cycles disperse energy should
temper the hope to gain more work by doing more cycles,
we consider the issue here, since work extraction from many cycles 
is one of the ways to express our unexpected results.

Let there be $\N$ cycles with $\alpha(\tau)=\am h(\tau)$ 
having $\N$ non-overlapping bell-shaped parts, like in Eq. (\ref{alshape}),
where $k(x)$ could be the Gaussian $\exp(-\half x^2)$.
Each cycle is characterized its typical inverse duration time 
$\Omega_n\ll a/\gamma$ and location $t_n^c$, and each new
cycles is slower than the previous one, $\Omega_{n+1}<\Omega_n$. 
For having cyclic behavior in the energy one 
finds from (\ref{DeltaU=}) that for low $T$ and small but almost
equal $\Omega_n$'s, one should choose $\bar x$ 
close to the inflection point $\xinfl$ of $\hc$:

\BEQ\label{xclose} \bar x_n-\xinfl=
\frac{a}{2\gamma\Omega_n}\,
\frac{|\hc'(\xinfl)|}{\hc'''(\xinfl)}
\left(\frac{2\pi\gamma T}{\hbar a}\right)^2+\frac{3\gamma\Omega_n}{2a}
+\O(\frac{\gamma^2\Omega_n^2}{a^2})
\EEQ
In order that this be small for all $n$
one needs that  $\Omega_\N\gg\gamma T^2/\hbar^2a$. 
Strictly speaking the $\alpha_n=\am h(\bar x_n)$ are now not exactly
equal; this can be healed by slightly adjusting the profile in the $n$'th 
cycle: $\hc(\Omega_n(t-t_n))\to \hc(\Omega_n(t-t_n^c))
\hc(\xinfl)/\hc(\bar x_n)$, 
yielding $\alpha(\tau_n)=\am\hc(\xinfl)$ for all $n$.
For small $\bar x_n-\xinfl$ this correction factor is close to unity
and can be omitted from the rest of the argument.

Let us define
\BEQ \tilde T=\frac{2\pi\gamma T}{\hbar a} \EEQ

Taking Eq. (\ref{DelPi=}) at $t=t_n$,
where $n$ cycles have been performed, and using that 
$\bar x-\xinfl\ll 1$ and that $|\Omega_n-\Omega_{n+1}|\ll\Omega_n$, 
brings for the yield of the $n$'th  cycle
\BEA\label{DDPi=} 
\Pi(t_{n})-\Pi(t_{n-1})&=&\frac{\amsq}{12\pi}
\hbar\left\{I_1\Omega_n \tilde T^2 
+I_2\frac{\gamma(\Omega_n^2-\Omega_{n-1}^2)} 
{2 a}+I_3\frac{\gamma^2\Omega_n^3}{a^2}\right\}
\EEA 
where 
\BEA 
I_1=\int_\minfty^{\infty} \d \xi[k'(\xi)]^2,\qquad
I_2=\tilde C_x^{(1)}[k'(\xinfl)]^2,\qquad 
I_3=\int_\minfty^{\infty}\d \xi[k''(\xi)]^2\EEA

For having an equal yield per cycle, one demands 
\BEA\label{DDPi=e} 
\Pi(t_{n})-\Pi(t_{n-1})=\W_{\rm cycle}=-\frac{\amsq\hbar a}{12\pi\gamma}w
=-\frac{\amsq\hbar a}{12\pi\gamma} \tilde T^3v\EEA
where $v>0$ and $w=\tilde T^3v$ are dimensionless.
It will turn out that there exists a 
consistent solution for $v$ in some definite range. 
Assuming that $\Omega(n)=\Omega_n$ is a smooth function of $n$
one obtains
\BEQ I_1\Omega \tilde T^2 
+I_2\frac{\gamma}{a}\Omega\Omega' 
+I_3\frac{\gamma^2}{a^2}\Omega^3=-w\EEQ
Solving for $\d n/\d\Omega$ and going to a new variable 
$y=\beta\hbar\Omega/(2\pi)$
one gets the total number of cycles
\BEQ \N=\frac{I_2}{\tilde T}\int_{\beta\hbar\Omega_\N/2\pi}
^{\beta\hbar\Omega_1/2\pi}\frac{\d y y}{v+I_1y+I_3y^3}
\EEQ 
The total yield is then
\BEQ \label{WTnot0}
\W_{\rm tot}=\N\W_{\rm cycle}=-\frac{\amsq I_2}{12 \pi}\,\frac{\hbar a}{\gamma}
\tilde T^2 \int_{\beta\hbar\Omega_\N/2\pi}^{
\beta\hbar\Omega_1/2\pi}\d y\frac{v\, y}{v+I_1y+I_3y^3} \EEQ
Here the minus sign indicates that
work is performed by the system on the environment. 
This is possible because Eq. (\ref{DelPi=}) expresses that,
in order to make the work extraction cycles, one had to start from the
equilibrium state and change $\alpha$ from $\alpha(-\infty)=0$
up to $\alpha(\tau_1)$. In this first part of the process
 energy was dispersed at an amount
\BEQ \Pi(t_1)=+\frac{\amsq I_2}{24\pi}\frac{\hbar\gamma}{a}\Omega_1^2\EEQ
Notice that for $v\gg (\beta\hbar\Omega_1)^3$
 the extracted work becomes according to (\ref{WTnot0})
\BEQ \label{WTnot0+}
\W_{\rm tot}=-\frac{\amsq I_2}{24\pi}\frac{\hbar\gamma}{a}
(\Omega_1^2-\Omega_\N^2)\EEQ
so for $\Omega_\N\ll\Omega_1$ there an almost perfect recovery,
which is possible since the number of cycles is still large.

For moderate $v$ more cycles are possible, but less work is recovered.
For the lower integration variable Eq. (\ref{xclose}) gives
\BEQ y_\N\equiv \frac{\beta\hbar\Omega_\N}{2\pi}\gg\frac{\gamma T}{2\pi\hbar a}
\EEQ 
For strong damping the physical timescale is $\tau_x=\gamma/a$. 
One assumes that $\Omega_1$ is  a large but finite number times $1/\tau_x$.
Choosing $T\ll \hbar a/\gamma$ means that the upper integration limit
$y_1=\beta\hbar\Omega_1/2\pi$ is much larger than unity.
But it is still possible to choose $y_\N\ll 1$, which is a useful condition
for achieving many cycles. One then has for small $v$
\BEQ \N=\frac{I_2}{4\sqrt{I_1I_3}}\,\frac{\hbar a}{\gamma T}\EEQ
which is indeed large. In the overdamped regime 
the yield can thus be expressed as 
\BEQ \label{WTnot1}
\W_{\rm tot}=-\frac{\amsq I_2}{24\sqrt{I_1I_3}}\,
\frac{\hbar a}{\gamma} v\tilde T^2=-\frac{\amsq\sqrt{ I_1I_3}}{6\pi^2I_2}
\,\frac{\hbar a}{\gamma} \frac{v}{\N^2}\EEQ

In the limit of weak damping we should notice that 
\BEQ I_2\equiv \tilde I_2\sqrt{\eps}\EEQ
where $\eps=am/\gamma^2\gg 1$ and where $\tilde I_2$ is a numerical constant 
of order unity. Thus the work dispersed for achieving the non-equilibrium
condition at $t_1$ is
\BEQ \Pi(t_1)=+\frac{\amsq \tilde I_2}{24\pi}\frac{\hbar\Omega_1^2}{\omega_0}\EEQ
where $\omega_0=\sqrt{a/m}$ is the free oscillation frequency.  Let us recall
that $\tau_d=2m/\gamma$ is the damping time. For $T\ll\hbar\Omega_1$
it holds that
\BEQ \N=\frac{\tilde I_2}{4\sqrt{I_1I_3}}\,\eps\frac{\hbar \omega_0}{T}\EEQ
At the typical temperature $T\lesssim\hbar\omega_0$ this carries an additional
large factor $\eps$. The yield per cycle carries a factor $1/\eps$,
so this total yield is independent of $\eps$. For small $v$ it reads
\BEQ \label{WTnot1+}
\W_{\rm tot}=-\frac{\amsq \pi^2\tilde I_2}{6\sqrt{I_1I_3}}\,v\,
\frac{ T^2}{\hbar\omega_0}=-\frac{\amsq \pi^2\tilde I_2^3}{96( I_1I_3)^{3/2}}
\,\hbar \omega_0 \frac{v\eps^2}{\N^2}\EEQ
which can be comparable to the dispersed work, but it 
is always less.

Summarizing this section, we have investigated the presence of many 
work extraction cycles both in the strongly and weakly damped regime. 
At low $T$ their maximal number can be large but it is finite.
The divergence $\N\sim 1/T$ is probably cut of at low enough $T$
when the amplitude $\alpha_0$ of our changes is small but finite.
When more than $\N$ cycles are made, the possibility of work 
extraction is lost, because of the dispersion inherent to cycles.
At moderate $T$ the possibility of work extraction by cyclic 
changes is quickly lost; it is a strictly quantum effect.

\renewcommand{\thesection}{\arabic{section}}
\section{On experiments to test the breakdown of
the second law} 
\setcounter{equation}{0}\setcounter{figure}{0} 
\renewcommand{\thesection}{\arabic{section}.}

In this section we will briefly comment on practical 
realizations of the low-temperature, non-weakly damped
quantum Brownian motion. We do not intend to make 
detailed proposals for experimental setups,
but we will mention certain fields, which according 
to commonly shared experimental views, display the 
above-mentioned strong-coupling and/or low-temperature regime.

\subsection{Once more: the characteristic timescales}

Let us first recall once more that
there are several important time-scales in the problem. 
$\tau_0$ is the characteristic time brought about by the external
potential, which the particle will have if there
is no interaction with thermal bath. For reasonably simple confining
potentials there is only one such a time. In particular,
for the harmonic external potential ${\cal V}(x)=\half ax^2$ 
it is read $\tau_0 =1/\omega_0=\sqrt{m/a}$.
Since no indications of damping are seen in this time, it can have
a physical meaning only for very weak damping: $\gamma \to 0$.

If damping is large, then the characteristic dynamical times
are $\tau _p = m/\gamma$, $\tau _x=\gamma /a$.
The overdamped regime appears with $\tau _p\ll\tau _x$,
and in this case $\tau _p$ and $\tau _x$ can be interpreted as the 
relaxation times of the momentum and coordinate, respectively.

In contrast, very weak damping means $\gamma \to 0$, and the damping time
$\tau_d\sim \tau _p=m/\gamma$ is the longest characteristic time.
For intermediate values of $\gamma$ the characteristic dynamical times in
the overdamped regime are $1/\omega_{1,2}$  defined in Eq. (\ref{mega1}),
and for the underdamped regime they are given in (\ref{tau0taud}).

The issue of this work is to consider the regime where another timescale,
the characteristic quantum timescale $\tau _{\hbar}={\hbar /T}$, 
plays a dominant role. In particular, this timescale governs quantum 
correlations of the bath ~\cite{superbath}. The high-temperature
classical case naturally corresponds to 
$\tau _{\hbar}\ll \tau_0 ,\tau _p, \tau_x$
and there quantum correlation effects can be neglected. 

Equilibrium quantum thermodynamics
is recovered in the limit $\tau _{\hbar}, \tau_0 \ll \tau _p$, which means
that the momentary motion of the Brownian particle practically does not notice
damping, though it does so at long times. Obviously, this condition
cannot be satisfied at low temperatures. 
In the present paper we are interested in the regime where both damping
and quantum correlation effects are important: 
$\tau _p \sim \tau _x  \sim \tau _{\hbar}$, including possibilities of
$\tau _p \ll \tau _x$ or $\tau _p, \tau _x\ll \tau _{\hbar}$, where our 
results are only strengthened. As noticed at the end of Section 7, 
new possibilities for work extraction has been found to occur 
in this last domain, where the inequality
$\tau _p\ll \tau _x\ll \tau _{\hbar}$ says that the noise is 
(anti-)correlated throughout the systems relaxation. 
It thus looks more like a quenched random variable than an annealed one,
thus not at all behaving like a white noise, the standard ingredient
needed to derive from a Langevin equation a Gibbsian equilibrium state.

Before proceeding with concrete examples, let us just notice 
that there is nothing exotic in the quantum time-scale itself: 
$\tau _{\hbar}= 7.6$ ps at $T=1$ K, which is fully in range 
of the modern technologies.

\subsection{Possible experimental realizations}

\subsubsection{Josephson junctions}

The first example to be discussed are Josephson junctions 
\cite{van,likho,likho1,clarke}. This well-known phenomenon
represents a standard example of quantum Brownian motion.
The Josephson junction consists of two superconductors separated
by a thin insulating barrier. Cooper pairs of electrons (or holes)
are able to tunnel through this barrier, thereby maintaining phase 
coherence in the process, and leading to a possibility to have
superconducting current. There is a direct map between properties
of this junction, and the standard model of the quantum Brownian 
motion \cite{van,likho,likho1,clarke}. 
In particular, the coordinate $x$ can be corresponded to the phase
difference of the Cooper pair wave functions, the friction founds
its place as resistance, mass is related to capacitance,  
and the current noise has the standard spectrum (\ref{K0t=},\ref{K0om=}), 
and can  be related to $\eta (t)$. 
Under certain well-defined conditions one can neglect tunneling
of the phase from one metastable state to another \cite{van},
and consider it in a confining, nearly-harmonic potential.
This system couples to the environment,
which acts as the bath of our theory.
In practice one can notice the occurrence of strong coupling
at low $T$ since then a careful shielding of the sample
is needed in order to prevent an influence of the environment
to the measuring apparatus.
It appears that the non-weakly damped and low-temperature limits 
are well-known for Josephson junctions, and were a subject of rather
long experimental activity \cite{van,likho,clarke}.
For example, the following regime
was explicitly realized as a condition of
``really-quantum effects'' \cite{likho}: $\tau _p \sim 0.1$ ps
which is smaller than $\tau _{\hbar}$ at $1$ K. The ratio
$\tau _p /\tau_p $ need not be of order one, but can vary significantly
(from $0.1$ to $10$) depending on the construction of the junction;
for details see \cite{likho}.
In experiments reported in \cite{van} the authors achieved 
$\tau _{\hbar}/\tau _x\sim 10$ at $T=1$ K, and $\tau _p/\tau _x \sim 0.1$,
which is a typical overdamped, low-temperature case. Notice that these
experiments were among the first ones, where the spectrum of the
low-temperature quantum noise was measured and found in the perfect
agreement with the assumed standard form of the quantum Langevin equations.

\subsection{Low temperature electrical circuits}
\label{circuits}

Experiments on mesoscopic, low-temperature electrical circuits
\cite{rlc1,rlc2} provide yet another example, where non-gibbsian values 
of $T_x$ were clearly observed, and found in a good agreement with the 
theoretical predictions. We recall that the linear RLC can be mapped to the 
harmonic brownian particle: the coordinate $x$ and the momentum $p$ of the 
particle correspond to the charge and the current of the circuit, and $m$ and 
$a$ are directly connected with the inductance $L$ and the inverse capacitance
$1/C$ of the circuit (see also our discussion after Eq.~(\ref{Hpq=})). 
Finally, the damping constant $\gamma$ corresponds to the ohmic resistance
$R$. One notices that the (quasi) Ohmic limit, where $\Gamma$ is
the largest characteristic frequency of the problem, is conveniently 
realized in the present context.

First of all we notice that for experiments described in
Refs.~\cite{rlc1,rlc2} all the 
relevant characteristic time-scales have basically similar values:
$\tau_{\hbar}\sim \tau_p\sim\tau_x=10^{-8}$ s, which 
makes the situation especially relevant for our purposes.

Here we will briefly discuss the possibilities of 
experimental detection the Clausius inequality violation at low
temperatures, 
since this seems to be the simplest possible issue.
Moreover, the most evident situation 
is realized upon a slow variation of the inductivity (mass)
$L$, where for $T\to 0$ | according to Eq.~(\ref{dEE1m}) and in the clear
contrast with the Clausius inequality $\dbarrm Q\le 0$ | one gets a 
finite positive heat provided that $\d m=\d L>0$.
One needs to observe $\langle x^2\rangle$ and $\langle p^2\rangle$
for several different values of the inductivity (mass) $L$. This
is sufficient to recover the corresponding changes of the average 
energy, as well as to recover
the work according to Eq.~(\ref{dEEm}). The heat is then obtained by
subtracting the work from the energy.
In the second step one can check the consistency of the results by
observing directly the work done by the external source. 
Altogether, the challenge of the main
experimental observation is in observation of the variances.

In Ref.~\cite{rlc1} the authors considered mesoscopic electrical circuits
 in the  context of single charge tunneling. The used circuits had thickness 
of the order 10 nm and wideness of the order 1 $\mu$m. The observations allowed
indirect determination of $\langle x^2\rangle$.
With the subsequent improvement made in \cite{rlc2} the correspondence with the
theoretical expression (\ref{Txexact}) is perfect. The observations were done
with $C=1/a=$4.5 fF, $L=$4.5 nH and for $R=\gamma$ in the 
range $10^{1}-10^{3}$ k$\Omega$. For damped circuits the relative 
importance of damping is quantified by the quality factor $\tau_p/\tau_0$, 
which in the above range of parameters varies from $10^{-1}$ to $10^{-3}$. 
To avoid thermal noises the circuits were cooled up to $20$ mK. At such a low
temperature quantum effects are dominating, since the quantum time-scale
$\tau_{\hbar}=\hbar/T\sim 10^{-8}$ s is larger than 
the other ones, $\tau_0\sim 10^{-9}-10^{-10}$, $\tau_p\sim 10^{-8}$ 
and $\tau_x\sim 10^{-9}$ s. To get an idea for the magnitude of the 
expected effect, let us estimate the outcome for 
$\Delta Q\simeq L~\dbarrm Q/\d L$. With the above parameters and 
$R=\gamma=10^{3}$ k$\Omega$ one gets from (\ref{dEE1m})
$\Delta Q\sim 10^{-19}$ J$=1$ eV. 

Since for mesoscopic circuits
the formula for $T_x$ was already verified, it is now a matter to
perform three measurements (the equivalents of $T_p$, $T_x$ and the
work production) on a single sample, to verify unambiguously the 
breakdown of the Clausius inequality.

\subsubsection{Trapped ions}

As another, more elementary example one can mention a trapped ion 
immersed in a photon bath. Taking as an estimate the mass of
proton ($m=10^{-26}$ kg), and $\gamma =10^{-15}$ kg s$^{-1}$
\cite{gardiner}, one gets
$\tau _d \sim\tau _{\hbar}$ at $1$ K, so the quantum coherence effects
are still active.
The ideal example of a harmonic Brownian particle 
will be an ion trapped in a so-called Paul trap \cite{paul}, or an
electron or ion in a Penning trap~\cite{Gabrielse}.
These electromagnetic traps are nowadays well realizable and suited
for variation of parameters. In particular, high quantum number
Rydberg states may have a long lifetime and a strong coupling to the 
radiation field.

\renewcommand{\thesection}{\arabic{section}}
\section{On the foundations of thermodynamics and perpetuum mobile} 
\setcounter{equation}{0}\setcounter{figure}{0} 
\renewcommand{\thesection}{\arabic{section}.}

This section intends to summarize to what extent the standard 
relations and laws of thermodynamics can be applied to a 
quantum Brownian particle. There are many formulations of the second
law, and some of them have been found invalid in previous discussion.
One may go to the extreme limit as saying that there is no
motivation to discuss a thermodynamics in the way we did. 
To show that there is justice in doing it, we summarize our results 
in the light of standard thermodynamic wisdom, and point at the
agreements and contradictions. 

For a general, pedagogic text on the history and
today's status of thermodynamics and the second law, 
we refer to the recent work by Uffink~\cite{Uffink}.
For a collection and discussion of the original papers, see
the book by Kestin~\cite{Kestin}. A very recent discussion on the base
of the axiomatic thermodynamics
was presented by Lieb and Yngvason~\cite{LiebYngvason}.
For a discussion of what can be meant by ``the'' entropy of a system,
see ~\cite{MaesLebowitz}.

\subsection{Has thermodynamics been violated or did it never apply?}

The conclusion of our analysis is that thermodynamics does not work
when, in the quantum regime, ones considers 
the Brownian particle in its reduced Hilbert space,
thus  summing out the bath variables of the total system.
This implies that the characteristics of the particle are
directly observable which is indeed the case with the standard
examples of the brownian motion. There are, however, situations, where
only some composite (system-plus-bath) quantities are measured, and the
need for a separation between particle and bath is questionable
(there still can be a possibility that such a separation can be done
on a different, more coarse-grained description of the overall
system, but we will not enter into that discussion here).
This is the case with a Kondo system, where the measured quantity
is for instance the magnetization, which is set by the magnetic 
impurity and the bath together (i.e. it lives in the common space of
the particle and the bath). Also for the dressing of a `bare' electron
by photons, it is the standard practice of the quantum field theory 
to consider the dressed mass and charge as directly observables quantities. 
However, when the system is a Josephson junction or a mesoscopic circuit, its own
characteristics are perfectly measurable, so there is an important case 
to make. When looking at the budget of the junction alone, one has to keep 
in mind that it can exchange energy with its environment.
At low enough temperature  this mechanism displays unexpected behavior 
and is responsible for non-thermodynamic characteristics.

If one accepts to consider the subsystem as a Brownian particle immersed
in a heat bath, then first it should be noticed that the particle 
acquires a cloud of bath modes around it. This dressing is a manifestation of
the (strong) damping of the particle by the bath. 
One can then ask the question: ``if thermodynamics does not apply,
where was it lost?'' If no technical errors have been made in our 
derivations, then the answer must be: ``It never applied''.

An argument in favor of this point of view is the fact that the 
interaction energy is non-vanishing, thus violating a well known 
condition for the derivation of the standard {\it equilibrium} 
thermodynamics.  The interaction energy reads

\BEQ U_{\rm int}=U_p-\half T_p-\half T_x \EEQ 
The result is given explicitly by Eqs. (\ref{Up=}) and (\ref{Tpexact}).
In a strict formulation of thermodynamics one requires that the
equilibrium value of the
interaction energy $U_{\rm int}$ is negligible, in order to separate
what is meant by the system from what is meant by the bath. 
In our case this would imply $\gamma\to 0$ or $T\to\infty$, 
and indeed in both limits standard thermodynamics is recovered. 
However, in general the same 
system has a non-zero interaction energy and, in the strict formulation,
should not be considered to be thermodynamic at finite $\gamma$ or $T$.
Whereas the limit of large $T$ can be naturally achieved in practice,
the weak-coupling limit $\gamma\to 0$ is much more difficult to
realize, since coupling constants are generally fixed numbers whose magnitude
cannot be manipulated at will.
Likewise, all physical systems having a non-vanishing interaction energy
with their baths should then not be considered as thermodynamic systems.
This would apply to very many systems at low enough temperature,
leaving an uncomfortable situation with respect to the well
behaved high temperature properties of the same systems.
Already at high $T$ the analysis of the Clausius inequality
in section \ref{procontra} gave a compelling argument in favor of
our choice for the Hamiltonian of the subsystem, leaving the rest
of the total Hamiltonian for the bath.

Let us inspect in some detail the weak damping limit $\gamma\ll\sqrt{am}$. 
Here it can be shown that, even at $T=0$, one has $U_{\rm int}\ll U$,
since, due to (\ref{Tpexact}),  $U_{\rm int}\sim\gamma$ but 
$U\to\half\hbar\omega_0$, implying
that the condition for the application of thermodynamics is almost 
fulfilled. Nevertheless, the Clausius inequality is typically violated,
by an amount of, again, order $\gamma$.
This argument somewhat weakens the point of view that thermodynamics
should never apply. One can also argue that it is compatible with it, 
since the violation of thermodynamics is of order of the 
small interaction energy.

Putting all arguments together, we reach the unavoidable conclusion: 
there are principle problems to define
thermodynamics at not very large temperatures and, in particular, in 
the regime of quantum entanglement. There is no resolution to this, 
and thermodynamically unexpected energy flows appear to be possible.

\subsection{Zeroth law}

The zeroth law is often said to state that in an equilibrium
situation there will be a unique temperature. A standard formulation
is that if two bodies are each in equilibrium with a third body,
then they are also in equilibrium with each other, and the three bodies 
have a common temperature.
Let us look, however, at a careful formulation out of equilibrium:
{\it If two parts of the system have an infinitesimally 
small temperature difference, then they will spontaneously 
equilibrate and reach a common temperature.}

For the (nearly) harmonic situation two different 
{\it effective} temperatures
$T_p$ and $T_x$ can be related to the momentum and the coordinate.
Recall that these temperatures arise from the generalized form
(\ref{asala1}-\ref{asala2}) of the Clausius inequality. 
The legitimation of such a definition
of effective temperatures is confirmed by their successful
use in glassy thermodynamics \cite{1,4}.

In our case, Eq. (\ref{TxTplargeT}) shows that $T_p$ 
deviates at large $T$ from $T_x$ by a term $\beta A \Gamma$, with
$A=\hbar^2\gamma/12 m$. So for any infinitesimal $\epsilon$, 
the regime $T>A\Gamma/\epsilon$ indeed has temperatures $T_p$ 
and $T_x$ that differ less than $\epsilon$. However, since they are 
parameters of the steady state, they will 
not equalize spontaneously, in conflict with the above 
formulation of the zeroth law. Instead, they become more and more
different from each other at lower temperatures, 
and at zero bath temperature
$T_p$ and $T_x$ are both finite but different from each other.
The fact that they remain finite just indicates that 
the corresponding quantum state does not have sharp values for 
$p$ and $x$; so this is a consequence of quantum complementarity.
The fact that these effective temperatures take non-gibbsian values
is a consequence of the quantum entanglement.
In the Gibbsian limit of weak coupling (i.e. $\gamma \to 0$) 
for the harmonic oscillator, both temperatures $T_x$, $T_p$ tend to
their common value $\frac{1}{2}\hbar \omega _0
{\rm cotanh}(\frac{1}{2}\hbar\beta \omega _0)$ 
of the harmonic oscillator weakly coupled to its bath.

We should mention that the existence of the zero law is frequently viewed
just as an axiom, but under certain conditions it can be derived from 
the second law (the entropy of a closed system never decreases) \cite{landau}.
As we mentioned already, this derivation is based on the use
of a weak interaction between the particle and its thermal bath. 
It confirms that if this weak-coupling condition is valid,
then the two effective temperatures are indeed 
approximately equal.

The fact that these unequal effective temperatures do not cause 
heat currents that equalize them, as would be required by 
standard thermodynamics, is reminiscent of the classical paradox 
that atoms should radiate, but, being in the quantum regime, they do not.

\subsection{First law}

The first law relates the change of system's energy into 
the heat added to it and the work done on it. 
It can not be broken, since it is a direct consequence of energy 
conservation, a central concept in quantum mechanics.
Nevertheless the formulation of this law is not merely a tautology, 
because it allows to separate clearly those ingredients of the energy 
change, which arise from non-observable degrees of freedom (heat obtained
by the Brownian particle from the thermal bath) and external 
sources (work done by them on the whole system).
Our identification of the energy of the subsystem 
as the expectation value of the Hamiltonian $\H$ was 
supported in section ~\ref{procontra} by requiring application of
standard thermodynamics at high $T$, and is imposed by the form 
of the Wigner function.
We stress that, given this identification of energy,
 our identification of the heat $\dbarrm {\cal Q}$ added
to the subsystem and the work $\dbarrm {\cal W}$ done on it, are well
accepted and widely discussed in literature, see e.g. the books by 
Keizer~\cite{keizerbook}, Balian~\cite{balian} and Klimontovich~\cite{klim}.

\subsection{Second law}
Let us stress that there are many formulations of the second law.
There are several formulations of the second law that are,
at least apparently, violated by the solution of our problem.

\subsubsection{Thomson's formulation, Kelvin's principle}
\label{Thomson=Kelvin}

The formulation by William Thomson, the later Lord Kelvin of Largs, is:
{\it It is impossible to perform a  cyclic process 
with no other result than  that heat  is absorbed from
a reservoir, and   work is performed}. An earlier and more particular 
version of this statement appeared due to Carnot. 

For general quantum systems starting from the equilibrium state, 
this can be proven mathematically~\cite{Lenard}; 
a simplified proof will be presented elsewhere~\cite{ANthomson}. 
In our setups it can always be verified, see
for example the fact that the energy dispersion 
(\ref{Picycle}) is non-negative. After finishing the cycle the bath is
not exactly in its Gibbsian state, but it is still very close to it,
because the bath is extensive. Basically the dispersed energy has run away
to infinity, leaving the system locally again in a Gibbsian state. This 
implies that also successive cycles will always disperse energy. 

However, out of equilibrium Thomson's formulation appears to be endangered.
The first point to notice is that this can already occur at the classical
level. The reason is simple. Consider, as we did in section ~\ref{Suddenwork},
a sudden weakening of the central spring. In doing so, energy is extracted 
from the system, but, due to the sudden nature, it is not the optimal amount.
One can improve on this by making the following cycle: quickly put the
spring  back at its original value, and then make the change slower.
This cycle, that started in a non-equilibrium state,
will yield work, and this work comes from the bath. We conclude that
the Thomson formulation can only refer to system changes on long
enough timescales, such that the initial state is practically in
equilibrium.

A more exciting violation of the non-equilibrium
Thomson formulation was observed for
smooth changes of the spring constant at low enough temperature. 
In section \ref{perpmob} we discussed the case of $\N\gg 1$ bell-shaped cycles
in the spring constant; each cycle has two inflection points, pre-peak 
and post-peak. Starting in the Gibbsian state, the first cycle up to 
the post-peak inflection point is considered as a mechanism that produces 
for us a proper `initial' non-equilibrium state. If the typical duration
of the successive cycles increases, parameters can be chosen such that
after each return of the spring constant to its  
post-peak inflection point value, the system has the same energy, while a
prescribed, fixed amount of work is extracted. There can be $\N\sim 1/T$ 
of these cycles, which can be large  at low enough $T$. They extract 
heat from the bath and convert it fully into work, forbidden by the
general (i.e. non-equilibrium) Thomson formulation. It could be checked 
that the total amount  of extracted work is less than the energy dispersed
in the first part of the first cycle, so energy conservation is not
endangered. 
The interesting fact is nevertheless that there can be $\N\sim 1/T$ 
of these cycles, which can be large  at low enough $T$.
Actually, making more cycles implies a smaller total 
extracted work $\sim 1/\N^2$, 
since these cycles themselves lead to additional dispersion.

In contrast to the violation of the Clausius inequality,
to be discussed below, the violation of Thomson's formulation is a
consequence of both quantum regime (low temperatures) and the non-equilibrium
character of the whole system (particle and bath). Indeed, any 
amount of work extracted by means of the particle is in fact extracted
from the whole system. If this global system is in equilibrium
(namely it is exactly described by the Gibbs distribution), 
there will be no possibility to extract work by making a cyclic change
of a system parameter, since this formulation of the second law applies as 
well to any closed equilibrium system~\cite{Lenard,ANthomson}.
On the other hand, the full account of quantum effects is
necessary to show our work extraction, since it disappears 
in the Gibbsian limit, namely both at moderate temperatures 
and/or for weak coupling to the thermal bath.

\subsubsection{Clausius' Principle} 

Clausius states: {\it It is impossible to perform a  cyclic process 
which has no other result than that  heat is absorbed from a reservoir
with a low temperature and emitted into a reservoir with a higher
temperature.}

The gained work could be used to drive some frictional process at
a higher temperature, which would turn it into heat,
in conflict with this principle.
Nevertheless, this principle is obeyed at high temperatures,
and only violated in the quantum regime at low $T$.

\subsubsection{Clausius inequality}
This formulation claims that in any thermodynamical process (in 
particular, for variation of a system parameter) the amount of heat received
from the thermal bath by the particle is limited from above by the bath 
temperature times the change of the von Neumann entropy of the particle:
\BEA
\dbarrm Q\le T\d S_{vN}.
\nonumber
\EEA
A particular formulation of this law is that no heat can be extracted 
from a zero-temperature thermal bath, it can only be dumped in it
(i.e., then it is impossible to have $\dbarrm Q\ge 0$).
This situation is particularly interesting, since it does not
employ in any way the concept of entropy, and therefore can be applied 
to situations, where entropy is not known, or not well-defined.
Physically it is also easy to understand. The energy of the cloud of bath 
modes around the subsystem will change if a system parameter is changed,
even at $T=0$. This change in energy of non-observable modes
is heat, and it can be positive or negative, depending on the sign
of the change. In one of the cases energy from
the cloud will increase the subsystem's energy, violating the 
Clausius inequality. For a closed system $\dbarrm Q$ goes to zero, and one
recovers from the above inequality the more standard formulation 
$\d S_{vN}\ge 0$, which appears to be a particular case of the Clausius
inequality. 

We have shown that both those general (all $T$) and particular 
($T=0$) formulations are 
violated in quantum case. Although at high temperatures these violations
are small, they nevertheless do exist. For researchers who are
reluctant to follow our identification of the effective temperatures,
it will perhaps be hard to agree on the violation of the zeroth
law, discussed above. However, the violation of the second law, which also
sets in at arbitrarily large temperatures, should be easier to
accept, since the Clausius inequality does not employ the notion
of effective temperatures. Moreover, in section ~\ref{Claustwosys} 
we have discussed a formulation which compares only equilibrium systems.

For our harmonic system we succeeded in generalizing the Clausius 
inequality, involving two temperatures and two entropies, in the very same 
way it was done for glassy systems and which applies to black holes.

In hindsight, the derivation of the Clausius inequality is 
nontrivial in the case under consideration.
In standard thermodynamics one formulation of the second law is that
the total entropy of a closed system cannot decrease.
When applied to a subsystem coupled to its equilibrium bath, this
immediately leads to the Clausius inequality. Hereto
one makes two assumptions: equilibration of the bath and additivity of
the entropy. Let us follow the subsequent steps.
Because of its equilibrium nature, the
heat received by the bath is associated with an entropy
change, $\dbarrm{\cal Q}_{bath}=T\d S_{bath}$. Energy conservation
says that $\dbarrm{\cal Q}_{bath}+\dbarrm{\cal Q}=0$, where the latter
is the heat received by the subsystem. This implies 

\BEQ \d S-\frac{\dbarrm{\cal Q}}{T}=\d S+\d S_{bath}=\d S_{\rm tot}\ge 0
\EEQ

In the world of quantum entanglement, however, both assumptions are
less obvious. First, it does not hold
that $S_{\rm tot}= S+ S_B$. We have shown this explicitly, since
at $T=0$ one has  $S_{\rm tot}=S_B=0$, but $S=S_{vN}>0$. 
Both the fact that energy is not quickly redistributed in the bath
and the non-additivity of the entropy imply that there is no 
a priori reason to expect that the Clausius inequality is satisfied.
In concordance with that, we have shown that it is indeed not valid.

\subsubsection{The rate of energy dispersion is non-negative}

In section \ref{incompletecyc} we have pointed out that at low enough
temperatures the rate of energy dispersion can easily be negative. 
This holds even when one starts in equilibrium. Thus non-negativity of energy 
dispersion cannot serve as a universal formulation of the second law.

\subsubsection{The total entropy of a closed system cannot decrease}

The most standard formulation of the second law is that the
(coarse grained) entropy of a closed system cannot decrease. In classical physics
for a subsystem in contact with a heat bath the equivalent is that 
the rate of entropy production of the subsystem is nonnegative.
For situations close to equilibrium it can often be expressed as
a bilinear expression in generalized currents, and the matrix elements
are called Onsager coefficients; this matrix is positive definite
in all known examples.

The rate of production of Boltzmann entropy was also considered by us.
In the case of weak damping there occur 
oscillations in the production rate around zero in each period; 
this sets in at moderate
temperatures, and is akin to the oscillations in the energy, that occur
already at any non-infinite temperature.
In sections \ref{weakentr} and \ref{strongentr}
we have pointed out that even at low temperatures and in the limit of strong
damping the rate of Boltzmann entropy production can be negative. 
So also  this criterion does not qualify as a solid definition of the second law.

We should stress that we did not find sensible production rates for
other entropies. Perhaps not accidentally, the Boltzmann entropies
of the coordinate and momentum sectors are the ones that  enter
 our generalized, two temperature version of the Clausius inequality.

In our setup the von Neumann entropy for the full closed system 
(fine grained entropy)
should not be confused with the von Neumann or Boltzmann entropies
of the subsystem,  which pertain to the Brownian particle only.
The von Neumann entropy of the full system is not altered by changing  
the strength of the spring constant. This entropy remains constant during the
overall unitary evolution of the whole system, and also remains constant
during variations of a parameter, since also there the overall 
evolution is still unitary.  The formulation of the second law in terms
of non-decrease of entropy definitely refers to the coarse grained entropy.
In the classical situation the fine-grained entropy is conserved
as well, by  the Liouville dynamics. 
For more definitions of entropy, see ~\cite{MaesLebowitz}.

In passing we notice that if one starts from a Gibbsian state of the
total system (central particle coupled to the bath), and changes a
system parameter, then the conservation of entropy prevents
the system to relax to a new Gibbsian state of the total system, since our
total system is isolated. 
Nevertheless, the subsystem (the central particle) does relax to
a state characterized by the parameters, which can be coded in the
effective temperatures, of that would-be global Gibbsian state.
It is the finite amount of energy dumped in the extensive bath that
does not relax, since our bath lacks anharmonic interactions, or 
coupling to an external superbath.  In contrast to a
superbath, anharmonic interactions do not change the 
essence of the argument on the overall unitary evolution,
conservation of both the von Neumann entropy and the energy. However,
they can widen the set of observables having would-be Gibbsian values.

\subsection{Third law}
This law claims that if the ground state 
of the Brownian particle is non-degenerate, 
then its von Neumann entropy is equal to zero. This is a direct consequence
of the quantum Gibbs distribution, which predicts the pure vacuum state at 
low temperatures. In our case
neither the von Neumann entropy nor the Boltzmann entropy vanishes when 
the bath temperature is zero. This occurs because of the quantum 
Brownian particle is in an entangled mixed state, and therefore cannot have
vanishing von Neumann entropy.

The third law is recovered when taking the weak-coupling limit.
In that case $T_p=T_x=\half\hbar\omega _0
{\rm cotanh}(\half\beta\hbar\omega _0)$,
implying that the parameter $v$ of Eq. (\ref{v=}) takes
the value $v=\half\, {\rm cotanh}\,\half\beta\hbar\omega _0$,
which causes the von Neumann entropy of the particle
(\ref{SvN=}) to vanish at $T= 0$. 
In a certain sense the violation of the third law reported here for
non-weak coupling is 
the most straightforward consequence of quantum entanglement.

\subsection{Perpetuum mobile of the first kind}

Taken literally, a perpetuum mobile performs perpetual, i.e., 
everlasting, motion. Nevertheless, rotational currents in ordinary 
superconductors, which may exist several days, are rarely
connected to perpetual motion. We shall therefore employ the word
``perpetuum mobile'' for any principle that yields work. 

One speaks of a perpetuum mobile of the first kind when the first 
law is violated, leading to an everlasting performance of work without
any cost. Such a setup is impossible in quantum mechanics, 
since it satisfies the first law by principle. 
So here is no issue in the question what ``perpetuum'' 
means precisely.

\subsection{Perpetuum mobile of the second kind}
\label{Sectperpmob}

Another formulation of the second law is:
{\it It is impossible to construct an engine which will work in a complete 
cycle, and convert all heat it absorbs from a reservoir, into mechanical 
work}~\cite{KondPrig}. A machine which would do so is called a 
{\it perpetuum mobile of the second kind}, and the second law states
 that such a machine is impossible.

\subsubsection{``Perpetuum'' mobile or perpetuum mobile?}
When the first law is respected but the second is violated, one
speaks of a ``perpetuum mobile of the second kind''. However, we wish to
make some remarks on the word `perpetuum'. Surely, in the eighteenth
century such a perpetuum
mobile was imagined, for instance, to cross the Atlantic by boat using
only the energy stored in the ocean water. As such, there would be a
basically infinite bath, and the mobile, if realized, could function 
perpetual, i.e. ``for ever''.
In general, when the bath is finite, it obviously 
has a finite energy at its disposal.
In many setups, such as those with a finite rate of energy extraction,
this implies also finite duration of the process.
Thus even in the classical situation, 
the term ``perpetuum'' need not be a precise adjective 
for this type of mobile, and the point of view could be taken that
a perpetuum mobile of the second kind need not
function arbitrarily long, but must only work many cycles.
In view of the failure to find so far any practical realization, 
this stretch of the definition seems allowable to us.

In quantum physics the situation is even more clear.
In a closed system the energy can never go below the ground state energy,
so the amount of  extracted work is always finite. As a result,
there can never be an infinite amount of cycles for which a definite
amount of work is extracted per cycle. One of the formulations of
the second law is: ``Perpetuum mobile of the second kind do not exist''.
Such a strong physical statement must, of course, be richer than the 
general statement on the existence of a ground state. 
The crux is that already one of such cycles, that extracts work 
from a thermal bath,  is forbidden. 
So, already in general, perpetual motion of the second kind does 
not have its literal meaning of everlasting motion; rather, 
it is a notion for a work extraction principle, and one cycle is therefore
good enough for the birth of a `baby perpetuum mobile'.

\subsubsection{The present situation}

For our aim the allowance of non-eternal duration of perpetuum 
mobile is relevant, since our effects only hold as
long as both the particle  does not relax, which happens
on the timescale $\tau_x$, and are  quantum coherent, which involves
the quantum timescale $\hbar/T$.

We have discussed a work extraction mechanism that cyclicly changes the
spring constant in a certain time interval. Each of these cycles is
slower than the relaxation time of the system. When the quantum timescale
$\tau_\hbar=\hbar/T$ is also slower than the relaxation time, there
occurs unexpected behavior: the contribution to the rate of dispersion 
inversely proportional to the duration of the cycle, normally the
leading term, has a small prefactor quadratic in temperature. Therefore 
quadratic and cubic terms in the inverse duration also play a role.
Out of equilibrium cycles have been designed where a constant amount
of heat is extracted from the bath is fully converted into work,
while the energy of the subsystem is at the end of each cycle back
at the value of the beginning.
{\it In this sense, systems described by our models, with 
parameters in the appropriate regime, present at low temperatures 
true realizations of perpetuum mobile of the second kind}.
Probably, it is also possible to extract work both from the bath
and from the subsystem (``efficiency larger than 100$\%$'').

In a more stringent definition of perpetuum mobile one requires
that in the cycles work is extracted ``without any further change''.
For our system this can be expressed as the requirement that the
Wigner function of the subsystem be back at its original value.
This would imply the requirements that $\Delta K=0$, 
$\Delta V=0$ and $ \Delta\dot V=0$ over each cycle, rather than only 
having $\Delta K+\Delta V=0$, where $K$ is the expectation of the kinetic
energy and $V$ of the potential energy. The question whether this extended 
constraint can satisfied by changing the spring constant and possibly
also the mass, is left for the future.

We should stress the conceptual difference between the present
situation and the well-known case, where work can be extracted due 
to a temperature difference between two thermal baths \cite{landau}. 
The latter is the standard setup for the thermodynamic heat engine: 
two baths are explicitly separated from each other, and therefore the whole 
system is in a nonequilibrium state, and can be used to perform work.
If those baths are kept in a direct contact for a sufficiently long time,
then they will go to equilibrium and after that no work can be extracted
any more~\cite{ANheatengine}. 
In contrast, here we are have presented a case with {\it one single} 
thermal bath. After a sudden increase of the
strength of the central spring, the central particle will 
go to equilibrium after one relaxation time $\tau_x$ or $\tau_d$. 
But in doing so, more than the initial surplus 
energy is transfered to the bath,
and in particular to the cloud of bath modes in its immediate surrounding.
After a certain moment, this heat flow towards the bath stops,
and then a smaller backflow occurs from the bath to the particle, 
before the whole comes into equilibrium. This backflow of heat is the mechanism
that makes it possible to extract work from the bath by manipulating the
particle, in a situation where this would be impossible classically.
In particular, for smooth changes at zero temperatures 
the integral of this relaxation function is needed, but it appears
to vanish, leading to a variety of new effects at low temperatures.

On a thermodynamic level, the analogy with the classical case 
was strengthened because we could 
identify effective temperatures, though we also stressed that by themselves
they do not tend to become equal to each other. 
All these intriguing aspects arise due to quantum effects, 
since we showed in detail that {\it the same system coupled to the same
bath} displays at high temperatures  the fully expected 
thermodynamical behavior.

\subsection{Perpetuum mobile of the third kind}

One can define a perpetuum mobile of the third kind when work is
performed at the cost of a diminishing, but still
non-vanishing zero-temperature entropy.
This can, in principle occur in systems, such as glasses, 
which are able to relax to equilibrium, but are temporarily 
stuck in certain metastable states. Then the zero-temperature entropy
can be used as a measure of this metastability~\cite{Nunp98}.

One could wonder whether our extraction of work is due to the
present non-vanishing zero-point entropy.
However, this is not the case,
since for the purely Gibbsian case of particle and bath,
the particle would have the same zero point entropy,
but no work could be extracted. 
Moreover, in our case the zero point entropy is an indication of 
quantum entanglement and not of metastability.

\renewcommand{\thesection}{\arabic{section}}
\section{Conclusion} 
\setcounter{equation}{0}\setcounter{figure}{0} 
\renewcommand{\thesection}{\arabic{section}.}

This paper is devoted to the statistical 
thermodynamics  of the quantum Brownian motion. 
The high-temperature case of this model can serve as a convenient 
pedagogic example, where almost all main statements of statistical
thermodynamics are derived exclusively from the first principles. 
Among other advantages, such an approach makes possible to reveal the conceptual
restrictions and limitations of the standard thermodynamical wisdom. 
With this aim in mind we focussed in the paper on the low-temperature
(quantum) situation of the Brownian motion model.

The stationary state of a quantum Brownian particle non-weakly
interacting with its thermal bath is non-Gibbsian. It is 
this property which makes the quantum Brownian motion a challenging 
problem, and classical thermodynamical wisdom appears 
to be inadequate even if the total state of the system and the bath
is Gibbsian. Both the classical and the quantum Gibbsian
thermodynamic theories emerge as particular limits 
in this more general setup. The classical Gibbs distribution with 
all its thermodynamic consequences is recovered for high temperatures, 
and the quantum Gibbs distribution is obtained for very weak damping.

In section \ref{langevin} we start from the quantum
Langevin equation. At low temperatures this equation
contains a colored Gaussian noise; because of quantum coherence, the
bath cannot generate white noise even in the limit where the friction 
has no memory. To achieve this interesting situation, 
no more is needed than the observation that quantum mechanics applies
(see detailed explanations after Eq.~(\ref{K0om=})).
Moreover, the quantum fluctuation-dissipation theorem
predicts different time-scales of noise and dissipation at low enough
temperatures \cite{landau,klim,klim-rev,weiss}. 
This is how quantum coherence enters into the considerations. Its
characteristic timescale is $\hbar/T$, where we had set the
Boltzmann constant $k_B$ equal to unity so far.
Restoring it, we have $t_\hbar = \hbar/(k_B T)=(6.23/T)\,K\,ps$.
At $T= 1 K$ one has $t_\hbar=6.23\, ps$,
in the range of typical microscopic processes in condensed matter. 

Since the stationary distribution is non-Gibbsian there are conceptual
differences compared with the equilibrium case.
For the harmonic potential one can define effective temperatures
(\ref{TT}, \ref{TTT}) for momentum and coordinate.
Both temperatures are different from the bath temperature $T$. 
Generalized thermodynamic relations can be introduced, which take
a quasi-Gibbsian form, Eq.~(\ref{dEE}-\ref{dEE2}),
and are closely related to the ones in the thermodynamics of
glasses ~\cite{1,Nhammer,4}. 

The inapplicability of standard thermodynamics is most clearly 
illustrated by the violation of the Clausius relation: $\dbarrm Q\le T\d
S$: heat received by the particle from the bath is restricted by 
temperature of the bath times the change of the particle's entropy. 
In section \ref{CC} we construct an explicit example which at low
temperatures realizes $\dbarrm Q> 0$.
This violation is significant at low temperatures,
where quantum effects are relevant, and is small for high temperatures.
It is important to notice that this violation
exists already for the totally equilibrium (Gibbsian) state of the
overall (particle plus bath) system (see our discussion after 
Eq.~(\ref{dQdUclT})). Since Thomson's formulation of
the second law is valid for such a state \cite{Lenard}, we have the
explicit counterexample showing that the very equivalence between
different formulations of the second law is broken at low temperatures.

For the dynamical consideration we start from a non-equilibrium initial
state obtained from the total Gibbsian by changing the width of
the confining potential. This change involves a small, controllable
energy input, and can be more realistic than the hitherto studied case
where particle and bath are initially uncorrelated.
After the non-equilibrium state has been prepared, the ensuing
relaxation of the particle presents a number of thermodynamical
anomalies at low temperatures of the bath. First, energy put into the
bath does not completely dissipates there (in contrast to the classical
situation), and thus work-extraction from a single thermal bath is
possible. This violates the second law in Thomson's formulation  as
applied to non-equilibrium. As a consequence of this, adiabatic changes
of parameters are not the most optimal ones anymore.
It is interesting to notice that at low temperatures a large (but
finite) number of work-extracting cycles is possible. 
According to our opinion, this explicitly realizes the basic
non-trivial content of the perpetuum mobile of the second kind,
because any possibility for infinite number of such cycles is ruled out merely by
the existence of the ground state for the overall system.

The second aspect of the low-temperature relaxation is that no
H-theorem exists at low temperatures, i.e. the properly defined
entropy production appears to be negative for some times. This holds even in the
moderately overdamped regime, when the brownian particle relaxes
monotonously. (Recall that without any bath those frequencies are purely
imaginary which leads to the known oscillatory behavior.) Within the
underdamped (weakly-coupled) situation negative entropy production
persists up to high temperatures, and disappears only in the
explicitly classical limit.

Let us recall that Thomson's formulation of the second law in 
its most general and universally applicable form \cite{landau}: 
{\it In cyclic processes no work can be extracted 
from a closed equilibrium system}, remains satisfied. 
This statement was derived in Ref.~\cite{Lenard},
and we will present a very simply proof elsewhere~\cite{ANthomson}.
It obviously applies to the analysis of this paper,
since we start from a Gibbsian with modified spring constant,
and the work extraction disappear when the change 
in the spring constant vanishes. Also for cyclic smooth
changes that start from equilibrium we could verify the 
non-negativity of the energy dispersion.

We now make some remarks concerning the definition 
of the thermal bath in our problem. 
The harmonic oscillator bath model, which was used by us, is technically
convenient, but at the same time it possesses all relevant properties of 
a thermal bath, which are typically postulated in the statistical
thermodynamics. The main of them is that the bath should have
infinite amount of degrees of freedom, a necessary 
condition to ensure relaxation of the Brownian particle. 
On the other hand, the quantum Langevin equation, which is the
starting point of our analysis, can be derived from rather 
different schemes (see e.g. \cite{weiss}), since in a sense they 
are more universal than the detailed properties of the 
considered thermal bath.

Finally, let us relax the conditions under which our results have 
been derived. We have already mentioned that they hold as well for 
$N\gg 1$ Brownian particles in an external potential. 
Though mutual interactions would complicate the analysis, it would not 
modify our basic statements. This can already be seen from the
case of non-interacting harmonically bound Brownian particles: 
under a change of variables they become interacting ones, 
while the characteristics of the bath remains basically
unaffected because it has very many degrees of freedom.

Our findings on the non-thermodynamical character of the
low-temperature brownian motion  may
have a wide scope of applications such as cooling, energy storage,
and thermodynamical limits of low-temperature computing.
Indeed, in the domain of information theory 
there is a large literature based on the fact that only the erasure 
of information must necessarily involve a dissipation of heat
(see extensive reviews in \cite{rex}), and the claim that the erasure of 
one bit of information costs at least an amount of energy
$k_BT\ln 2$, the so-called Landauer bound.  It is well known that 
this bound is based on a straightforward application of the Clausius
inequality. From our observations it is clear that the Landauer bound
can also be broken, and strong effects may occur at low temperatures. 
This may have implications for computing in the quantum
regime~\cite{ANLand}.

For spins coupled to a bath the quantum nature expresses itself in
off-diagonal elements of the density matrix. These decay after
the time ${\cal T}_2$, which can range up to seconds.
In this regime related  work extraction setups are possible.

Our results can be phrased in the statement that Maxwell's demon
exists: it is the property of quantum entanglement
in quantum mechanics~\cite{demon}.
They may further have implications for thermodynamics in high-energy
physics and the early Universe.

The aim of the paper has been to show that
violations of the second law have a natural place in the 
physics of quantum particles that are non-weakly coupled 
to quantum baths.  In this domain
we have given conditions for the realization of the most 
notorious objects in the history of physics: 
perpetuum mobile of the second kind.

\section*{Acknowledgments}
We thank R. Balian, H. van Beijeren, H. Knops, R. Lipowsky,
C. Pombo, H. Spohn, L. Suttorp and M. Wagner for discussion 
and R. Lipowsky also for hospitality at the 
Max Planck Institute for Colloids and Interfaces in Golm, 
where a part of this work was done.
We stress that the responsibility for the presented results 
and interpretations lies with the authors alone.
The research of A.E. A. was supported by FOM (The Netherlands) 
and by NATO.

 \renewcommand{\thesection}{\arabic{section}}
\section*{Appendix A: Smooth changes of the spring constant}
\renewcommand{\thesection}{\arabic{section}.}
\renewcommand{\theequation}{A.\arabic{equation}}
\setcounter{equation}{0}

In this appendix we derive the work for continuous changes of the
spring constant. Hereto we first notice that 
perturbative expression (\ref{betai=}), (\ref{bnu0=}) 
of the exact result (\ref{betanu=}) can be derived directly
by perturbation theory. Let us first 
denote $a$ by $a_1$ and expand also $\phi$
to first order in $a_1-a_0$. We may use
\BEQ \sin\phi\,e^{i\phi_0-i\phi}=\sin\phi_0+(a_1-a_0)
e^{-i\phi_0}\frac{\d\phi}{\d a}
=\sin\phi_0-(a_1-a_0)\sin^2\phi_0\,e^{-i\phi_0}\frac{\Gamma^2+\nu^2}
{\gamma\Gamma^2\nu}\EEQ 
Now inserting (\ref{phifrel}) and taking
the large $\Gamma$ limit of $\hat f(i\nu)$ from (\ref{hatf=}),
we get from Eq. (\ref{bnu0=})
\BEQ \label{bnu01=}
\beta(\nu)=\sin\phi_0(\nu)\,e^{i\nu t}\,\,
\left[1+\frac{a_0-a_1}{\gamma w}
\left(\frac{1-e^{-\omega_1t-i\nu t}}{\omega_1+i\nu}
-\frac{1-e^{-\omega_2t-i\nu t}}{\omega_2+i\nu}\right)\right]\EEQ
Now let us remember that for $t<0$ the spring constant was $a_0$,
while for $t>0$ it is $a_1$.
Thus one could write in (\ref{e1}): $a\to a(t)=a_0+(a_1-a_0)\theta(t)$. By 
treating the term $(a_0-a_1)\theta(t)x(t)$ together with $\eta(t)$, one 
can read off the formal solution from the analog of Eq. (\ref{xt=}), and 
solve it perturbatively to first order in $(a_1-a_0)$. It can be verified
that the result coincides with (\ref{bnu01=}).

This first order perturbation theory can immediately be 
generalized for many steps,
\BEQ a(t)=a_k=(1-\alpha_k)a,\qquad (t_k<t<t_{k+1}) \EEQ
where $t_0=-\infty$ and $t_1$ was taken equal to zero so far, 
but can be arbitrary.
One writes $a(t)=a_0+\sum_{k\ge 1}(a_k-a_{k-1})\theta(t-t_k)$ and
gets \BEQ \label{bnu0k=}
\beta(\nu)=\sin\phi_0(\nu)\,e^{i\nu t}\,\,
\left[1-\sum_{k\ge 1}
\frac{a_k-a_{k-1}}{\gamma w}\,\theta(t-t_k)
\left(\frac{1-e^{-(\omega_1+i\nu) (t-t_k)}}{\omega_1+i\nu}
-\frac{1-e^{-(\omega_2+i\nu) (t-t_k)}}{\omega_2+i\nu}\right)\right]\EEQ
At a given instant of time the sum has a finite number 
of terms because of the $\theta$ functions.

In Eq. (\ref{bnu0k=}) we considered that effect of many small
changes in the spring constant.
When we make many changes with small $\alpha(t)=\alpha_k$ 
in the domain $t_k<t<t_{k+1}$, we get 
\BEQ \langle x^2\rangle=\frac{T_x(a)}{a}+\frac{\hbar }{\pi\gamma}
[\alpha_kC_x(0)-
\sum_{j=1}^{k}(\alpha_j-\alpha_{j-1})C_x(\frac{\gamma(t-t_j)}{2m})\,]\EEQ
where $t_1=t_i$ is the moment of the first change, taken equal to zero so far. 
Let us write $\alpha_k=\alpha(\tau_k)$ assume
and that the changes are small. Then the sum can be replaced by an integral, 
\BEQ \label{Vtcont}
2V(t)=a\langle x^2(t)\rangle=T_x(a)+\frac{\hbar a}{\pi\gamma }
[\alpha(\frac{\gamma t}{2m})C_x(0)-
\int_0^\infty\d \tau\,\alpha'(\frac{\gamma t}{2m}-\tau)
C_x(\tau)\,]\EEQ
where the upper integration border could be put equal to $\infty$,
since $\alpha'$ vanishes for times less than $t_i$.

The work needed to make the change $a_{k-1}\to a_k$ at time $t_k$ is
equal to $(a_{k}-a_{k-1})\langle x^2(t_k)\rangle$, so it equals
\BEQ \d \W_k=(\alpha_{k-1}-\alpha_k)\left\{
\half T_x(a)+\frac{\hbar a}{2\pi\gamma }\left[\alpha_kC_x(0)-
\sum_{j=1}^{k}(\alpha_j-\alpha_{j-1})C_x(\frac{\gamma(t_k-t_j)}{2m})
\right]\right\}\EEQ
If there are many steps with small increments, we can go to a continuum 
limit. Replacing the sum by an integral, we obtain
 the rate of work added to the system
\BEQ \frac{\d \W}{\d t}=\frac{\d \W_{\rm rev}}{\d t}+\frac{\d \Pi}{\d t}
\EEQ
where
\BEQ \frac{\d \W_{\rm rev}}{\d t}=
-\frac{\gamma}{2m}\, \frac{\d \alpha(\tau)}{\d \tau}
\left[\half T_x+\frac{\hbar a}{2\pi\gamma }\alpha(\tau)C_x(0)\right]
\EEQ
is the adiabatic (recoverable) rate of work and
\BEQ \label{dPIdtau=} \frac{\d \Pi}{\d t}=
\frac{\hbar a}{4\pi m}\,\frac{\d \alpha(\tau)}{\d \tau}\,
\int_0^\infty\d \tau\,\alpha'(\frac{\gamma t}{2m}-\tau) C_x(\tau)
\EEQ
is the rate of energy dispersion.

\renewcommand{\thesection}{\arabic{section}}
\section*{Appendix B: Moderate cutoff frequency and finite change of 
spring constant}
\renewcommand{\thesection}{\arabic{section}.}
\renewcommand{\theequation}{B.\arabic{equation}}
\setcounter{equation}{0}

In this section we address the vanishing of work dispersion at $T=0$,
without making the approximation of large Debye frequency $\Gamma$.
Then the full equation (\ref{betanut=}) has to be 
employed, rather than the approximation (\ref{bnu0=}).
With help of an algebraic manipulation program we have checked 
that at $T=0$ the important findings $C_x^{(0)}=0$,
see Eq. (\ref{kobalt}), and 
$C_p^{(0)}=C_E^{(0)}=0$ (c.f. Eq. (\ref{CE012=})) remain valid then.
So a negative rate of energy dispersion 
occurs also for a finite cutoff. 

Let us mention, however,
that the effect is weakened when $\alpha_0$, the amplitude of 
the change of the spring constant, is not very small. This probably
affects the maximal number of work extraction cycles.

If one changes the mass and not the spring constant, the system 
does not exhibit this interesting behavior, since the 
analog is $C_x^{(0)}$ does not
vanish then, implying that the leading term in the energy
dispersion does not vanish at low $T$.


\begin{thebibliography}{99}
\bibitem{superbath} Notice that in the present paper we completely neglected
situations, where the bath is an open system itself. Then it can have other
correlation times different compared with $\tau _{\hbar}$. In particular,
those times can stay finite even for $T\to 0$. The simplest realization
is to couple each bath oscillator to a `superbath', consisting again of
harmonic oscillators. This approach was worked out for 
a model used for the quantum measurement process, see ~\cite{b}.
\bibitem{nonaddS}
A  celebrated example to verify the non-additivity of the 
entropy is the case of two
spin $\half$ particles, in the pure $s=0$ state, $|\psi\rangle=
\frac{1}{\sqrt{2}}
(|\uparrow\uparrow\rangle-|\downarrow\downarrow\rangle)$. 
The density matrix reads
$$ \rho=\frac{1}{2}
|\uparrow\uparrow\rangle\langle\uparrow\uparrow|-
\frac{1}{2}|\uparrow\uparrow\rangle\langle\downarrow\downarrow|-
\frac{1}{2}|\downarrow\downarrow\rangle\langle\uparrow\uparrow|+
\frac{1}{2}|\downarrow\downarrow\rangle\langle\downarrow\downarrow|
$$
When summing out spin 2 one has
$$ \rho^{(1)}={\rm tr}_2\,\rho=
\frac{1}{2}
|\uparrow\cdot\rangle\langle\uparrow\cdot|+
\frac{1}{2}|\downarrow\cdot\rangle\langle\downarrow\cdot|
$$
The entropy of that state is 
$S_{1}=-{\rm tr}\rho^{(1)}\ln\rho^{(1)}=\ln 2$. 
Likewise, it is seen that $S_2=\ln 2$, so $S_1+S_2=2\ln 2$.
But $S=-{\rm tr}\rho\ln\rho=0$, since $\rho=$ 
diag$(1,0,0,0)$ in the basis of the total spin
$(s=m=0)$, $(s=1; m=0,\pm 1)$. This non-additivity 
remains when both systems become large.
\end{thebibliography}
\end{document}